 \documentclass[aps,prd,preprint,floatfix,nofootinbib,superscriptaddress,showpacs,amssymb]{revtex4}
 \usepackage{graphicx}
 \usepackage{graphicx}
 \usepackage{type1cm}
 \def\deg{^\circ}
\def\deg{^\circ}

\def\Z0{${\em Z^0\/}$}

\def\r#1 {$^{#1}$}

\newcommand{\gevc} { {\rm GeV/}c}
\newcommand{\gevcc}{ {\rm GeV/}c^2}

\def\gepsfcentered#1{
  \def\testit{#1}
  \def\lbracket{[}
  \ifx\testit\lbracket
    \let\dofilecmd=\gepsfwithopt
  \else
    \let\dofilecmd=\gepsfnoopt
  \fi
  \dofilecmd}

\def\gepsfnoopt#1{
  \begin{center}
  \leavevmode
  \epsffile{#1}
  \end{center}}

\def\gepsfwithopt#1 #2 #3 #4]#5{
  \begin{center}
  \leavevmode
  \gepsfmaxx=0.94\textwidth
  \epsffile[#1 #2 #3 #4]{#5}
  \end{center}}

\newdimen\gepsfmaxx
\def\epsfsize#1#2{
  \ifnum \epsfxsize=0
    \ifnum \epsfysize=0
      \ifnum #1 > \gepsfmaxx
        \gepsfmaxx
      \else
        #1
      \fi
    \else
      \epsfxsize
    \fi
  \else
    \epsfxsize
  \fi
}

\begin{document}

%\pagewiselinenumbers
%\doublespace

 \bibliographystyle{apsrev}

%%%%%%%%%%%%%%%%%%%%%%%%%%%%%%%%%%%%%%%%%%%%%%%%%%%%%
 \title {
 Study of multi-muon events produced in {\boldmath $p\bar{p}$}
 collisions  at {\boldmath $\sqrt{s}=1.96$} TeV 
}
%%%%%%%%%%%%%%%%%%%%%%%%%%%%%%%%%%%%%%%%%%%%%%%%%%%%%
\affiliation{Institute of Physics, Academia Sinica, Taipei, Taiwan 11529, Republic of China} 
\affiliation{Argonne National Laboratory, Argonne, Illinois 60439} 
\affiliation{University of Athens, 157 71 Athens, Greece} 
\affiliation{Baylor University, Waco, Texas  76798} 
\affiliation{Istituto Nazionale di Fisica Nucleare Bologna, $^w$University of Bologna, I-40127 Bologna, Italy} 
\affiliation{Brandeis University, Waltham, Massachusetts 02254} 
\affiliation{University of California, Davis, Davis, California  95616}
\affiliation{University of California, Santa Barbara, Santa Barbara, California 93106} 
\affiliation{Instituto de Fisica de Cantabria, CSIC-University of Cantabria, 39005 Santander, Spain} 
\affiliation{Carnegie Mellon University, Pittsburgh, PA  15213} 
\affiliation{Enrico Fermi Institute, University of Chicago, Chicago, Illinois 60637} 
\affiliation{Comenius University, 842 48 Bratislava, Slovakia; Institute of Experimental Physics, 040 01 Kosice, Slovakia} 
\affiliation{Joint Institute for Nuclear Research, RU-141980 Dubna, Russia} 
\affiliation{Duke University, Durham, North Carolina  27708} 
\affiliation{Fermi National Accelerator Laboratory, Batavia, Illinois 60510} 
\affiliation{University of Florida, Gainesville, Florida  32611} 
\affiliation{Laboratori Nazionali di Frascati, Istituto Nazionale di Fisica Nucleare, I-00044 Frascati, Italy} 
\affiliation{University of Geneva, CH-1211 Geneva 4, Switzerland} 
\affiliation{Glasgow University, Glasgow G12 8QQ, United Kingdom} 
\affiliation{Harvard University, Cambridge, Massachusetts 02138} 
\affiliation{Division of High Energy Physics, Department of Physics, University of Helsinki and Helsinki Institute of Physics, FIN-00014, Helsinki, Finland} 
\affiliation{University of Illinois, Urbana, Illinois 61801} 
\affiliation{The Johns Hopkins University, Baltimore, Maryland 21218} 
\affiliation{Center for High Energy Physics: Kyungpook National University, Daegu 702-701, Korea; Seoul National University, Seoul 151-742, Korea; Sungkyunkwan University, Suwon 440-746, Korea; Korea Institute of Science and Technology Information, Daejeon, 305-806, Korea; Chonnam National University, Gwangju, 500-757, Korea} 
\affiliation{University College London, London WC1E 6BT, United Kingdom} 
\affiliation{Centro de Investigaciones Energeticas Medioambientales y Tecnologicas, E-28040 Madrid, Spain} 
\affiliation{Massachusetts Institute of Technology, Cambridge, Massachusetts  02139} 
\affiliation{University of Michigan, Ann Arbor, Michigan 48109} 
\affiliation{Michigan State University, East Lansing, Michigan  48824}
%\affiliation{Institution for Theoretical and Experimental Physics, ITEP, Moscow 117259, Russia} 
\affiliation{University of New Mexico, Albuquerque, New Mexico 87131} 
\affiliation{Northwestern University, Evanston, Illinois  60208} 
\affiliation{The Ohio State University, Columbus, Ohio  43210} 
\affiliation{Okayama University, Okayama 700-8530, Japan} 
\affiliation{Osaka City University, Osaka 588, Japan} 
\affiliation{Istituto Nazionale di Fisica Nucleare, Sezione di Padova-Trento, $^x$University of Padova, I-35131 Padova, Italy} 
\affiliation{LPNHE, Universite Pierre et Marie Curie/IN2P3-CNRS, UMR7585, Paris, F-75252 France} 
\affiliation{Istituto Nazionale di Fisica Nucleare Pisa, $^y$University of Pisa, $^z$University of Siena and $^{aa}$Scuola Normale Superiore, I-56127 Pisa, Italy} 
\affiliation{University of Pittsburgh, Pittsburgh, Pennsylvania 15260} 
\affiliation{Purdue University, West Lafayette, Indiana 47907} 
\affiliation{University of Rochester, Rochester, New York 14627} 
\affiliation{The Rockefeller University, New York, New York 10021} 
\affiliation{Istituto Nazionale di Fisica Nucleare Trieste/Udine, $^{cc}$University of Trieste/Udine, Italy} 
\affiliation{Tufts University, Medford, Massachusetts 02155} 
\affiliation{Waseda University, Tokyo 169, Japan} 
\affiliation{Wayne State University, Detroit, Michigan  48201} 
\author{T.~Aaltonen}
\affiliation{Division of High Energy Physics, Department of Physics, University of Helsinki and Helsinki Institute of Physics, FIN-00014, Helsinki, Finland}
\author{J.~Adelman}
\affiliation{Enrico Fermi Institute, University of Chicago, Chicago, Illinois 60637}
\author{B.~\'{A}lvarez~Gonz\'{a}lez}
\affiliation{Instituto de Fisica de Cantabria, CSIC-University of Cantabria, 39005 Santander, Spain}
\author{S.~Amerio$^x$}
\affiliation{Istituto Nazionale di Fisica Nucleare, Sezione di Padova-Trento, $^x$University of Padova, I-35131 Padova, Italy} 
\author{D.~Amidei}
\affiliation{University of Michigan, Ann Arbor, Michigan 48109}
\author{A.~Anastassov}
\affiliation{Northwestern University, Evanston, Illinois  60208}
\author{J.~Antos}
\affiliation{Comenius University, 842 48 Bratislava, Slovakia; Institute of Experimental Physics, 040 01 Kosice, Slovakia}
\author{G.~Apollinari}
\affiliation{Fermi National Accelerator Laboratory, Batavia, Illinois 60510}
\author{A.~Apresyan}
\affiliation{Purdue University, West Lafayette, Indiana 47907}
\author{T.~Arisawa}
\affiliation{Waseda University, Tokyo 169, Japan}
% ...

\collaboration{CDF Collaboration\footnote{With visitors from $^a$University of Massachusetts Amherst, Amherst,
 Massachusetts 01003,
$^b$Universiteit Antwerpen, B-2610 Antwerp, Belgium, 
$^c$University of Bristol, Bristol BS8 1TL, United Kingdom,
$^d$Chinese Academy of Sciences, Beijing 100864, China, 
$^e$Istituto Nazionale di Fisica Nucleare, Sezione di Cagliari, 09042 Monserrato (Cagliari), Italy,
$^f$University of California Irvine, Irvine, CA  92697, 
$^g$University of California Santa Cruz, Santa Cruz, CA  95064, 
$^h$Cornell University, Ithaca, NY  14853, 
$^i$University of Cyprus, Nicosia CY-1678, Cyprus, 
$^j$University College Dublin, Dublin 4, Ireland,
$^k$Royal Society of Edinburgh/Scottish Executive Support Research Fellow,
$^l$University of Edinburgh, Edinburgh EH9 3JZ, United Kingdom, 
$^m$Universidad Iberoamericana, Mexico D.F., Mexico,
$^n$Queen Mary, University of London, London, E1 4NS, England,
$^o$University of Manchester, Manchester M13 9PL, England, 
$^p$Nagasaki Institute of Applied Science, Nagasaki, Japan, 
$^q$University of Notre Dame, Notre Dame, IN 46556,
$^r$University de Oviedo, E-33007 Oviedo, Spain, 
$^s$Simon Fraser University, Vancouver, British Columbia, Canada V6B 5K3,
$^t$Texas Tech University, Lubbock, TX  79409, 
$^u$IFIC(CSIC-Universitat de Valencia), 46071 Valencia, Spain,
$^v$University of Virginia, Charlottesville, VA  22904,
$^{dd}$On leave from J.~Stefan Institute, Ljubljana, Slovenia, 
}}
\noaffiliation

\noaffiliation
%\collaboration{CDF collaboration}
%%%%%%%%%%%%%%%%%%%%
 \begin{abstract}
 We report a study of multi-muon events produced at the
 Fermilab Tevatron collider and recorded by the CDF~II detector. In a data 
 set acquired with a dedicated dimuon trigger and corresponding to an 
 integrated luminosity of 2100 pb$^{-1}$, we isolate a significant sample of 
 events in which at least one of the muon candidates is produced 
 outside of the beam pipe of radius 1.5 cm. The production cross section
 and kinematics of events in which both muon candidates are produced inside
 the beam pipe are successfully modeled by known QCD processes which
 include heavy flavor production. In contrast, we are presently unable to 
 fully account for the number and properties of the remaining events, in which
 at least one muon candidate is produced outside of the beam pipe, in terms
 of the same understanding of the CDF~II detector, trigger, and event 
 reconstruction. Several topological and kinematic properties of these 
 events are presented in this paper. These events offer a plausible 
 resolution to long-standing inconsistencies related to $b\bar{b}$
 production and decay.     
 \end{abstract} 
%%%%%%%%%%%%%%%%%%%%
 \pacs{13.85.-t, 14.65.Fy,  14.80.-j }
 \preprint{FERMILAB-PUB-08-046-E}
 \maketitle
%%%%%%%%%%%%%%%%%%%%%%%%%%%%%%%%%%%%%%%%%%%%%%%%%%
 \section {Introduction}  \label{sec:ss-intro}
%%%%%%%%%%%%%%%%%%%%%%%%%%%%%%%%%%%%%%%%%%%%%%%%%%
 This article presents the study of events, acquired with a dedicated
 dimuon trigger, that we are currently unable to fully explain with our
 understanding of the CDF~II detector, trigger, and event reconstruction.
 We are continuing detailed studies with a longer timescale for completion,
 but we present here our current findings.
 
 This study was  motivated by the presence of several inconsistencies that
 affect or affected measurements of the $b\bar{b}$ production at the Tevatron:
 (a) the ratio of the observed  $b\bar{b}$ correlated production cross
 section to the exact next-to-leading-order (NLO) QCD prediction~\cite{mnr}
 is measured to be $R=1.15 \pm 0.21 $ when  $b$ quarks are selected via 
 secondary vertex identification, whereas this ratio is found to be
 significantly larger than two when identifying $b$ quarks through their
 semileptonic decays~\cite{bstatus}; (b) sequential semileptonic decays
 of single $b$ quarks are supposedly the main source of dileptons with
 invariant mass smaller than that of $b$ quarks, but the observed invariant
 mass spectrum is not well modeled by the simulation of this
 process~\cite{dilb}; and (c) the value of $\bar{\chi}$, the average
 time-integrated mixing probability of $b$ flavored hadrons, derived from
 the ratio of muon pairs from semileptonic decays of $b$ and $\bar{b}$
 quarks with opposite and same sign charge, is measured at hadron colliders
 to be significantly larger than that measured by the LEP 
 experiments~\cite{bmix,pdg}.
 
 The first inconsistency (a) has been addressed in a recent study
 of the CDF collaboration~\cite{bbxs}. That study uses a data sample
 acquired with a dedicated dimuon trigger to re-measure the correlated  
 $\sigma_{b\rightarrow\mu,\bar{b}\rightarrow \mu}$ cross section. As in
 previous studies~\cite{2mucdf,bmix}, Ref.~\cite{bbxs} makes use of the
 precision tracking provided by the CDF silicon microvertex detector to
 evaluate the fractions of muons due to the decays of long-lived $b$- 
 and $c$-hadrons, and to the other background contributions. The new 
 measurement is in good agreement with theoretical expectations
($R=1.20 \pm 0.21$), as well as
 with analogous measurements that identify $b$ quarks via secondary vertex
 identification. However, it is also substantially smaller than previous
 measurements of this cross section~\cite{2mucdf,d0b2}.
 The new
 CDF measurement~\cite{bbxs} requires that both trigger muons arise from
 particles that have decayed inside the beam pipe of 1.5 cm radius. According
 to the simulation, approximately 96\% of the known sources of dimuons, 
 such as Drell-Yan, $\Upsilon$, $Z^0$, and heavy flavor production, satisfy
 this condition. We will show that not only the rate, but also the kinematic
 properties of the events that satisfy this condition are correctly modeled
 by the simulation of known processes. However, this article also presents
 the observation of a much larger than expected sample of events that does
 not satisfy this condition. This component, which was present in previous
 measurements in which this decay-radius requirement was not made, will be
 described and investigated at length in this article.

 We utilize the same dimuon data set,  simulated samples, and  
 analysis tools described in  Ref.~\cite{bbxs}. Section~\ref{sec:ss-det}
 describes the detector systems relevant to this analysis. The data selection
 and Monte Carlo simulation are briefly summarized in Sec.~\ref{sec:ss-anal}.
 Section~\ref{sec:ss-inv} investigates differences in the experimental methods
 used to derive  $\sigma_{b\rightarrow\mu,\bar{b}\rightarrow \mu}$
 in Ref.~\cite{bbxs} and in previous measurements, and isolates a larger
 than expected sample of events in which at least one muon candidate
 is produced beyond the beam pipe. Section~\ref{sec:ss-3mu} connects the
 presence of these events to the discrepancy between the observed and predicted
 invariant mass spectrum of lepton pairs produced by single $b$ quark 
 sequential decays. The properties of these events are explored in 
 Secs.~\ref{sec:ss-a0} and~\ref{sec:ss-inter}. Our conclusions are summarized
 in Sec.~\ref{sec:ss-concl}. 
%%%%%%%%%%%%%%%%%%%%%%%%%%%%%%%%%%%%%%
 \section{CDF II detector and trigger} \label{sec:ss-det}
%%%%%%%%%%%%%%%%%%%%%%%%%%%%%%%
 CDF~II is a multipurpose detector, equipped with a charged particle
 spectrometer and a finely segmented calorimeter. In this section, we 
 describe the detector components that are relevant to this analysis.
 The description of these subsystems can be found in
 Refs.~\cite{det1,det2,det3_0,det3,det4_0,det4,det5,det6,det7,det8}.
 Two devices inside the 1.4 T solenoid are used for measuring the momentum
 of charged particles: the silicon vertex detector (SVXII and ISL) and the
 central tracking chamber (COT). The SVXII detector consists of 
 microstrip sensors arranged in six cylindrical shells with radii between
 1.5 and 10.6 cm, and with a total $z$ coverage~\footnote{
 In the CDF coordinate system, $\theta$ and $\phi$ are the polar and
 azimuthal angles of a track, respectively, defined with respect to the
 proton beam direction, $z$. The pseudorapidity $\eta$ is defined as 
 $-\ln \;\tan (\theta/2)$. The transverse momentum of a particle is 
 $p_T= p \; \sin (\theta)$. The rapidity is defined as 
 $y=1/2 \cdot \ln ( (E+p_z)/(E-p_z) )$, where $E$ and $p_z$ are the 
 energy and longitudinal momentum of the particle associated with the track.} 
 of 90 cm. The first SVXII layer, also referred to as the L00 detector,
 is made of single-sided sensors mounted on the beryllium beam pipe.
 The remaining five SVXII layers are made of double-sided sensors and 
 are divided into three contiguous five-layer sections along the beam 
 direction $z$. The vertex $z$-distribution for $p\bar{p}$ collisions is
 approximately described by a Gaussian function with a rms of 28 cm.
 The transverse profile of the Tevatron beam is circular and has a rms
 spread of $\simeq 25\; \mu$m in the horizontal and vertical directions.
 The SVXII single-hit resolution is approximately $11\; \mu$m and allows
 a track impact parameter~\footnote{
 The impact parameter $d$ is the distance of closest approach of a track
 to the primary event vertex in the transverse plane.}
 resolution of approximately $35\; \mu$m, when
 also including the effect of the beam transverse size. The two additional 
 silicon layers of the ISL help to link tracks in the COT to hits in the
 SVXII. The COT is a cylindrical drift chamber containing 96 sense wire
 layers grouped into eight alternating superlayers of axial and stereo
 wires. Its active volume covers $|z| \leq 155$ cm and 40 to 140 cm in
 radius. The transverse momentum resolution of tracks reconstructed using
 COT hits is $\sigma(p_T)/p_T^2 \simeq 0.0017\; [\gevc]^{-1}$. The trajectory
 of COT tracks is extrapolated into the SVXII detector, and tracks are
 refitted with additional silicon hits consistent with the track extrapolation.

 The central muon detector (CMU) is located around the central electromagnetic
 and hadronic calorimeters, which have a thickness of 5.5 interaction lengths
 at normal incidence. The CMU detector covers a nominal pseudorapidity range
 $|\eta| \leq 0.63$ relative to the center of the detector, and  is segmented
 into two barrels of 24 modules, each covering 15$^\circ$ in $\phi$. Every
 module is further segmented  into three  submodules, each covering
 4.2$^\circ$ in $\phi$ and consisting of four layers of drift chambers.
 The smallest drift unit, called a stack, covers a 1.2$^\circ$ angle in
 $\phi$. Adjacent pairs of stacks are combined together into a tower.
 A track segment (hits in two out of four layers of a stack) detected in
 a tower is referred to as a CMU stub. A second set of muon drift chambers
 (CMP) is located behind an additional steel absorber of 3.3 interaction
 lengths. The chambers are 640 cm long and are arranged axially to form a
 box around the central detector. The CMP detector covers a nominal
 pseudorapidity range $|\eta| \leq 0.54$ relative to the center of the
 detector. Muons which produce a stub in both the CMU and CMP systems are
 called CMUP muons. The CMX muon detector consists of eight drift chamber
 layers and scintillation counters positioned behind the hadron calorimeter.
 The CMX detector extends the muon coverage to $|\eta| \leq 1$ relative to
 the center of the detector.

 The luminosity is measured using gaseous Cherenkov counters (CLC) that
 monitor the rate of inelastic $p\bar{p}$ collisions. The inelastic 
 $p\bar{p}$ cross section at $\sqrt{s}=1960$ GeV is scaled from measurements
 at $\sqrt{s}=1800$ GeV using the calculations in Ref.~\cite{sigmatot}.
 The integrated luminosity is determined with a 6\% systematic
 uncertainty~\cite{klimen}.
 
 CDF uses a three-level trigger system. At Level 1 (L1), data from every
 beam crossing are stored in a pipeline capable of buffering data from 42
 beam crossings. The L1 trigger either rejects events or copies them into
 one of the four Level 2 (L2) buffers. Events that pass the L1 and L2 
 selection criteria are sent to the Level 3 (L3) trigger, a cluster of
 computers running  speed-optimized reconstruction code.  

 For this study, we select events with two muon candidates identified by the
 L1 and L2 triggers. The L1 trigger uses tracks with $p_T \geq 1.5 \; \gevc$
 found by a fast track processor (XFT). The XFT examines COT hits from the
 four axial superlayers and provides $r-\phi$ information in azimuthal
 sections of 1.25$^\circ$. The XFT passes the track information to a set of
 extrapolation units that determine the CMU towers in which a CMU stub  
 should be found if the track is a muon. If a stub is found, a L1 CMU
 primitive is generated. The L1 dimuon trigger requires at least two CMU
 primitives, separated by at least two CMU towers. The L2 trigger 
 additionally requires that at least one of the muons also has a CMP stub
 matched to an XFT track with $p_T \geq 3 \;\gevc$. All these trigger
 requirements are emulated by the detector simulation on a run-by-run basis.
 The L3 trigger requires a pair of CMUP muons with invariant mass larger
 than $5 \; \gevcc$, and $|\delta z_0| \leq 5$ cm, where $z_0$ is the $z$
 coordinate of the muon track at its point of closest approach to the beam 
 line in the $r-\phi$ plane. These requirements define the dimuon trigger
 used in this analysis.

 Two  other triggers are also utilized to acquire calibration samples
 used in this analysis. We use events acquired requiring a L1 CMUP
 primitive with $p_T \geq 4 \; \gevc$ accompanied by a L2 requirement
 of an additional track with $p_T \geq 2\; \gevc$ and impact parameter
 $0.12\leq d \leq1$ mm as measured by the Silicon Vertex Trigger 
 (SVT)~\cite{svt}. The SVT calculates the impact parameter of each XFT
 track, with respect to the beam line, with a 50 $\mu$m resolution that
 includes the 25 $\mu$m contribution of the beam transverse width. Events
 selected with this trigger, referred to as $\mu-$SVT, are used to verify
 the muon detector acceptance and the muon reconstruction efficiency. 
 We use an additional trigger, referred to as {\sc charm}, that acquires
 events with two SVT tracks with $p_T\geq 2\; \gevc$ and with impact
 parameter $0.12\leq d\leq 1.00$ mm. In this data sample, we reconstruct
 $D^0 \rightarrow K^- \pi^+$ decays to measure the probability that the
 punchthrough of a charged hadron mimics a muon signal.
%%%%%%%%%%%%%%%%%%%%%%%%%%%%%
%%%%%%%%%%%%%%%%%%%%%%%%%%%%%%%
 \section{Data selection and Monte Carlo simulations} \label{sec:ss-anal}
%%%%%%%%%%%%%%%%%%%%%%%%%%%%%%%
 This study starts using the same data set and analysis selection criteria
 employed in the measurement of the correlated $b\bar{b}$ cross 
 section~\cite{bbxs}, that corresponds to an integrated luminosity of 
 742  pb$^{-1}$. When extending the scope of that analysis, we also use larger
 data sets corresponding to integrated luminosities of 1426 and 2100 pb$^{-1}$.
 The correlated $b\bar{b}$ cross section measurement selects events acquired 
 with the dimuon trigger and which contain at least two CMUP muons with same
 or opposite sign charge. If events contain more than two muons that pass our
 selection cuts, the two with the highest transverse momenta, referred to as
 initial muons, are considered. Events are reconstructed offline taking 
 advantage of more refined calibration constants and reconstruction
 algorithms than those used by the L3 trigger. COT tracks are extrapolated 
 into the SVXII detector, and refitted adding hits consistent  with the 
 track extrapolation. Stubs reconstructed in the CMU and CMP detectors are
 matched to tracks with  $p_T \geq 3 \; \gevc$. A track is identified as a
 CMUP muon if $\Delta r\phi$, the distance in the $r-\phi$ plane between 
 the track projected to the CMU (CMP) chambers and a CMU (CMP) stub, is 
 less than 30 (40) cm. We require that muon-candidate stubs correspond to 
 a L1 CMU primitive, and correct the muon momentum for energy losses in the
 detector. We also require the $z_0$ distance between two muon candidates to
 be smaller than 1.5 cm. We reconstruct primary vertices using all tracks 
 with SVXII hits that are consistent with originating from a common vertex.
 In events in which more than one interaction vertex has been reconstructed,
 we use the one closest in $z$ to the average of the muon track 
 $z_0$-positions and within a 6 cm distance. We evaluate the impact
 parameter of each muon track with respect to the primary vertex.
 The primary vertex coordinates transverse to the beam direction are
 measured with an accuracy of approximately 3 $\mu$m~\cite{bbxs}.
 Cosmic rays are removed by requiring that the azimuthal angle between muons
 with opposite charge is smaller than 3.135 radians. Muon pairs arising from
 cascade decays of a single $b$ quark are removed by selecting dimuon 
 candidates with invariant mass greater than 5 $\gevcc$. We also reject
 muon pairs with invariant mass larger than 80 $\gevcc$ that are mostly
 contributed by $Z^0$ decays. The data sample that survives these 
 selection criteria consists of 743006 events.

 In this study, data are compared to different simulated samples. The
 heavy flavor production is  simulated with the {\sc herwig} Monte Carlo
 program~\cite{herwig}, the settings of which are described in Appendix~A
 of Ref.~\cite{bbxs}. Hadrons with heavy flavors  are subsequently 
 decayed using the {\sc evtgen} Monte Carlo program~\cite{evtgen}. 
 The detector response to particles produced by the above generators
 is modeled with the CDF~II detector simulation that in turn is based on 
 the {\sc geant} Monte Carlo program~\cite{geant}.

%%%%%%%%%%%%%%%%%%%%%%%%%%%%%%%%%%%%%%%%%%%
\section{Study of the data sample composition}\label{sec:ss-inv}
\begin{figure}[htb]
 \begin{center}
 \vspace{-0.2in}
 \leavevmode
 \includegraphics*[width=\textwidth]{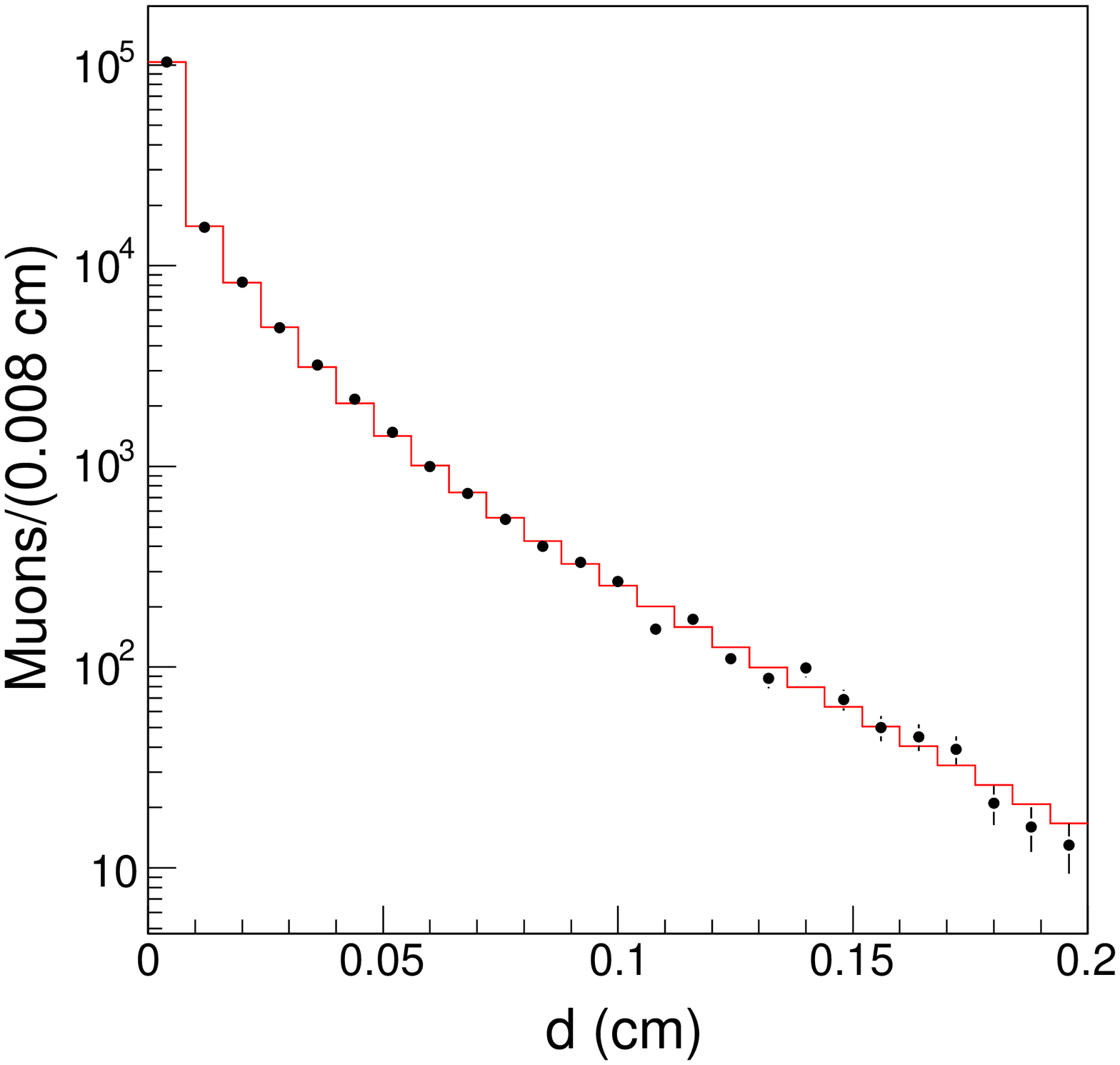}
 \caption[]{The projection of the  two-dimensional impact parameter
            distribution of muon pairs onto one of the two axes is 
            compared to the fit result (histogram).}
 \label{fig:figbb_6}
 \end{center}
 \end{figure}

%%%%%%%%%%%%%%%%%%%%%%%%%%%%%%%%%%%%%%%%%%%
 The procedure to extract $\sigma_{b\rightarrow\mu,\bar{b}\rightarrow \mu}$
 from the data is to fit the observed impact parameter distributions of the 
 selected muon pairs with the expected impact parameter distributions of muons
 from various sources. To ensure an accurate impact parameter measurement,
 Ref.~\cite{bbxs} requires that each muon track is reconstructed in the 
 SVXII detector with hits in the two inner layers and in at least two of
 the remaining four external layers. These SVXII quality requirements 
 reduce the data sample to 143743 events. After this selection, the dominant
 sources of reconstructed muons are semileptonic decays of bottom and
 charmed hadrons, prompt decays of quarkonia, Drell-Yan production, and
 muons mimicked by prompt hadrons or hadrons arising from heavy flavor 
 decays. In the following, the sum of these contributions will be 
 referred to as QCD production. Monte Carlo simulations are used to model
 the impact parameter distributions of muons from $b$- and $c$-hadron decays. 
 The impact parameter distribution of muons from prompt sources, such as
 quarkonia decays and Drell-Yan production, is constructed using muons
 from $\Upsilon(1S)$ decays. The sample composition determined by the fit
 is shown in Table~\ref{tab:tab_1}.
%%%%%%%%%%%%%%%%%%%%%%%%%%%%%%%%%%
 \begin{table}
 \caption[]{Number of events attributed to the different dimuon sources by
           the fit to the muon impact-parameter distribution in the range
           $0-0.2$ cm.  The fit parameters $BB$, $CC$, and $PP$ represent 
           the $b\bar{b}$, $c\bar{c}$, and prompt dimuon contributions, 
           respectively. The component $BC$ represents events containing
           $b$ and $c$ quarks. The fit parameter $BP$ ($CP$) estimates
           the number of events in which there is only one $b$ ($c$) quark
           in the detector acceptance and the second muon is produced by  
           misidentified prompt hadrons. The data correspond to an
           integrated luminosity of 742 pb$^{-1}$. }
 \begin{center}
 \begin{ruledtabular}
 \begin{tabular}{lc}
 Component  &  No. of Events      \\
  $BB$      & $54583 \pm ~678$    \\ 
  $CC$      & $24458 \pm 1565$    \\
  $PP$      & $41556 \pm ~651$    \\
  $BP$      & $10598 \pm ~744$    \\
  $CP$      & $10024 \pm 1308$    \\
  $BC$      & $~2165 \pm ~693$    \\
 \end{tabular}
 \end{ruledtabular}
 \end{center}
 \label{tab:tab_1}
 \end{table}
%%%%%%%%%%%%%%%%%%%%%%%%%%%%%%%%%%%%%%%%%%%
 The projection of the two-dimensional impact parameter distribution is 
 compared to the fit result in Fig.~\ref{fig:figbb_6}.
%%%%%%%%%%%%%%%%%%%%%%%%%%
%%%%%%%%%%%%%%%%%%%%%%%%%%%%%%%%%%%%%%%%%%%%%%%%
 After removing the contribution of muons mimicked by hadrons from heavy 
 flavor decays, the study in Ref.~\cite{bbxs} determines the size of
 $b\bar{b}$ production to be $52400 \pm 2747$ events. For muons with
 $p_T \geq 3 \; \gevc$ and $|\eta| \leq 0.7$, Ref.~\cite{bbxs} reports
  $\sigma_{b\rightarrow\mu,\bar{b}\rightarrow \mu}= 1549 \pm 133$ pb.
 The ratio of this cross section to the NLO prediction ($1.20 \pm0.21$) is
 appreciably smaller than that reported in previous
 measurements~\cite{2mucdf,d0b2}, and in agreement with the correlated 
 $b \bar{b}$ cross section measurements that select $b$ quarks via
 secondary vertex identification ($1.15 \pm 0.21$)~\cite{ajets,shears}.
 This result mitigates previous inconsistencies between measurements and 
 theoretical predictions of the correlated $b\bar{b}$ cross section. 
 
\begin{figure}
 \begin{center}
 \vspace{-0.3in}
 \leavevmode
 \includegraphics*[width=\textwidth]{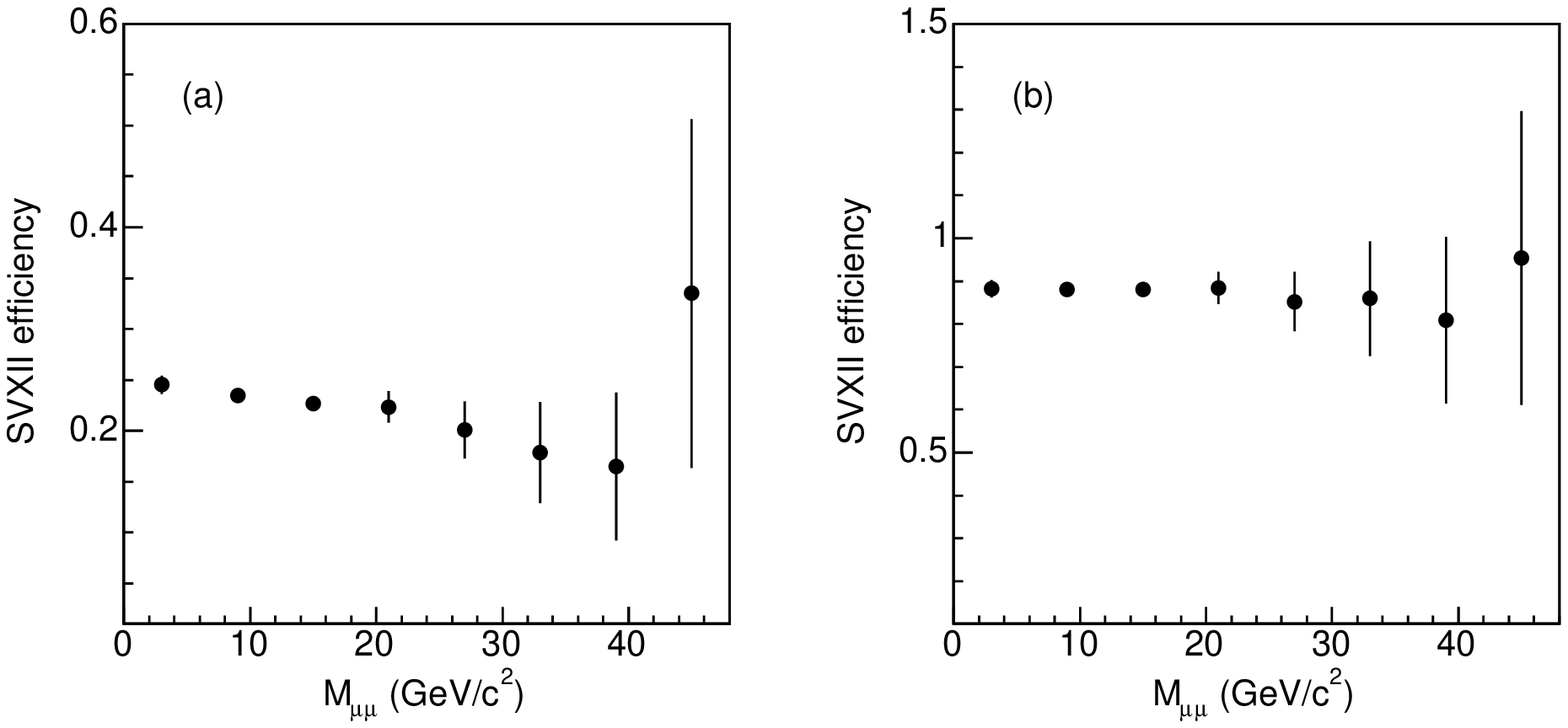}
 \caption[]{Efficiency of SVX tight (a) and loose (b) selection in simulated
            dimuon events due to heavy flavor production (see text). The 
            efficiency is shown as a function of the dimuon invariant mass.}
 \label{fig:fig_1}
 \end{center}
 \end{figure}

 However, a new problem arises that concerns the sample composition when
 the requirement that muons are accurately measured in the SVXII detector
 is released.
 The study in Ref.~\cite{bbxs} uses very strict selection criteria, referred to as
 tight SVX selection in the following, by requiring muon tracks with hits
 in the first two layers of the SVXII detector, and at least in two of 
 the remaining four outer layers. This requirement selects muon parent
 particles which decayed within a distance of $\simeq 1.5$ cm from the
 nominal beam line, or in other words inside the beam pipe. According
 to the simulation, approximately 96\% of dimuons due to known QCD
 processes, such as Drell-Yan, $\Upsilon$, $Z^0$, and heavy flavor 
 production, satisfy this latter condition. The efficiency of the tight
 SVX requirements for prompt dimuons is purely geometrical, and is
 measured to be $0.257 \pm 0.004$ by using $\Upsilon(1S)$ candidates
 ~\cite{bbxs}. For dimuons arising from heavy flavor production, the
 efficiency of the tight SVX selection is determined to be
 $0.237 \pm0.001$ by using muons from $J/\psi$ decays after reweighting
 their $p_T$ distribution to be equal to that of muons from simulated
 decays of heavy flavors. As shown by  Fig.~\ref{fig:fig_1}~(a), the
 7\% decrease of the efficiency for heavy flavors is due to a small 
 fraction of high-$p_T$ $b$ hadrons decaying after the first SVXII layer.
 Using the sample composition determined by the fit to the muon impact
 parameter distribution, listed in Table~\ref{tab:tab_1}, we estimate 
 that ($24.4 \pm 0.2$)\% of the initial data sample survives the tight
 SVX requirements.
%%%%%%%%%%%%%%%%%%%%%%%%%%
%%%%%%%%%%%%%%%%%%%%%%%%%

\begin{figure}[]
 \begin{center}
 \vspace{-0.2in}
 \leavevmode
 \includegraphics*[width=\textwidth]{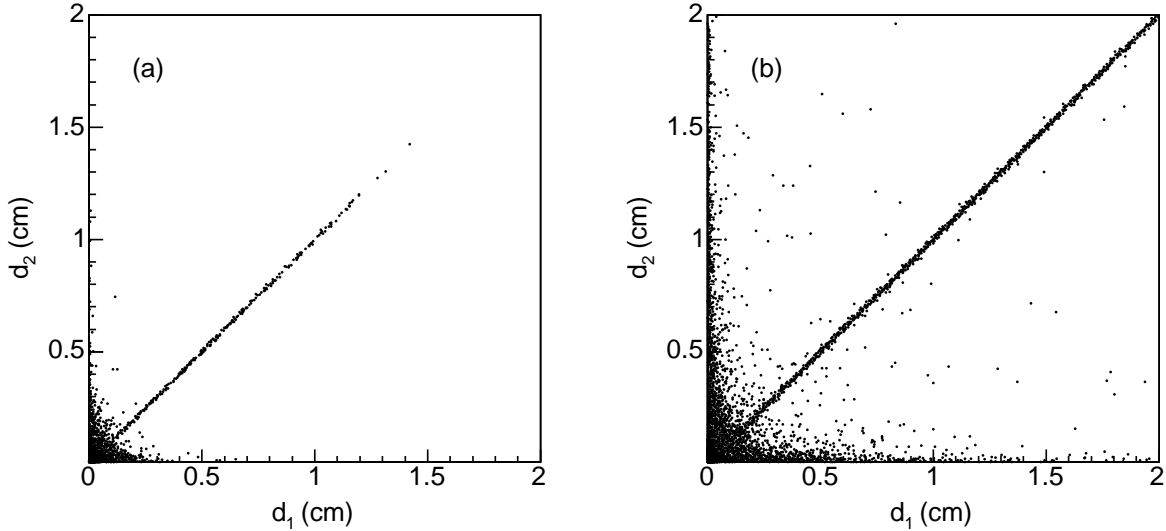}
 \caption[] {Two-dimensional  impact parameter distributions of muons that
             pass the (a) tight and (b) loose SVX requirements. Cosmic muons
             are reconstructed as two back-to-back muons of opposite charge 
             and cluster along the $d_1=d_2$ diagonal.}
 \label{fig:fig_cosm}
 \end{center}
 \end{figure}

 Analyses performed by the CDF collaboration customarily select tracks for
 secondary vertexing  purposes with less stringent requirements, such as 
 tracks with hits in at least three out of the eight layers of the SVXII 
 and ISL detectors (referred to as loose SVX selection in the following). 
 The latter selection accepts muons from parent particles with a decay 
 length as long as $\simeq 10$ cm. As shown by Fig.~\ref{fig:fig_1}~(b),
 in this case the SVX selection efficiency is much higher and does not 
 depend on the dimuon invariant mass. By using $\Upsilon(1S)$ and $J/\psi$ 
 candidates, we measure the efficiency of the loose SVX requirements to be 
 $0.88 \pm 0.01$. The acceptance of the different SVX selections as a 
 function of the decay length of the muon parent particle is verified
 using cosmic muons that overlap in time with a $p\bar{p}$ collision (for
 this purpose we remove the request that the azimuthal angle between two
 initial muons be less than 3.135 radians). Cosmic muons, which are 
 reconstructed as two back-to-back muons of opposite charge, cluster along
 the diagonal of the two-dimensional distribution of the muon impact  
 parameters. As shown in Fig.~\ref{fig:fig_cosm}, the loose SVX selection
 accepts larger decay lengths than the tight SVX selection. As shown by the
 scatter of the points along the $d_1=d_2$ diagonal, both SVX selections
 yield rms resolutions that are negligible on a scale of
 the order of centimeters.
%%%%%%%%%%%%%%%%%%%%%%%%%%
%%%%%%%%%%%%%%%%%%%%%%%%%%

 If the dimuon sample before the tight SVX selection (743006 events) had
 the same composition of the sample listed in Table~\ref{tab:tab_1} 
(143743 events), the average efficiency of the tight SVX requirements in
 this data set would be $0.244\pm0.002$, whereas it is found to be 
 $0.1930\pm0.0004$. This feature suggests the presence of a large
 background
 that, unlike the QCD contribution, is significantly suppressed by the 
 tight SVX selection.
 Because it went unnoticed for a long time, this background will be
 whimsically referred
 to as the ghost contribution in the following.
 In the assumption that the contribution of ghost
 events to the dimuon sample selected with tight SVX
 requirements is negligible, the size of the ghost sample can be estimated as
 the difference between the number of muon pairs prior to any SVX 
 requirements and the number of muons passing the tight SVX selection
 divided by the efficiency of the tight SVX requirements (see
 Table~\ref{tab:tab_2}). In Table~\ref{tab:tab_2}, the contribution of
 ghost events to dimuons that pass the loose SVX
 requirements is determined as the difference between the numbers of
 events that pass the loose SVX requirements and of events that pass 
 the tight SVX requirements, divided by the efficiency of the tight SVX
 requirements and multiplied by that of the loose SVX requirements.
 The size of the ghost sample (153895 $\pm$ 4829 events) is of
 a magnitude comparable to $b\bar{b}$ production ($221564 \pm 11615$ events).
 When using the loose SVX requirements, the size of the ghost
 sample is reduced by a factor of two, whereas 88\% of the dimuons
 due to known processes survive (the ghost size is 72553 $\pm$ 7264 events,
 whereas the $b\bar{b}$ contribution is 194976 $\pm$ 10221 events).  
 Muon pairs in the ghost sample are equally split in opposite
 and same sign charge combinations. 
%%%%%%%%%%%%%%%%%%%%%%%%%
 \begin{table}
 \caption[]{Number of events that pass different SVX requirements.
            QCD indicates the sum of the various components listed
            in Table~\ref{tab:tab_1}. Ghost indicates the additional
            background in the data. Dimuons are also split into pairs
            with opposite ($OS$) and same ($SS$) sign charge.}
 \begin{center}
 \begin{ruledtabular}
 \begin{tabular}{lccc}
  Type       & Total            & Tight SVX   & Loose SVX           \\
  All        & 743006             & 143743      & 590970            \\ 
  All $OS$   &                    &  98218      & 392020            \\
  All $SS$   &                    &  45525      & 198950            \\
  QCD        & 589111 $\pm$ 4829  & 143743      & 518417 $\pm$ 7264 \\    
  QCD $OS$   &                    &  98218      & 354228 $\pm$ 4963 \\
  QCD  $SS$  &                    &  45525      & 164188 $\pm$ 2301 \\
  Ghost      & 153895 $\pm$ 4829  &   0         & 72553  $\pm$ 7264 \\
  Ghost $OS$ &                    &   0         & 37792  $\pm$ 4963 \\
  Ghost $SS$ &                    &   0         & 34762  $\pm$ 2301 \\
 \end{tabular}
 \end{ruledtabular}
 \end{center}
 \label{tab:tab_2}
 \end{table}
%%%%%%%%%%%%%%%%%%%%%%%%%%%%%%%%%%%%%%%%%%%%%%%%%%%%%%%%%%%%%%

\begin{figure}
 \begin{center}
 \vspace{-0.3in}
 \leavevmode
 \includegraphics*[width=\textwidth]{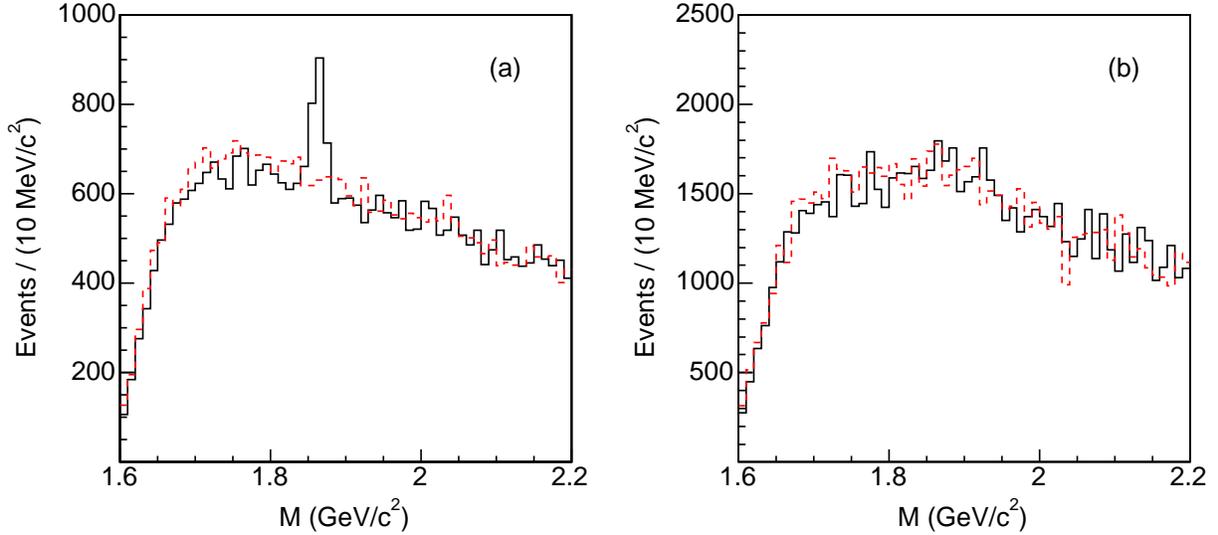}
 \caption[]{Invariant mass, $M$, distributions of $RS$ (histogram) and
            $WS$ (dashed histogram) $D^0$ candidates in (a) QCD and (b)
            ghost events.}
 \label{fig:fig_40}
 \end{center}
 \end{figure}

 We have investigated at length the possibility that ghost muons
 are a consequence of the experimental conditions of the present study.
 The appearance of ghost events does not depend on the instantaneous
 luminosity nor  the presence of multiple $p\bar{p}$ interactions.
 We have investigated in many ways the possibility that ghost events
 are ordinary QCD events in which one of the initial muons appears to
 originate beyond the beam-pipe radius because of pattern recognition
 problems in the SVX or COT detectors (see Appendix~A). As an example,
 we compare yields of $D^0 \rightarrow K^-\pi^+$ (and charge-conjugate)
 decays in QCD and ghost events. We search for $D^0$ candidates by using
 tracks of opposite sign charge, with $p_T\geq 1.0\; \gevc$,
 $|\eta| \leq 1.1$, and contained in a  60$^{\deg}$ cone around the
 direction of each initial muon. The two-track systems are constrained
 to arise from a common space point. Track combinations are discarded if
 the three-dimensional vertex fit returns a $\chi^2$ larger than 10 or if
 the vertex is not in the hemisphere containing the $D^0$ candidate.
 We attribute the kaon mass to the track with the same charge as the initial
 muon ($RS$ combination, as expected for $B \rightarrow \mu^- D^0$ decays).
 We also study wrong sign combinations ($WS$) attributing the kaon mass to
 the track with opposite charge. A $D^0$ signal found in the $WS$ combinations
 is a measure of the fraction of fake muons, whereas a $D^0$ signal in 
 $RS$ combinations found in ghost events indicates a heavy flavor contribution.
 As shown in Figs.~\ref{fig:fig_40}~(a) and~(b), a clear $D^0$ signal is
 observed in QCD but not in ghost events. It is our conclusion that ghost
 events are not due to track reconstruction failures in normal QCD events.
%%%%%%%%%%%%%%%%%%%%%%%%%%
%%%%%%%%%%%%%%%%%%%%%%%%

 The unnoticed presence of a ghost contribution of this size, that is
 incrementally reduced by stricter SVX requirements, can help explain
 the inconsistencies mentioned in the introduction. The general
 observation is that the measured
 $\sigma_{b\rightarrow\mu,\bar{b}\rightarrow \mu}$ increases as
 the SVX requirements are made looser and is almost a factor of two
 larger than that measured in Ref.~\cite{bbxs} when no SVX requirements
 are made~\cite{d0b2}. As mentioned above, the magnitude of the ghost
 contribution is comparable to the $b\bar{b}$ contribution when no 
 SVX selection is made and in combination would account for the
 measurement reported in~Ref.\cite{d0b2}. Similarly, for the loose SVX 
 criteria, the magnitude of the ghost contribution ($72553\pm 7264$ events
 equally split in $OS$ and $SS$ combinations), when added to
 the expected $b\bar{b}$ contribution of $194976 \pm 10221$ events,
 coincides with the cross section measurement reported in~Ref.\cite{2mucdf}
 and the $\bar{\chi}$ value reported in~Ref.\cite {bmix} since these
 measurements use similar sets of silicon criteria. 
%%%%%%%%%%%%%%%%%%%%%%%%%%%%%%%%%%%%%%%%%%%%
\subsection{Ordinary sources of ghost events}\label{sec:ss-origin}
\begin{figure}
 \begin{center}
 \vspace{-0.3in}
 \leavevmode
 \includegraphics*[width=\textwidth]{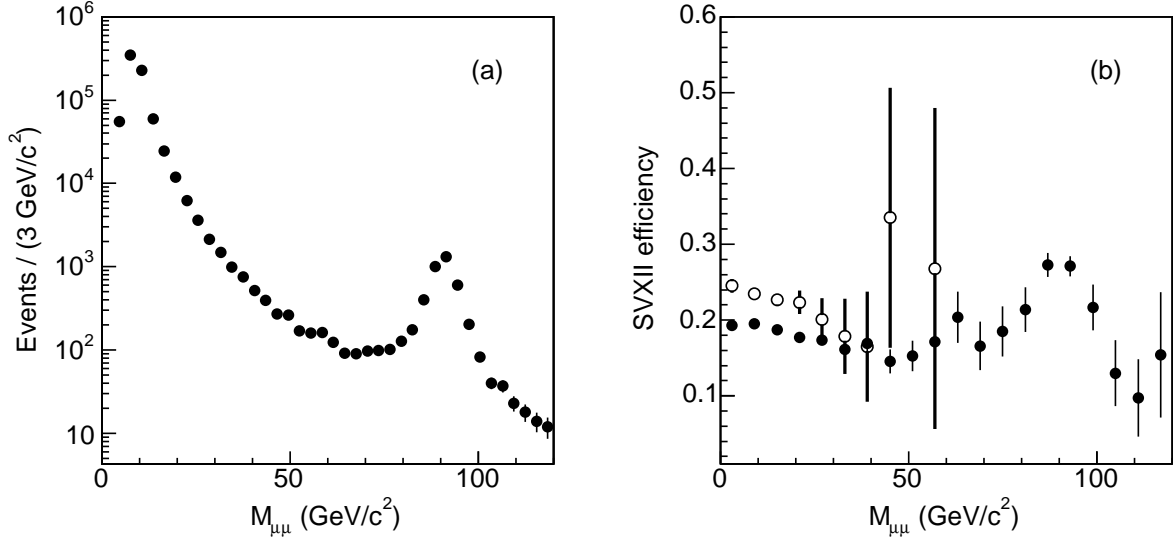}
 \caption[]{Invariant mass distribution (a) of the dimuon pairs used in the
            study. The efficiency (b) of the tight SVX requirements as a 
            function of the dimuon invariant mass in the data ($\bullet$)
            is compared to that in the heavy flavor simulation ($\circ$).}
 \label{fig:fig_2}
 \end{center}
 \end{figure}

%%%%%%%%%%%%%%%%%%%%%%%%%%%%%%%%%%%%%%%%%%%%
 In the following, we investigate several sources of ghost
 events that might not have been properly accounted
 for by previous experiments. Possible sources are:
 (a) semileptonic decays of hadrons with an unexpectedly large Lorentz boost;
 (b) muonic decays of particles with a lifetime longer than that of heavy
 flavors, such as $K$ and $\pi$ mesons; (c) decays of  $K^0_S$ mesons and
 hyperons; and (d) secondary interactions of prompt tracks that occur in 
 the detector volume.
 In the last two cases, muons are predominantly produced by punchthrough
 of the secondary prongs that hit the muon detectors.
 Figure~\ref{fig:fig_2} shows the invariant mass of dimuon pairs before
 the tight SVX selection, and the efficiency of this selection as a
 function of the dimuon invariant mass. The efficiency of tight SVX
 requirements in the data is below that in the simulation
 only for dimuon invariant masses smaller than $40\; \gevcc$, and then
 rises to the expected value of 0.257 where events are mostly contributed
 by prompt $Z^0$ decays. This feature does not favor the first hypothesis (a).
%%%%%%%%%%%%%%%%%%%%%%%%%%
%%%%%%%%%%%%%%%%%%%%%%%%%

\begin{figure}[]
 \begin{center}
 \vspace{-0.2in}
 \leavevmode
 \includegraphics*[width=\textwidth]{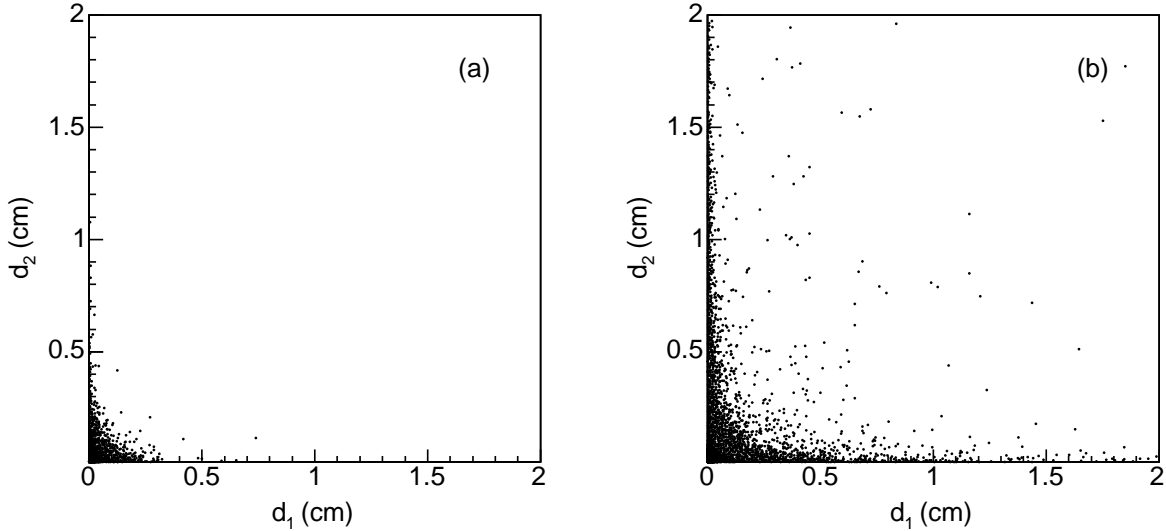}
 \caption[]{Two-dimensional impact parameter distribution of dimuons
            that pass the (a) tight and (b) loose SVX requirements.}
 \label{fig:fig_3}
 \end{center}
 \end{figure}

\begin{figure}[]
 \begin{center}
 \vspace{-0.2in}
 \leavevmode
 \includegraphics*[width=\textwidth]{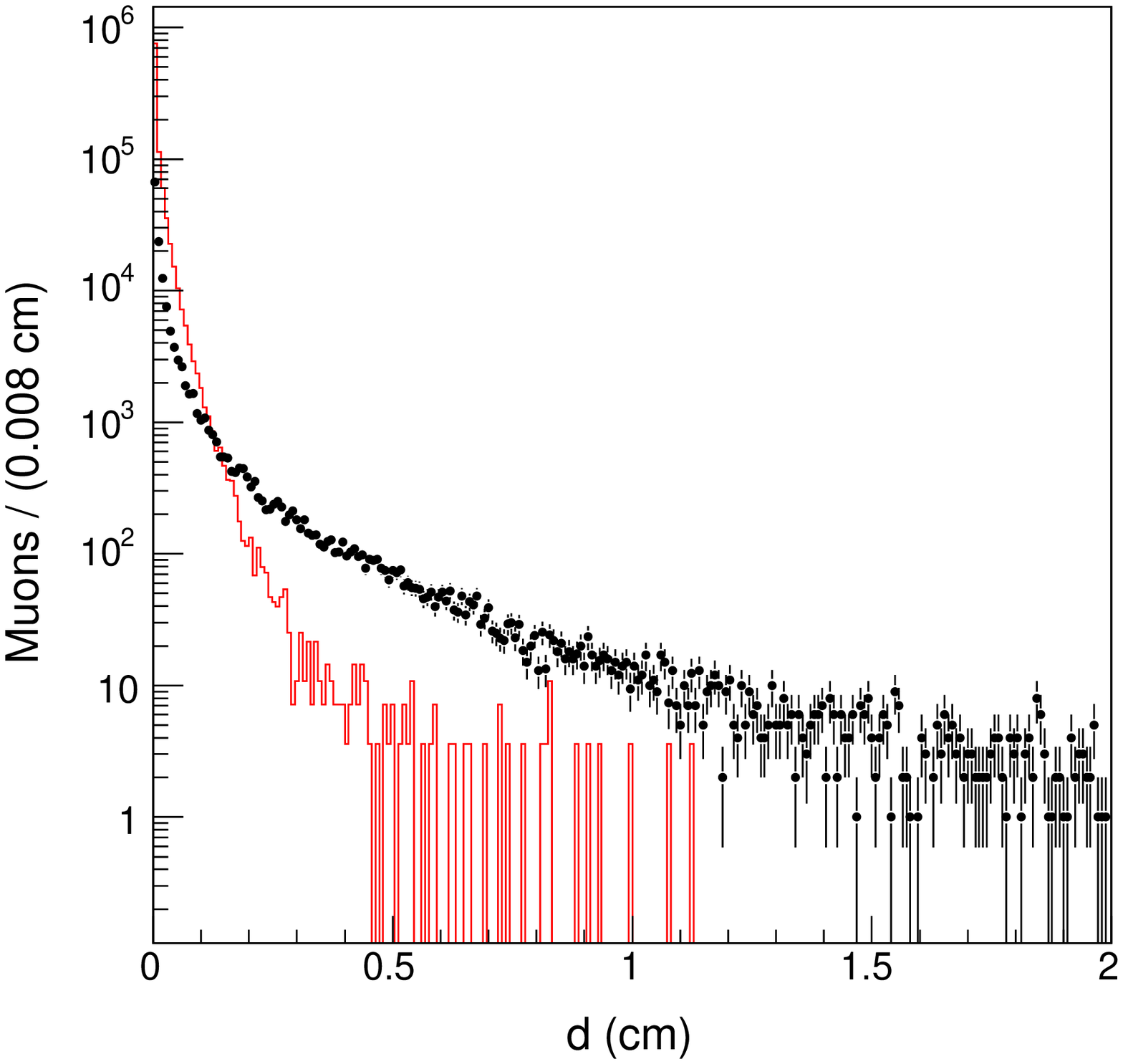}
 \caption[] {Impact parameter distribution of muons contributed by ghost
             ($\bullet$) and QCD (histogram) events. Muon tracks are
             selected with loose SVX requirements. The detector resolution
             is $\simeq 30 \; \mu$m, whereas bins are 80 $\mu$m wide.} 
 \label{fig:fig_4bis}
 \end{center}
 \end{figure}

\begin{figure}[]
 \begin{center}
 \vspace{-0.2in}
 \leavevmode
 \includegraphics*[width=\textwidth]{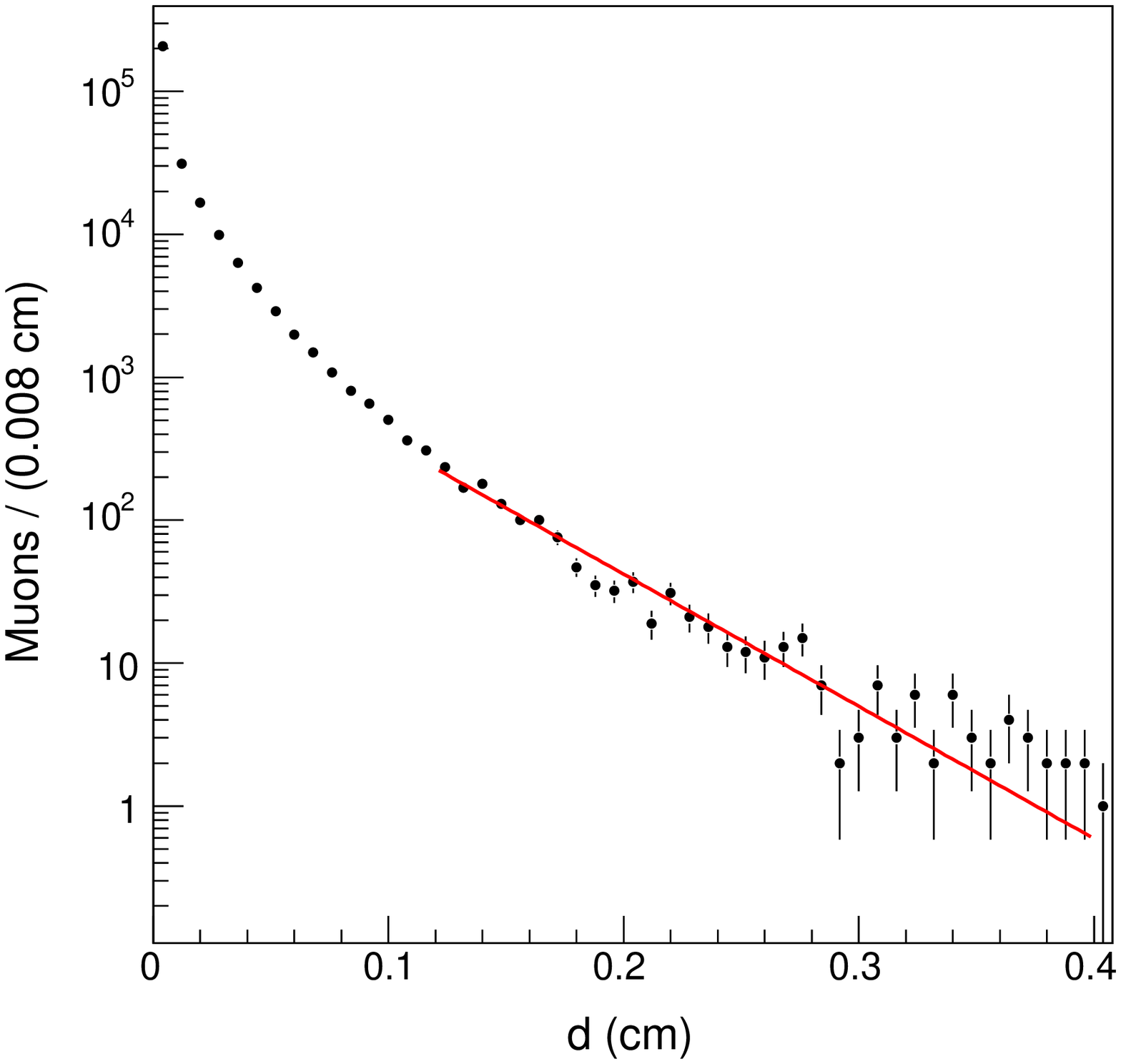}
 \caption[] {Impact parameter distribution of muons that pass the tight
             SVX requirements. The line represents the fit described in 
             the text.}
 \label{fig:fig_4}
 \end{center}
 \end{figure}

 A long-lived particle contribution is suggested by the comparison of
 the impact parameter distribution of dimuons that pass the loose and
 tight SVX requirements. The request that muons pass loose SVX
 requirements is momentarily used to reduce the possible contribution
 of muons  from secondary interactions occurring beyond the SVXII detector.
 We note that loose SVX requirements sculpt the impact parameter
 distribution of muons arising from the decay of objects with a lifetime
 much longer than that of $b$ hadrons, such as $\pi$, $K$, or $K^0_S$ mesons.
 Two-dimensional impact parameter distributions are shown in
 Fig.~\ref{fig:fig_3}. One-dimensional distributions are shown in 
 Fig.~\ref{fig:fig_4bis}. The impact parameter distribution of muons
 in ghost events  differs from that of the QCD contribution.
%%%%%%%%%%%%%%%%%%%%%%%%%%
%%%%%%%%%%%%%%%%%%%%%%%%%%%%%%%%%%%%%%%%%%%%%%%%%%%%%%%%%%%%%%
%%%%%%%%%%%%%%%%%%%%%%%%%%
%%%%%%%%%%%%%%%%%%%%%%%%%%%%%%%%%%%%%%%%%%%%%%%%%%%%%%%%%%%%%%%%%%%%
%%%%%%%%%%%%%%%%%%%%%%%%%%
 According to the heavy flavor simulation~\cite{bbxs}, dimuons with impact
 parameter larger than 0.12 cm only arise from $b\bar{b}$ production. We
 fit the impact parameter distribution in Fig.~\ref{fig:fig_4} with the 
 function $A \exp (-d/(c \tau))$ in the range $0.12 - 0.4$ cm. The best
 fit returns $c\tau= 469.7 \pm 1.3 \; \mu$m in agreement with the value
 $470.1 \pm 2.7$ $\mu$m  expected for the $b$-hadron mixture at the
 Tevatron~\cite{pdg}. We conclude that the data sample selected with the
 tight SVX selection is not appreciably contaminated by ghost 
 events. This supports our procedure for estimating the ghost size
 by assuming that the ghost contribution to events selected with tight
 SVX requirements is negligible. It also follows that the $b\bar{b}$
 contribution to dimuons with impact parameter larger than 0.5 cm is
 negligible.
%%%%%%%%%%%%%%%%%%%%%%%%%%
%%%%%%%%%%%%%%%%%%%%%%%%%%

\begin{figure}[]
 \begin{center}
 \vspace{-0.2in}
 \leavevmode
 \includegraphics*[width=\textwidth]{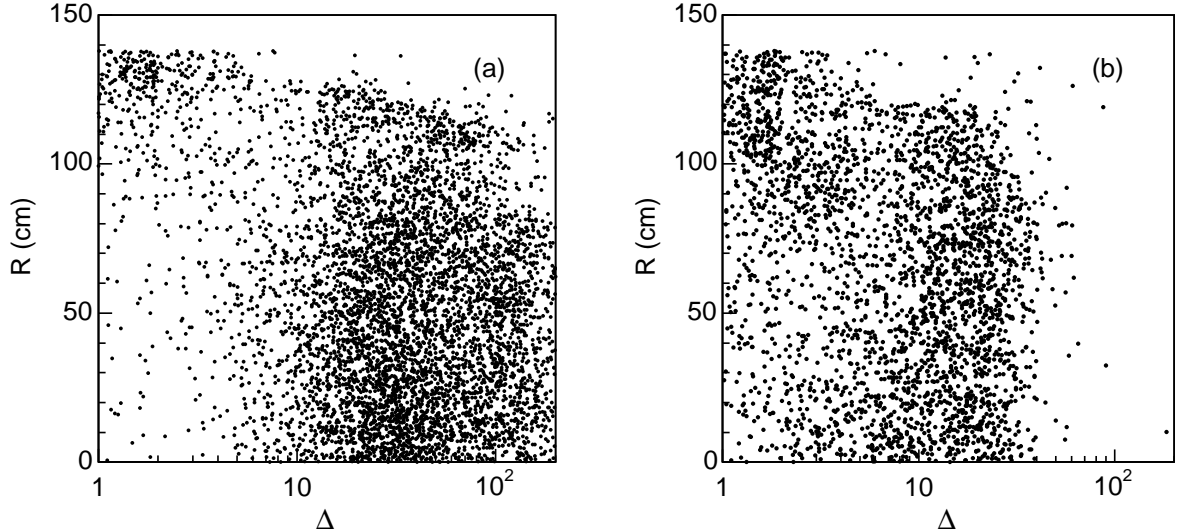}
 \caption[] {Distribution of $\Delta$ (see text) as a function of the
             distance $R$ of the (a) $K$ and (b) $\pi$ decay vertices
             from the beamline. For comparison, the analogous distribution
             for real muons from heavy flavor decays does not extend beyond
             $\Delta=9$. } 
 \label{fig:fake_1}
 \end{center}
 \end{figure}

\begin{figure}[]
 \begin{center}
 \vspace{-0.2in}
 \leavevmode
 \includegraphics*[width=\textwidth]{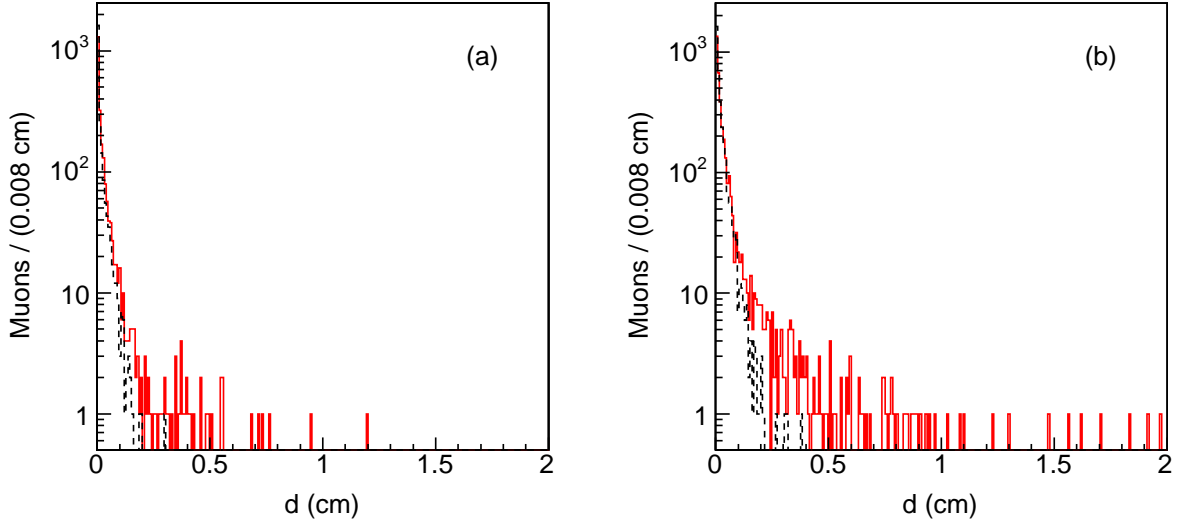}
 \caption[] {Impact parameter distributions of simulated CMUP muons
             (histogram) that pass all analysis requirements, including
             the loose SVX  selection, and arise from (a) pions and (b)
             kaon in-flight decays. The dashed histograms show the
             impact parameter of the parent pions and kaons.} 
 \label{fig:fake_2}
 \end{center}
 \end{figure}

 In ghost events, the presence of a large tail extending to high impact
 parameters suggests the contribution of particles with a lifetime much
 longer than that of $b$ quarks, such as $K^0_S$, $K$ and  $\pi$ mesons,
 and hyperons. We first investigate muons produced by pion and kaon
 in-flight-decays [source  (b)]. As reported in Ref.~\cite{bbxs}, after having selected
 muon pairs with the tight SVX requirements, approximately 30\% of the QCD
 contribution is due to prompt hadrons mimicking a muon signal.
 The size of the ghost sample has been estimated assuming that the
 efficiency of the tight SVX requirements for these tracks is the same as 
 that for real muons. This is a reasonable assertion when fake muons are
 generated by hadronic punchthroughs. However, muons arising from $\pi$ 
 or $K$ decays inside the tracking volume may yield misreconstructed tracks
 that are linked to hits in the SVXII detector less efficiently than real 
 muons. We estimate this efficiency using pions and kaons
 produced in the large statistics heavy flavor simulation used to derive 
 the dimuon acceptance for the $\sigma_{b\rightarrow\mu,\bar{b}\rightarrow\mu}$
 measurement~\cite{bbxs}. We use the quantity
  $\Delta^2 =1/3 \cdot [(\eta^h- \eta^{\rm track})^2/\sigma_{\eta}^2+
                        (\phi^h-\phi^{\rm track})^2/\sigma_{\phi}^2+
                        (1/p_T^h -1/p_T^{\rm track})^2/\sigma_{1/p_T}^2]$
 to measure the difference between the momentum vectors of the undecayed
 pions or kaons ($h$) and that of the closest reconstructed tracks~\footnote{
 The assumed experimental resolutions are 
     $\sigma_{\phi} [{\rm rad}] = \sigma_{\eta}=10^{-3}$ 
 and $\sigma_{1/p_T}=2\cdot 10^{-3}\; [\gevc]^{-1}$. }.
 Figure~\ref{fig:fake_1} shows the $\Delta$ distribution as a function of $R$,
 the decay distance from the beamline. One notes that most of the decays 
  at radial distances $R \leq 120$ cm yield misreconstructed tracks.
 The numbers of in-flight-decays that produce CMUP muons with 
 $p_T \geq 3\; \gevc$, a L1 primitive, and which pass different SVX selections
 are listed in Table~\ref{tab:tab_2bis}.
 %%%%%%%%%%%%%%%%%%%%%%%%%%
%%%%%%%%%%%%%%%%%%%%%%%%%%%%%%%%%%%%%%%%%%%%%%%%%%%%%%%%%%%%%%%%%%%%
%%%%%%%%%%%%%%%%%%%%%%%%%
 \begin{table}
 \caption[]{Number of pions and kaons corresponding to a misreconstructed
            track ($\Delta \geq 5$) with $p_T \geq 3 \; \gevc$ and 
            $|\eta| \leq 0.7$, that decay inside the tracking volume,
            produce CMUP muons with a L1 primitive, and pass different SVX
            selections.}
 \begin{center}
 \begin{ruledtabular}
 \begin{tabular}{lcc}
  Selection        &   $\pi$   &    K    \\
  Tracks           &  2667199  & 1574610 \\ 
  In-flight-decays &    14677  &   40561 \\
  CMUP+L1          &     1940  &    5430 \\
  Loose SVX        &      897  &    3032 \\
  Tight SVX        &      319  &    1135 \\    
 \end{tabular}
 \end{ruledtabular}
 \end{center}
 \label{tab:tab_2bis}
 \end{table}
%%%%%%%%%%%%%%%%%%%%%%%%%%%%%%%%%%%%%%%%%%%%%%%%%%%%%%%%%%%%%%
 The efficiency of the tight SVX requirements for a single muon
 due to in-flight-decays
 (0.16 and 0.21 for $\pi$ and $K$ decays, respectively) is smaller
 than that for muons in QCD events  ($\simeq 0.5$). The contributions
 of muons due to in-flight-decays to ghost events is evaluated using 
 simulated events produced in   generic-parton hard scattering~\footnote{
 We use option 1500 of the {\sc herwig} program to generate final states
 produced by hard scattering of partons with transverse momentum larger
 than 3 $\gevc$~\cite{bbxs}.}.
 In the simulation, there are 44000 track pairs per CMUP pair due to
 $b\bar{b}$ production with the same kinematic acceptance ($p_T\geq 3\; \gevc$
 and $|\eta|\leq 0.7$). The ratio of the number of pions to that of  kaons
 is approximately 5/1. Each simulated track in the kinematic acceptance
 is weighted with the corresponding in-flight-decay probabilities 
 of producing CMUP muons listed in Table~\ref{tab:tab_2bis}. Tracks are  
 also weighted with the probabilities, measured in Ref.~\cite{bbxs},
 that $\pi$ or $K$ punchthrough mimics a CMUP signal. In the latter case,
 the efficiency of the SVX requirement is the same as for real muons, and
 we ignore the cases in which both muons arise from hadronic punchthrough.
 Having normalized this simulation to the number of observed initial muons 
 arising from $b\bar{b}$ production, we predict a contribution to ghost
 events due to in-flight-decays of pions and kaons that is 57000 events, 
 44\% and 8\% of which pass the loose and tight SVX selection, respectively.
 In the 25000 simulated events that pass the loose SVX selection, 
 approximately 15000 muons arise from kaon in-flight-decays. These predictions
 depend on how well the {\sc herwig} generator models generic parton hard
 scattering and its uncertainty is difficult to estimate.
 Figure~\ref{fig:fake_2} shows the impact parameter distribution of muons 
 arising from in-flight-decays of pions and kaons produced in simulated
 $b\bar{b}$ and $c\bar{c}$ events. The number of events in 
 Fig.~\ref{fig:fake_2} has to be multiplied by five in order to
 be compared with the data in  Fig.~\ref{fig:fig_4bis}. Our estimate of the
 number of muons arising from in-flight-decays accounts for 35\% of the 
 ghost muons, but  for less than 10\% of those with $d \geq 0.5$ cm.
%%%%%%%%%%%%%%%%%%%%%%%%%%
%%%%%%%%%%%%%%%%%%%%%%%%%%%%%%%%%%%%%%%%%%%%%%%%%%%%%%%%%%%%%%%%%%%%

\begin{figure}[]
 \begin{center}
 \vspace{-0.2in}
 \leavevmode
 \includegraphics*[width=\textwidth]{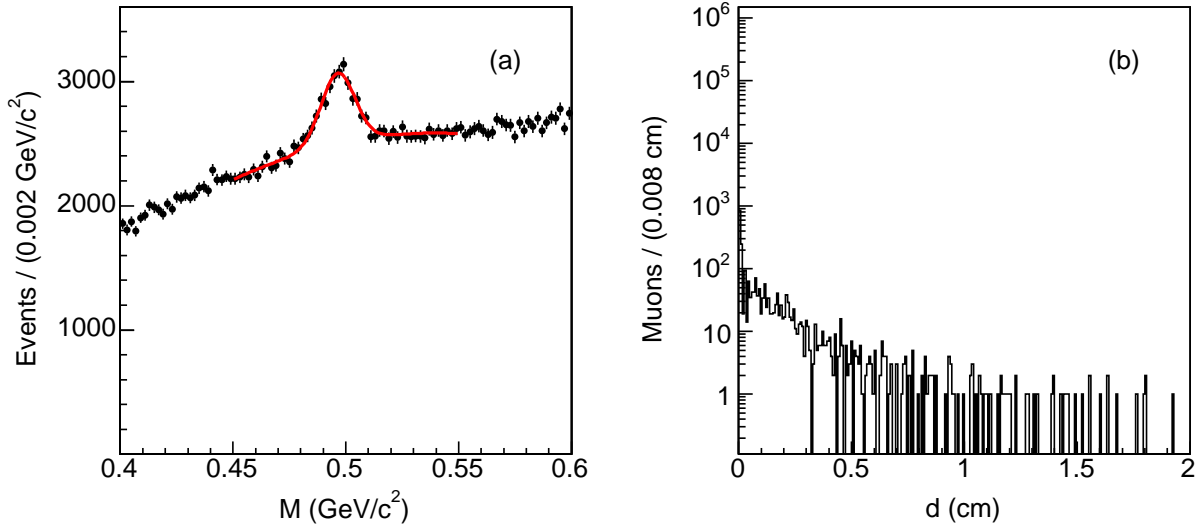}
 \caption[] {Distributions of (a) the invariant mass of pairs of initial
             muons and opposite sign tracks and of (b) the impact parameter
             of initial muons, produced by $K^0_S$ decays, that pass the
             loose SVX selection. The solid line represents a fit described
             in the text. In the impact parameter distribution, the
             combinatorial background is removed with a sideband subtraction
             method. For comparison, the vertical scale in (b) is kept the
             same as in Fig.~\ref{fig:fig_4bis}.}
 \label{fig:fake_3}
 \end{center}
 \end{figure}

\begin{figure}[]
 \begin{center}
 \vspace{-0.2in}
 \leavevmode
 \includegraphics*[width=\textwidth]{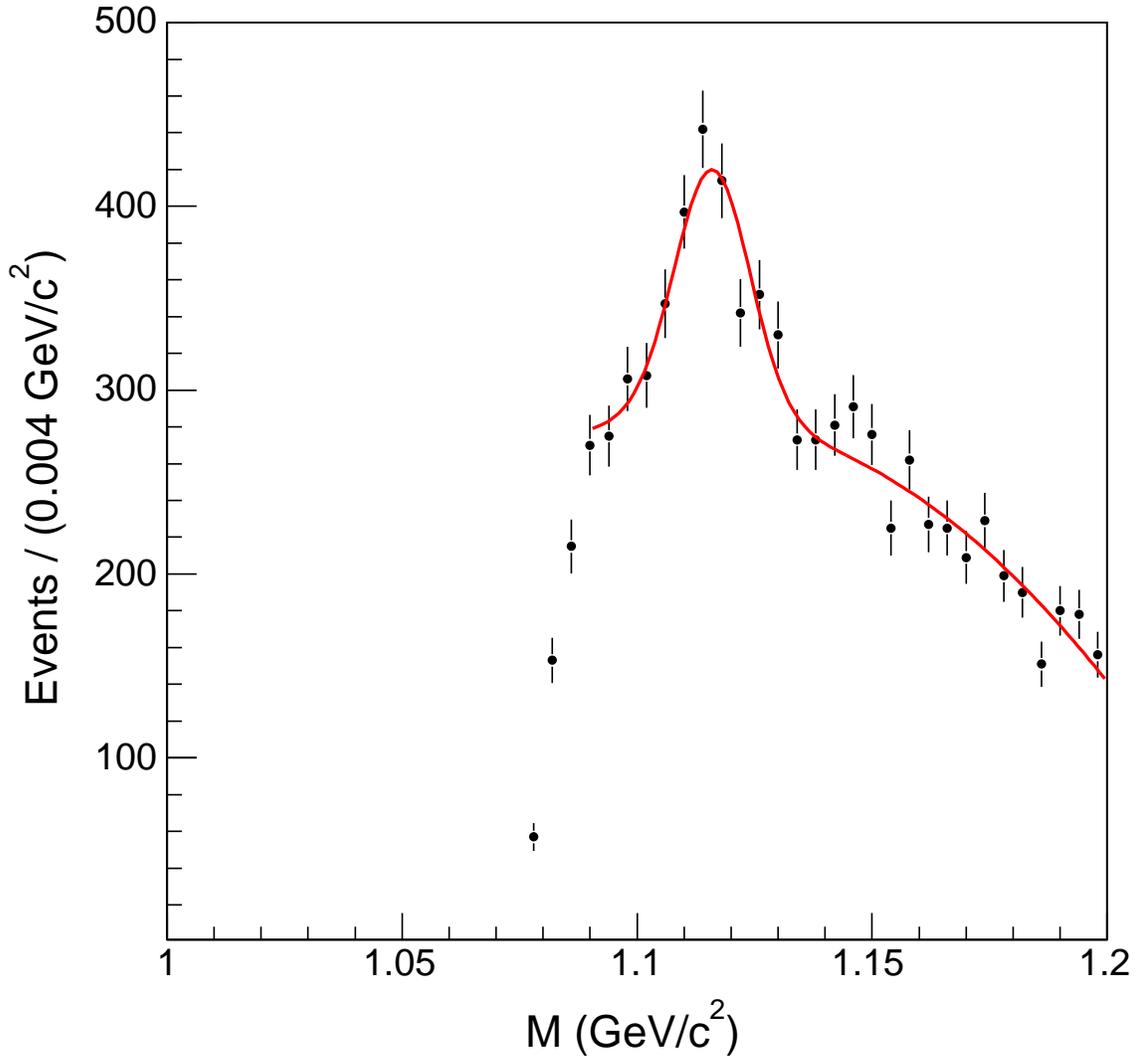}
 \caption[]{Distributions of  the invariant mass of pairs of initial
            muons and opposite sign tracks produced by $\Lambda$ decays.
            We attribute the proton (pion) mass to the track with positive 
            (negative) charge. The solid line represents a fit that uses
            a Gaussian function to model the signal and a fourth order
            polynomial to model the combinatorial background.}
 \label{fig:fig_lam}
 \end{center}
 \end{figure}

 In addition, muons in ghost events can be mimicked by the punchthrough
 of hadrons arising from the decay of $K^0_S$ mesons or hyperons [source (c)]. We have
 searched the dimuon data set for $K^0_S \rightarrow \pi^+\pi^-$ decays in
 which a pion punchthrough mimics the muon signal. We combine all initial
 muon tracks with all opposite sign tracks with $p_T \geq 0.5 \; \gevc$ 
 contained in a 40$^{\deg}$ cone around the direction of the initial muons. 
 Muon-track combinations are constrained to arise from a common space point.
 They are discarded if the three-dimensional vertex fit returns a $\chi^2$
 larger than 10. Figure~\ref{fig:fake_3}~(a) shows the invariant mass
 distribution of the $K^0_S$ candidates reconstructed assuming that both
 tracks are due to pions. A fit of the data with a Gaussian function to model
 the signal plus a second order polynomial to model the background yields
 a signal of $5348 \pm 225$ $K^0_S$ mesons. The impact parameter
 distribution of initial muons produced by $K^0_S$ decays is shown in
 Fig.~\ref{fig:fake_3}~(b). The data also contain a smaller number of cases
 in which the initial muon is mimicked by the products of hyperon decays.
 Using a similar technique, we have searched the data for
 $\Lambda \rightarrow p \pi^-$ decays and we find a signal of 
 $678 \pm 60$ $\Lambda$ baryons (see Fig.~\ref{fig:fig_lam}).
 Since in both case the kinematic acceptance times reconstruction efficiency
 is approximately 50\%,  source (c)
 ($\simeq 12000$ events)  explains $\simeq 8$\% of the ghost events.
%%%%%%%%%%%%%%%%%%%%%%%%%%
%%%%%%%%%%%%%%%%%%%%%%%%%%%%%%%%%%%%%%%%%%%%%%%%%%%%%%%%%%%%%%%%%%%%
%%%%%%%%%%%%%%%%%%%%%%%%%%
%%%%%%%%%%%%%%%%%%%%%%%%%%%%%%%%%%%%%%%%%%%%%%%%%%%%%%%%%%%%%%%%%%%%

\begin{figure}[]
 \begin{center}
 \vspace{-0.2in}
 \leavevmode
 \includegraphics*[width=\textwidth]{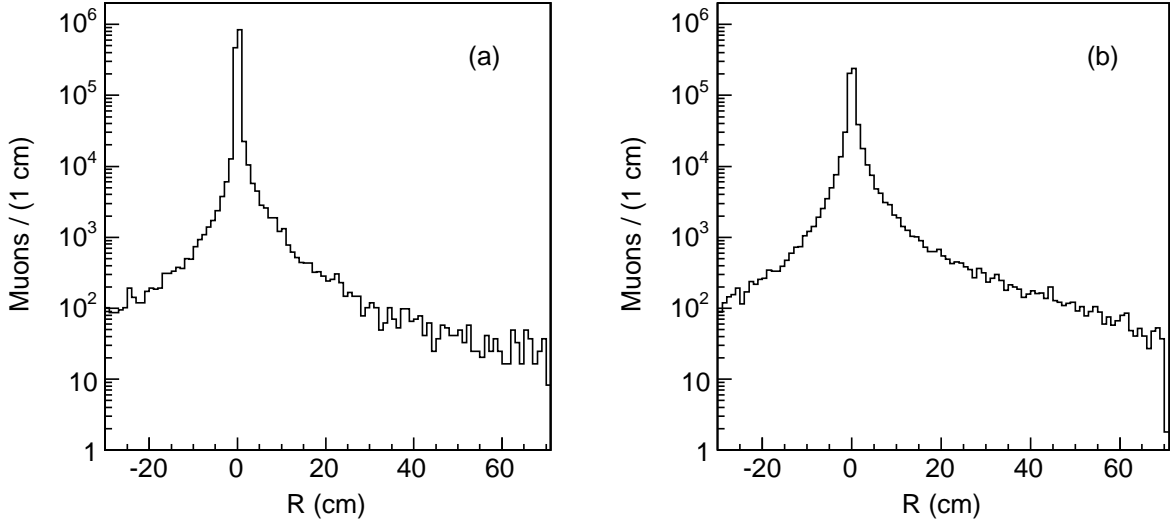}
 \caption[]{Distributions of $R$, the signed distance of muon-track
            vertices from the nominal beam line for (a) QCD and
            (b) ghost events (see text).}
 \label{fig:fig_sint}
 \end{center}
 \end{figure}

 The final source (d) of ghost events, secondary interactions in the
 detector volume, is investigated using the data. We search for
 secondary interactions by combining initial muons with all  tracks
 with $p_T\geq 0.5 \; \gevc$ contained in a 40$^{\deg}$ cone around 
 the muon direction. Muon-track combinations are constrained to arise
 from a common space point. They are discarded if the three-dimensional
 vertex fit returns a $\chi^2$ larger than 10.
 The distribution of $R$, the distance of a reconstructed secondary 
 vertex from the detector origin in the plane transverse to the beam
 line, is shown in Fig.~\ref{fig:fig_sint}. The distance $R$ is 
 negative when the secondary vertex is  in the hemisphere opposite to
 that containing the momentum of the muon-track system. Secondary
 interactions are characterized by spikes at $R$ values where the
 detector material is concentrated, such as SVX supports or the COT
 inner support cylinder. From the absence of visible spikes, we conclude 
 that the contribution of multi-prong secondary interactions to initial 
 muons in ghost events is negligible. At the same time, we cannot exclude
 some contribution to ghost events from elastic or quasi-elastic nuclear
 scattering of hadronic tracks in the detector material.
%%%%%%%%%%%%%%%%%%%%%%%%%%
%%%%%%%%%%%%%%%%%%%%%%%%%%%%%%%%%%%%%%%%%%%%%%%%%%%%%%%%%%%%%%%%%%%%

 Our estimate  of the size of possible sources of ghost events 
 underpredicts the observed number of ghost events by approximately a 
 factor of two (154000 observed and 69000 accounted for).
 However, given the possible large uncertainty of the
 in-flight-decay prediction and the possible contribution of elastic
 or quasi-elastic nuclear scattering in the detector material, at this
 point of our study we cannot exclude that the ghost sample can 
 be completely accounted for by a combination of all the above-studied 
 background sources. Were ghost events all due to these ordinary 
 sources, they would not contain a significant number of additional
 real muons. Therefore, these sources are unlikely to be the origin of the
 excess of low-mass dileptons reported in Ref.~\cite{dilb}. That study is
 repeated in the next section.
%%%%%%%%%%%%%%%%%%%%%%%%%%%%%%%%%%%%%%%%%%%%%%%%%%%%%%%%%%%%%%
 \section{Study of events that contain an additional muon}
 \label{sec:ss-3mu}
%%%%%%%%%%%%%%%%%%%%%%%%%%%%%%%%%%%%%%%%%%%%%%%%%%%%%%%%%%%%%%
 We begin this study with events that contain a pair of initial muons 
 passing our analysis selection  without any SVX requirements.
 We then search for additional tracks with  $p_T \geq 2 \; \gevc$ and a
 matching stub in the CMU, CMX, or CMP muon detectors (the three detectors
 cover the pseudorapidity region $|\eta| \leq 1.1$ ). No SVX requirements
 are made on these additional muons. For muons with $p_T \geq 2 \; \gevc$
 and $|\eta| \leq 1.1$, the muon detector efficiency in the heavy flavor 
 simulation is $0.805 \pm 0.008$. We measure the muon detector efficiency
 in the data by using $J/\psi$ candidates acquired with the $\mu$-SVT trigger
 (see  Ref.~\cite{bbxs} for more details). After reweighting the kinematics
 of the muons from $J/\psi$ decays to reproduce that of simulated muons from 
 heavy flavor decays, the efficiency is measured to be $0.838 \pm 0.004$.

 According to the heavy flavor simulation, additional real muons predominantly 
 arise from sequential decays of single $b$ hadrons
 (the $g \rightarrow b \bar{b}$ and $g \rightarrow c \bar{c}$ contributions
 are suppressed by the request of two initial muons with $p_T \geq 3\; \gevc$,
 $|\eta| \leq 0.7$, and invariant mass larger than 5 $\gevcc$). In addition,
 one expects a contribution of additional muons from hadrons mimicking
 the muon signal. In the data, 9.7\% of the dimuon events contain an
 additional muon (71835 out of 743006 events). In events containing an
 $\Upsilon(1S)$ candidate, that are included in the dimuon sample, the
 probability of finding an additional muon is ($0.90 \pm 0.01$)\%.
 Of the $5348 \pm 225$ events with an identified $K^0_S$
 meson only $94 \pm 41$,  ($1.7 \pm0.8$)\%, survive the request of an additional muon in 
 the event.

 Our investigation starts with measuring the efficiency of the tight 
 SVX requirements for initial muon pairs in events that also contain
 at least one   additional muon. The efficiency drops from 0.193 to 0.166.
 If ghost events were all due to initial muons arising from $\pi$ or
 $K$ decays, or secondary interactions in the detector volume, this
 efficiency would have increased back to 0.244 because these types of source
 contain fewer additional muons than events with heavy flavors.
 For example, this is the case for events containing an $\Upsilon(1S)$ 
 or $K^0_S$ candidate. This observation anticipates that a fraction of
 the ghost events contains more additional muons than QCD events.

\begin{figure}
 \begin{center}
 \vspace{-0.3in}
 \leavevmode
 \includegraphics*[width=0.5\textwidth]{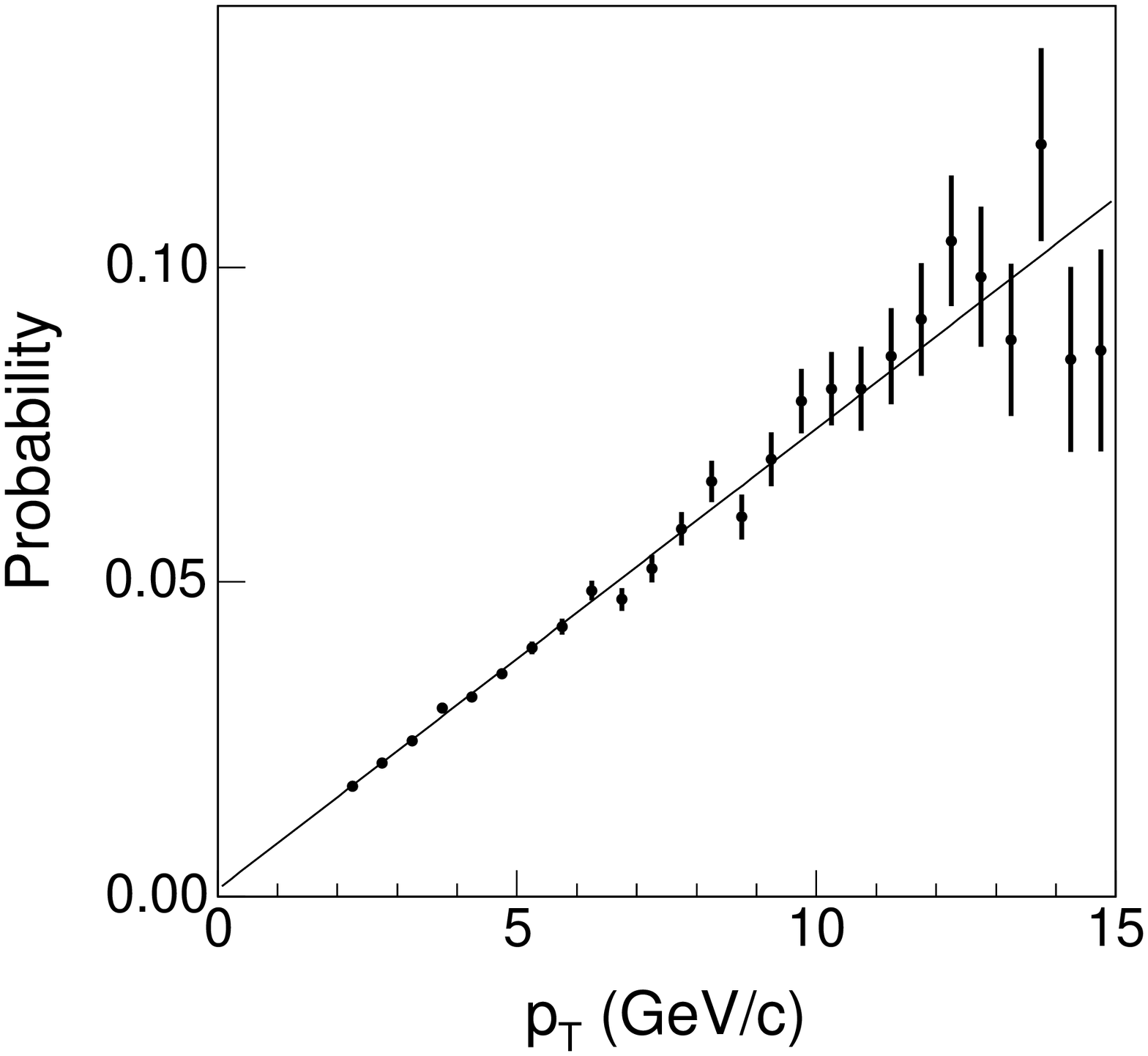}\includegraphics*[width=0.5\textwidth]{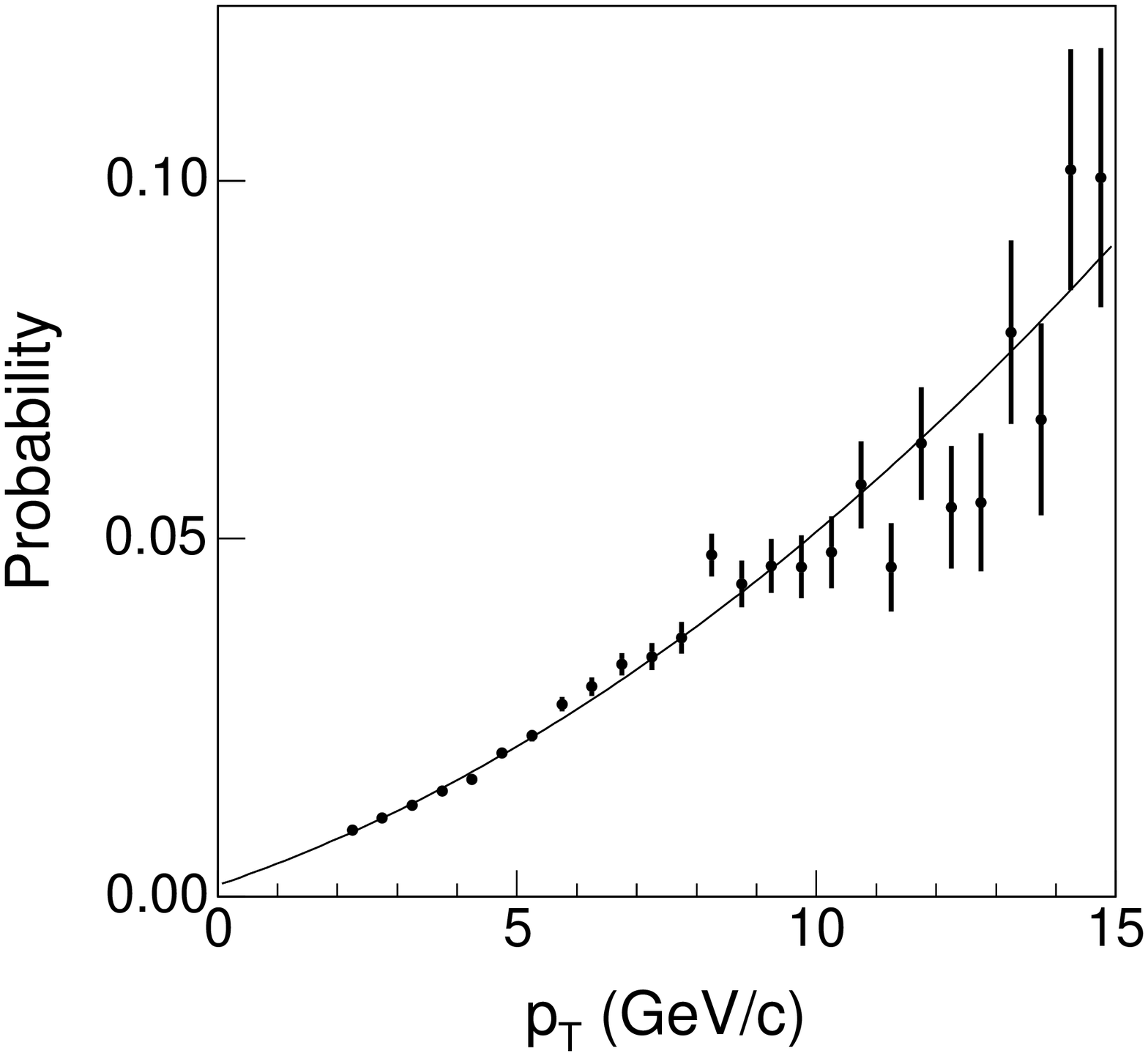}
 \caption[]{Probability that a track with $|\eta| \leq 1.1$ mimics a muon 
            signal in the CMU, CMX, or CMP detectors as a function of the
            kaon (left) or pion (right) transverse momentum. We have verified
            that these fake probabilities do not depend on the SVX
            requirements applied to the tracks.}
 \label{fig:fig_6}
 \end{center}
 \end{figure}

 Following the study in Ref.~\cite{dilb}, additional muons are paired
 with one of the initial muons if their invariant mass is smaller than
 $5 \; \gevcc$. For this study, we use a larger statistics data
 sample~\footnote{
 The correlated $b\bar{b}$ cross section measurement uses 742 pb$^{-1}$
 of data in which the dimuon trigger is not prescaled as a function of
 the instantaneous luminosity. From the rate of dimuon events that pass
 the analysis selection, the luminosity of the larger data sample is
 estimated to correspond to 1426 pb$^{-1}$.}.
 Following the analysis procedure of Ref.~\cite{dilb}, we retain muon 
 combinations with charges of opposite sign ($OS$). As in Ref.~\cite{dilb},
 we estimate the contribution of fake muons from the number of observed
 same sign ($SS$) muon pairs. In the case of Drell-Yan or quarkonia
 production, fake additional muons arise from the underlying event and
 one expects no charge correlation between initial and additional fake
 muons. However, in the simulation of heavy flavor decays, the numbers of
 $OS$  and $SS$ tracks surrounding an initial muon are not equal. These tracks
 come from the $b$- and $c$-quark fragmentation and decay. In the
 simulation, the ratio of $OS$ to $SS$ combinations as well as the
 ratio of the numbers of pion to kaon tracks is a function of the 
 invariant mass of the muon-track pair. The CDF~II detector simulation  
 does not describe the punchthrough of hadrons. Therefore, we evaluate 
 the fake muon contribution by weighting pion and kaon tracks in the
 heavy flavor simulation with the probability that hadronic punchthrough
 mimics a muon signal. These fake probabilities, shown in Fig.~\ref{fig:fig_6}
 as a function of the track $p_T$, have been measured using a sample
 of $D^0 \rightarrow K^- \pi^+$ decays acquired with the {\sc charm}
 and $\mu$-SVT triggers. The procedure for determining these probabilities
 is described in detail in Appendix~B of Ref.~\cite{bbxs}.

%%%%%%%%%%%%%%%%%%%%%%%%%%
%%%%%%%%%%%%%%%%%%%%%%%%%
 The rate of real plus fake muon pairs with small invariant mass is
 evaluated after rescaling the parton level cross section predicted
 by the {\sc herwig} generator to match the measurements
   $\sigma_{b\rightarrow\mu,\bar{b}\rightarrow \mu}=1549 \pm 133$ pb and
   $\sigma_{c\rightarrow\mu,\bar{c}\rightarrow\mu}=624\pm 104$ pb~\cite{bbxs}.
 In the simulation, $SS$ combinations due to either real or fake muon pairs
 are subtracted from $OS$ combinations. In the simulation, 
 the initial pair of muons is always arising from heavy-flavor semileptonic 
 decays. In the data,  9\% of the initial muons recoiling
 against a small mass dimuon are due to prompt hadrons
 mimicking the muon signal (relative
 size of the $BB$ and $BP$ components in  Table~\ref{tab:tab_1}). 
 In addition, 2\% of these recoiling muons are due to hadrons from 
 heavy flavor decays~\cite{bbxs}. We account for this by increasing the 
 rates predicted by the simulation by a factor of 1.12.

\begin{figure}
 \begin{center}
 \vspace{-0.3in}
 \leavevmode
 \includegraphics*[width=\textwidth]{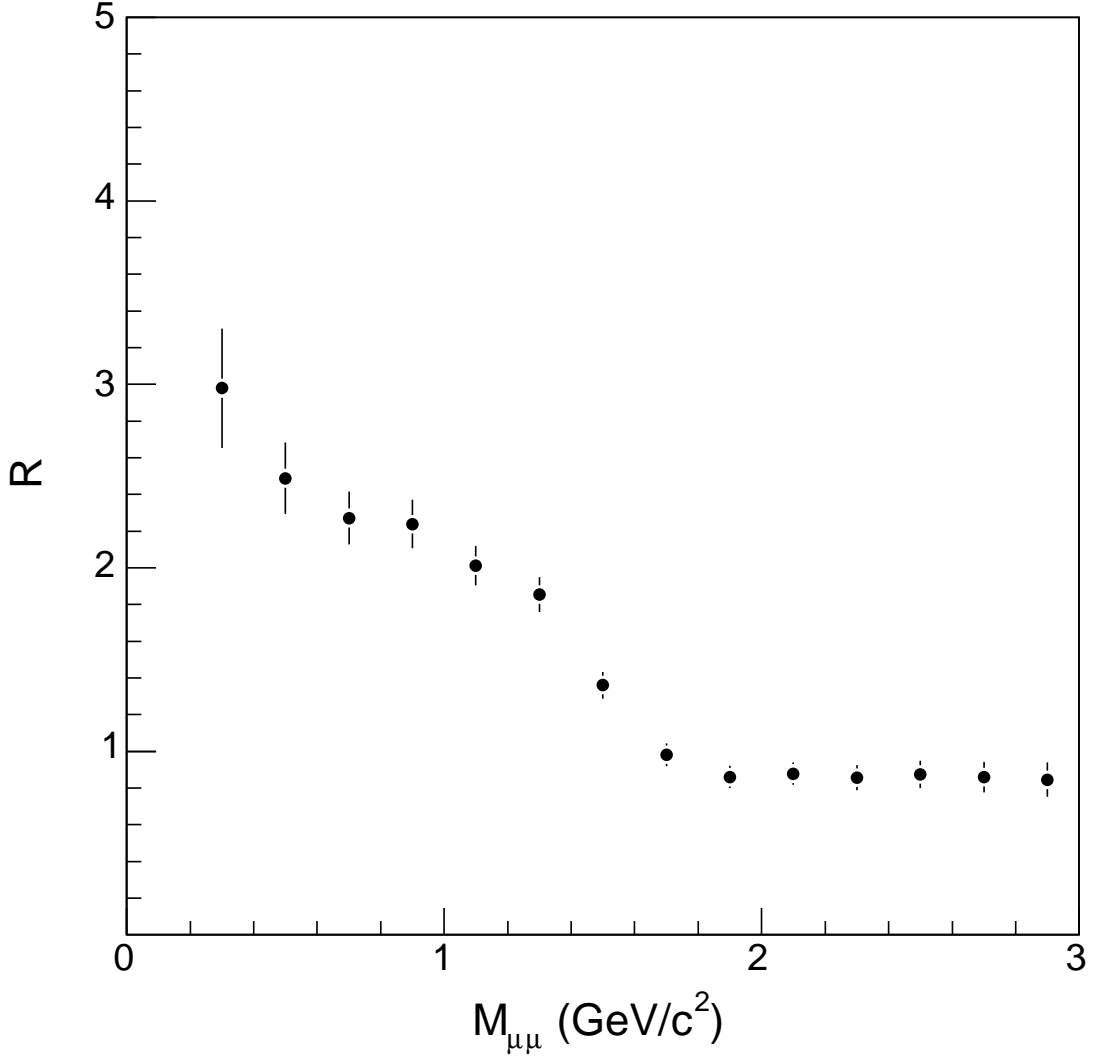}
 \caption[]{Ratio $R$ of total number of $OS-SS$ muon pairs to that of real
            $OS-SS$ pairs arising from heavy flavor decays as a function of 
            the dimuon invariant mass. We use simulated events generated 
            with the {\sc herwig} Monte Carlo program. The generator 
            parton-level cross sections have been scaled to match the 
            data~\cite{bbxs}. The number of fake muon pairs has been
            evaluated by weighting simulated hadronic tracks with the
            probability of mimicking a muon signal as measured with data.
            Errors are statistical only.}
 \label{fig:fig_7}
 \end{center}
 \end{figure}

\begin{figure}
 \begin{center}
 \vspace{-0.3in}
 \leavevmode
 \includegraphics*[width=\textwidth]{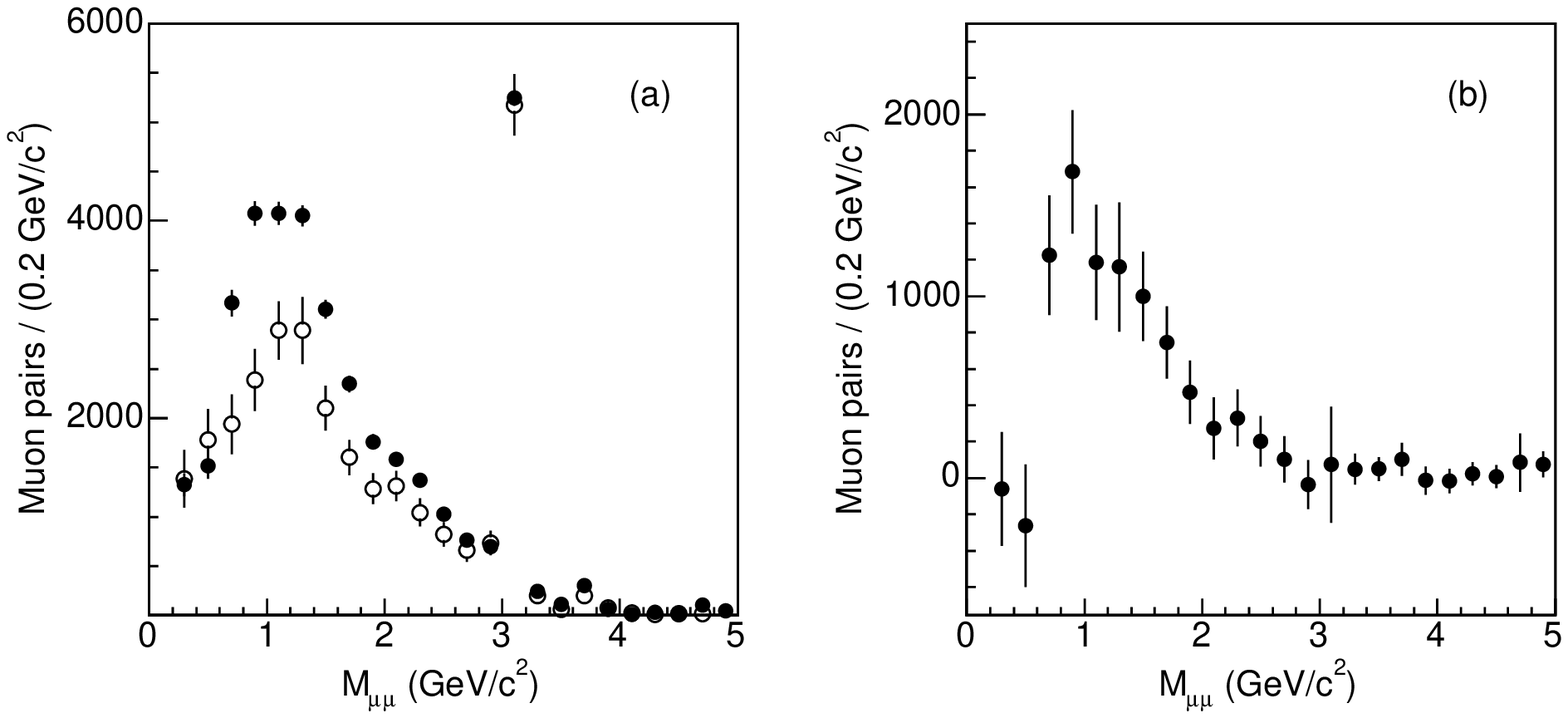}
 \caption[] {The invariant mass distribution of (a) $OS-SS$ muon pairs in 
             the data ($\bullet$) is compared to the simulation prediction
             ($\circ$). One of the two initial muons in the event is combined
             with an additional muon if their invariant mass is smaller than
             5 $\gevcc$. The difference (b) between data and prediction is
             also shown.}
 \label{fig:fig_8}
 \end{center}
 \end{figure}

\begin{figure}
 \begin{center}
 \vspace{-0.3in}
 \leavevmode
 \includegraphics*[width=\textwidth]{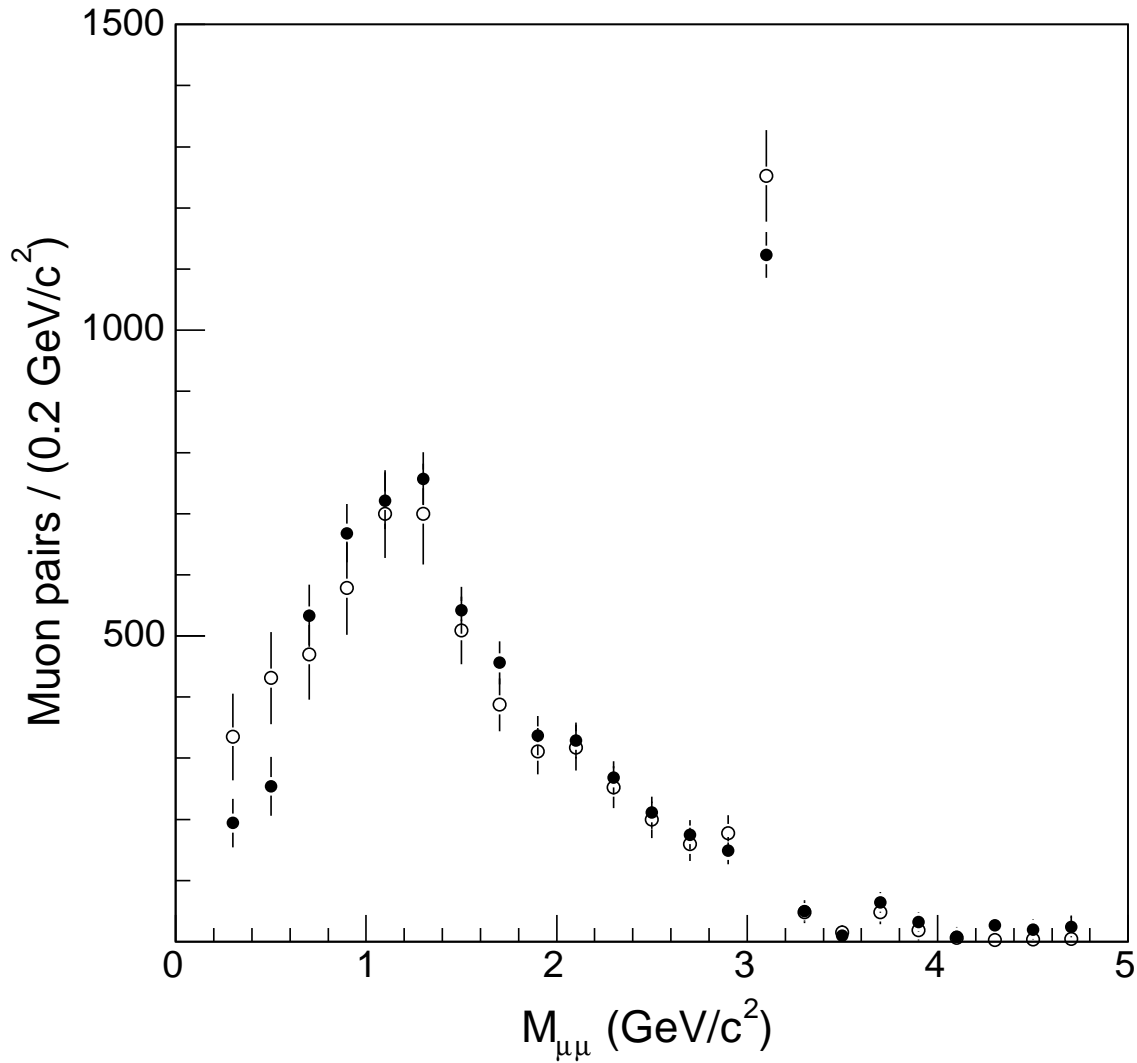}
 \caption[]{The invariant mass distribution of $OS-SS$ muon pairs in the
            data ($\bullet$) is compared to the simulation prediction
            ($\circ$). Initial muons are selected using the
            tight SVX requirements.}
 \label{fig:fig_9}
 \end{center}
 \end{figure}

 Figure~\ref{fig:fig_7} shows the ratio of the total number of $OS-SS$
 muon pairs predicted by the above calculation to that of real $OS-SS$
 dimuons from heavy flavor decays. The  fake contribution is approximately
 33\% of that of real muons from sequential decays of single $b$ quarks.
 %%%%%%%%%%%%%%%%%%%%%%%%%%
%%%%%%%%%%%%%%%%%%%%%%%%% 
 Figure~\ref{fig:fig_8} compares the invariant mass spectrum of $OS-SS$ muon 
 pairs in the data and in the heavy flavor simulation. Since the simulation
 is effectively normalized to the observed number of initial muon pairs,
 the prediction has a 3\% systematic error due to the branching ratio 
 $b \rightarrow c \rightarrow \mu$ plus a 3\%  uncertainty due to the 
 absolute pion and kaon rates predicted by the simulation~\cite{bbxs}
 (the systematic uncertainty of the muon detector efficiency is negligible).
 %%%%%%%%%%%%%%%%%%%%%%%%%%
%%%%%%%%%%%%%%%%%%%%%%%%%
 This systematic uncertainty is not shown in Fig.~\ref{fig:fig_8}.
 The number of $J/\psi$ mesons in the data is correctly
 modeled by the simulation in which  $J/\psi$ mesons only arise from
 $b\bar{b}$ production. The agreement between the number of observed 
 and predicted $J/\psi$ mesons selected without any SVX requirement
 supports the estimate of the efficiency of the tight SVX requirement
 and the resulting value of the correlated $b\bar{b}$ cross section
 reported in Ref.~\cite{bbxs}. However, the data are underestimated
 by the simulation for invariant masses smaller than 2 $\gevcc$.
 The excess of $8451 \pm 1274$ events results from an observation
 of $37042 \pm 389$ and a prediction of $28589\pm 1213$ events.
 The size and shape of the excess is consistent with what was first
 reported in Ref.~\cite{dilb}, in which the excess was mostly observed
 in a high statistics $e \mu$ sample. We have an advantage with 
 respect to the previous observation. The robustness of the prediction
 can be verified by comparing the observed and predicted invariant
 mass spectrum of dimuon pairs when the initial muons are selected
 with the tight SVX requirements. In this case we observe $6935\pm 154$
 events, whereas $ 6918 \pm 293$ are predicted. The corresponding
 invariant mass distribution is shown in Fig.~\ref{fig:fig_9}.
 %%%%%%%%%%%%%%%%%%%%%%%%%%
%%%%%%%%%%%%%%%%%%%%%%%%%
\subsection{Kinematics of additional muons in ghost events}
\label{sec:ss-addkin}
 The excess of $8451 \pm 1274$ $OS-SS$ pairs with invariant mass smaller
 than $ 5\; \gevcc$ is a measure of the charge asymmetry of additional
 muons as a function of the invariant mass of the muon pair. For
 1,426,571 initial dimuons, we find 94148 $OS$ and 57106 $SS$ combinations
 with an additional muon with $m_{\mu\mu} \leq 5 \; \gevcc$. A qualitative
 estimate predicts that 14200 $SS$ and $OS$ fake muon combinations are
 produced by the underlying event~\footnote{
 These numbers are derived from the 1\% probability of finding an
 additional muons in events with $\Upsilon(1S)$ candidates and assuming
 that the underlying event is the same for all processes.}.
 The heavy flavor simulation, which also accounts for fake muons, predicts
 40899 $OS$ and $12309$ $SS$ real plus fake combinations for a grand total
 of 55100 $OS$ and 26500 $SS$ pairs. This approximate prediction 
 underestimates the data by 39000 $OS$ and 30500 $SS$ pairs. The number
 of the $OS$ and $SS$ pairs in ghost events is determined more precisely
 as the difference between the data and the QCD expectation. The QCD
 expectation is the number of muon combinations found in events in which
 the initial dimuons pass the tight SVX requirements divided by the SVX
 requirement efficiency. This study is summarized in Table~\ref{tab:tab_3}.
%%%%%%%%%%%%%%%%%%%%%%%%%
 \begin{table}
 \caption[]{Number of events as a function of $N_c$, the number of
            combinations of initial and additional muons. Additional muons
            are combined with initial muons if the pair invariant mass is
            smaller than 5 $\gevcc$. The numbers of events with at least one 
            combination are split according to opposite ($OS$) or same ($SS$) 
            charge sign. ``SVX'' are numbers of events that pass the tight
            SVX  selection. QCD is the latter number divided by the 
            efficiency of the tight SVX requirements. Ghost is the
            difference between the total  and the QCD contributions.}
 \begin{center}
 \begin{ruledtabular}
 \begin{tabular}{lcccc}
  Topology     &  Total      &  SVX   &     QCD            &   Ghost    \\
 $N_c \geq 0$  &  1426571  & 275986 & $1131090 \pm 9271$ & $295481 \pm 9271$\\ 
 $N_c \geq 1$  &  141039   &  22981 &   $94184 \pm  772$ & $ 46855 \pm 772$ \\
 $OS$          &   94148   &  15372 &   $63000 \pm  516$ & $ 31148 \pm 516$ \\
 $SS$          &   57106   &   8437 &   $34578 \pm  283$ & $ 22528 \pm 283$ \\
 $N_c \geq 2$  &   10215   &    828 &    $3393 \pm   28$ & $ 6822  \pm 28$  \\
 \end{tabular}
 \end{ruledtabular}
 \end{center}
 \label{tab:tab_3}
 \end{table}
%%%%%%%%%%%%%%%%%%%%%%%%%%%%%%%%%%%%%%%%%%%%%%%%%%%%%%%%%%%
 In ghost events, the fraction of events that carries an
 additional real or fake muon with any charge is ($15.8 \pm 0.3$)\%, 
 approximately a factor of two higher than in QCD events. The fraction 
 of additional muons due to tracks mimicking a muon signal is estimated
 in the next section. 

\begin{figure}
 \begin{center}
 \vspace{-0.3in}
 \leavevmode
 \includegraphics*[width=\textwidth]{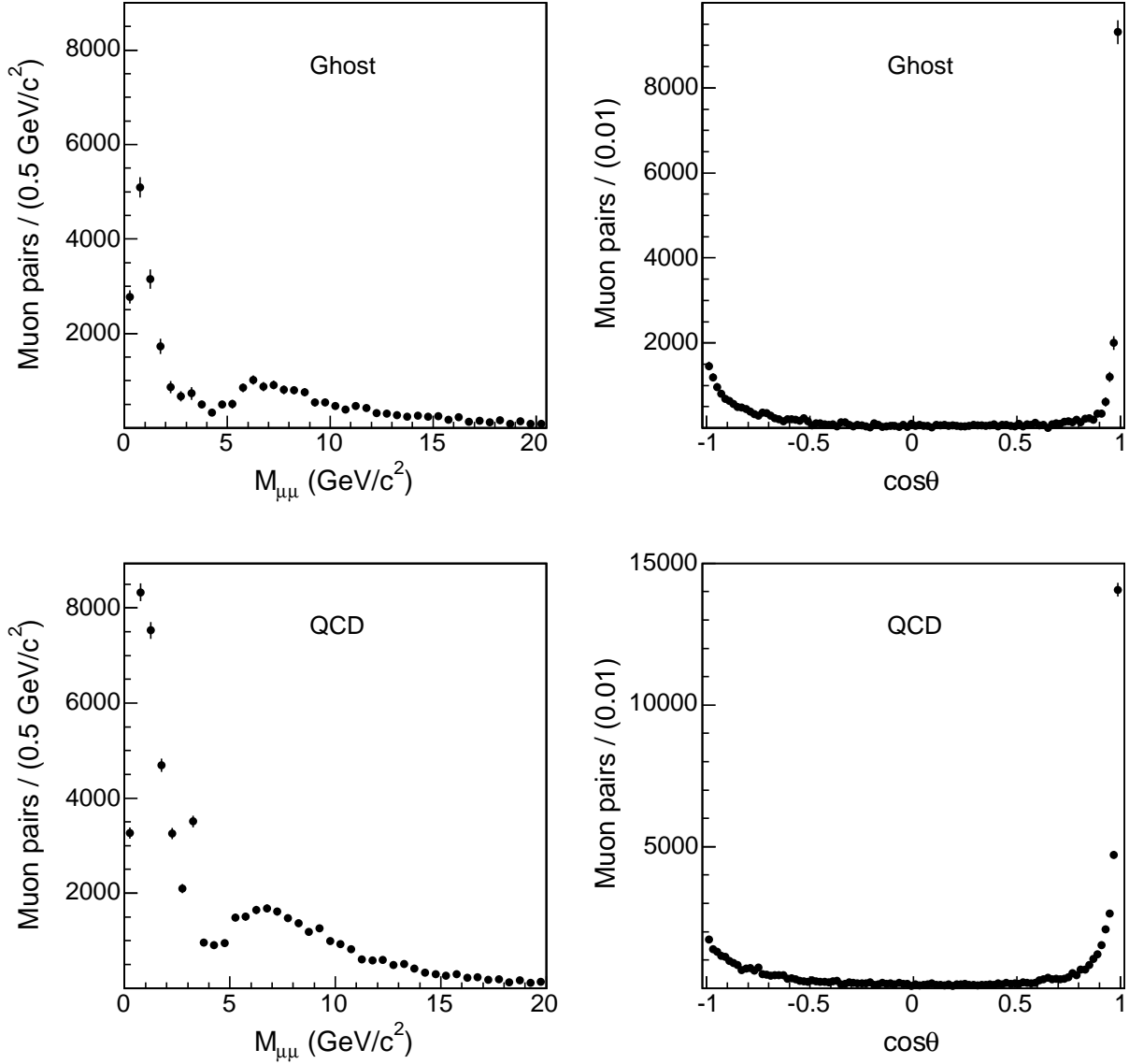}
 \caption[]{Events with $OS$ initial muon pairs and an additional muon
            combined with the opposite-charge initial muon. We
            show the invariant mass, $M_{\mu\mu}$, and opening angle,
            $\theta$, distributions of these combinations for the QCD
            and ghost contributions.}
 \label{fig:fig_10}
 \end{center}
 \end{figure}

 In order to compare with the previous measurement~\cite{dilb}, we have
 analyzed dimuon pairs with $ m_{\mu^+ \mu^-} \leq 5 \; \gevcc $.
 This requirement is appropriate for selecting dimuons produced by
 sequential semileptonic decays of single $b$-quarks, but could bias the
 investigation of ghost events. Therefore, we search dimuon
 events for additional muons without any invariant mass cut. If the initial
 dimuon pair has opposite charge, we combine the additional muon with the
 initial muon of opposite charge ($OSO$ combinations). If the initial muons
 have  same  charge, we randomly combine the additional muon with one of
 the initial muons ($SSO$ and $SSS$ combinations). The QCD contribution is 
 estimated as the number of combinations in events in which initial dimuons
 pass the tight SVX requirements (SVX contribution) divided by the efficiency
 of the tight SVX requirements. As before, the ghost contribution is the 
 difference between the data and the QCD contribution. The number of
 three-muon combinations is listed in Table~\ref{tab:tab_4}.
 %%%%%%%%%%%%%%%%%%%%%%%%%
 \begin{table}
 \caption[]{Numbers and types of three-muon combinations. We separate
            events according to the charge of the initial muons. The  
            topology $OSO$ is that of two opposite-charge initial dimuons;
            by definition, the third muon has opposite charge with respect
            to one of them. When initial dimuons have same sign charge, 
            the third muon charge can have either the same ($SSS$) or
            opposite sign ($SSO$).}
\begin{center}
\begin{ruledtabular}
 \begin{tabular}{lcccc}
 Topology  &  All   &  SVX    &        QCD       &      Ghost      \\
 $OSO $    &  90022 &  14497  & $59414 \pm 487$  & $30608 \pm 487$ \\
 $SSO $    &  48220 &   7708  & $31590 \pm 259$  & $16630 \pm 259$ \\
 $SSS $    &  28239 &   4139  & $16963 \pm 139$  & $11276 \pm 139$ \\
 \end{tabular}
 \end{ruledtabular}
 \end{center}
 \label{tab:tab_4}
 \end{table}
%%%%%%%%%%%%%%%%%%%%%%%%%%%%%%%%%%%%%%%%%%%%%%%%%%%%%%
 Figure~\ref{fig:fig_10} shows the invariant mass and opening angle
 distribution of $OSO$  combinations for the QCD and ghost contributions. 
 %%%%%%%%%%%%%%%%%%%%%%%%%%
%%%%%%%%%%%%%%%%%%%%%%%%%
%%%%%%%%%%%%%%%%%%%%%%%%%%
 \begin{figure}
 \begin{center}
 \vspace{-0.3in}
 \leavevmode
 \includegraphics*[width=\textwidth]{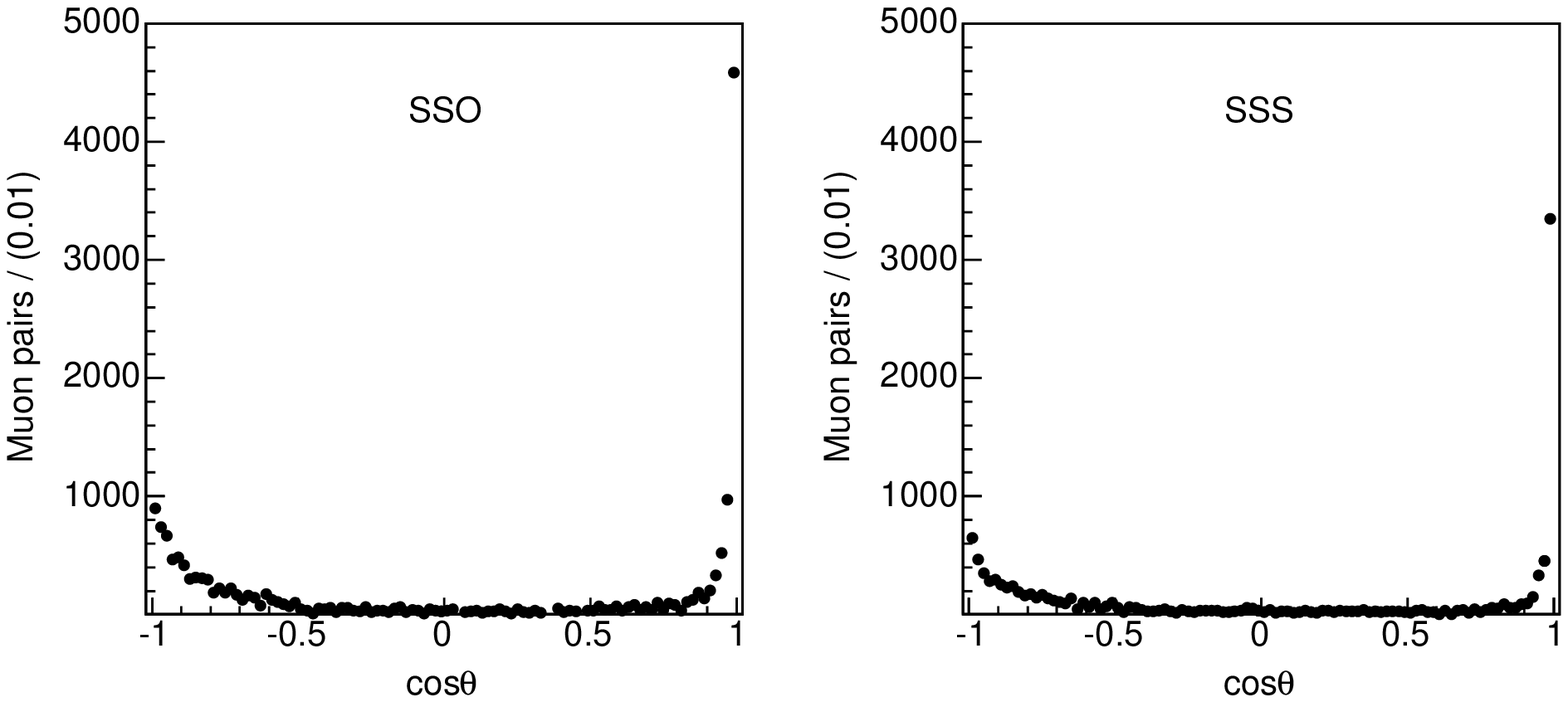}
 \caption[]{Opening angle distributions of dimuon combinations for ghost events. 
            The initial dimuons have same
            sign charge, and combinations of an additional and initial muons 
            are split according to the charge of the additional muon. The 
            plots are the projection of two-dimensional distributions in 
            which the additional muon is combined with both initial muons.}
 \label{fig:fig_11}
 \end{center}
 \end{figure}
%%%%%%%%%%%%%%%%%%%%%%%%%
 Muon pairs due to $b$ sequential decays, which account for most of the QCD
 contribution, peak at small invariant masses and small opening angles. 
 The tail at large masses and opening angles results from fake muons
 with  wrong charge. The distributions of analogous pairs in the
 ghost sample have a quite similar behaviour. However, it is
 important to note that combinations of initial and additional muons
 in ghost events have a smaller opening angle than those from sequential
 $b$ decays. As shown in Fig.~\ref{fig:fig_11}, $SSO$ and $SSS$
 combinations have similar opening angle distributions. Therefore, it seems
 reasonable to restrict the study of ghost events to muons and tracks 
 contained in a cone of angle $\theta \leq 36.8^{\deg}$, corresponding to
 $\cos \theta \geq 0.8$, around the direction of each initial muon.
%%%%%%%%%%%%%%%%%%%%%%%%%%
 \section{Study of muon and track properties in ghost events}~\label{sec:ss-a0}
\begin{figure}
 \begin{center}
 \vspace{-0.3in}
 \leavevmode
 \includegraphics*[width=\textwidth]{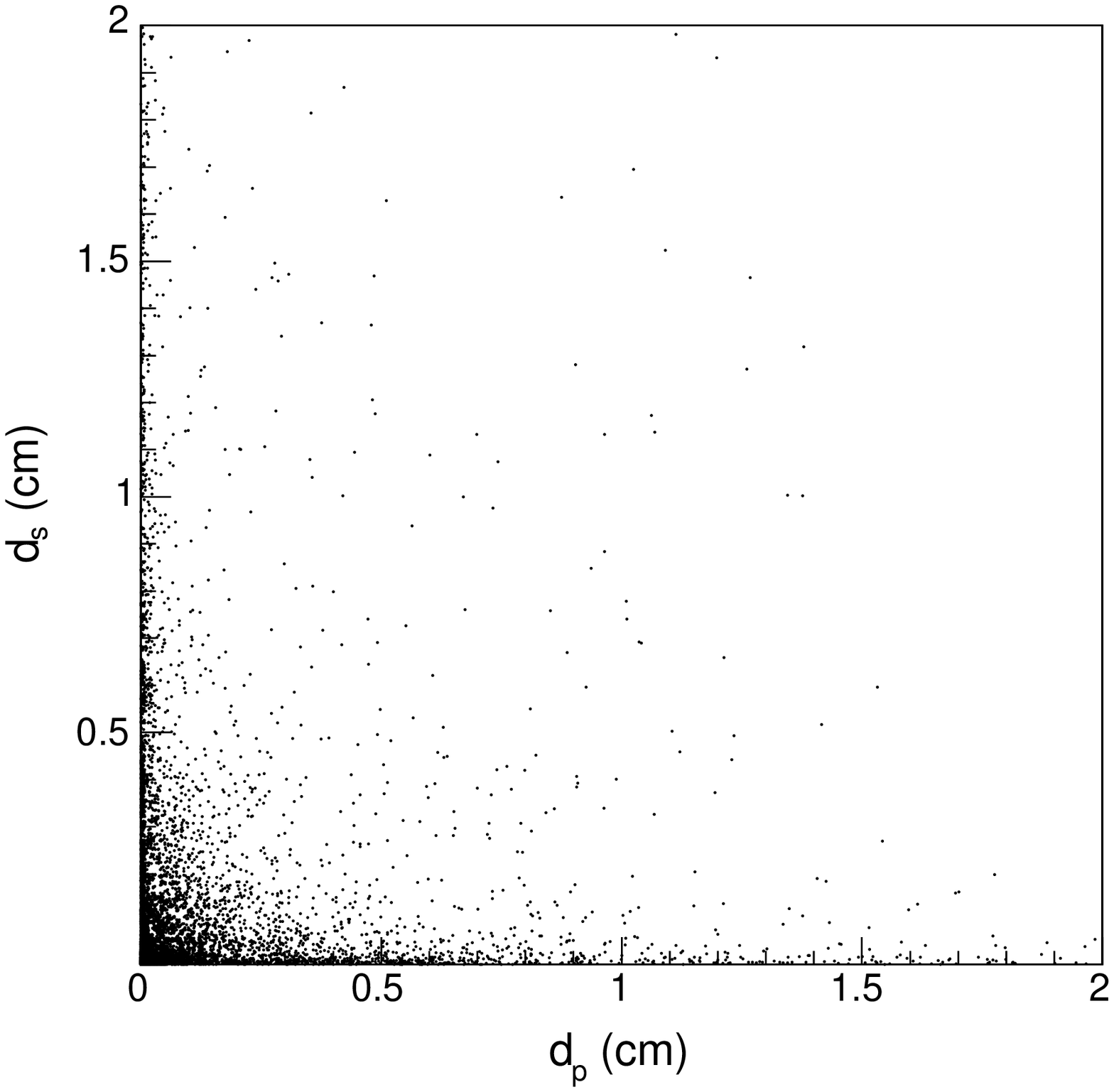}
 \caption[]{Two-dimensional distribution of the impact parameter of 
            additional muons, $d_s$, versus that of initial muons, $d_p$,
            for ghost events. Muons are selected with loose
            SVX requirements. The QCD contribution has been removed.}
 \label{fig:fig_13}
 \end{center}
 \end{figure}

\begin{figure}
 \begin{center}
 \vspace{-0.3in}
 \leavevmode
 \includegraphics*[width=\textwidth]{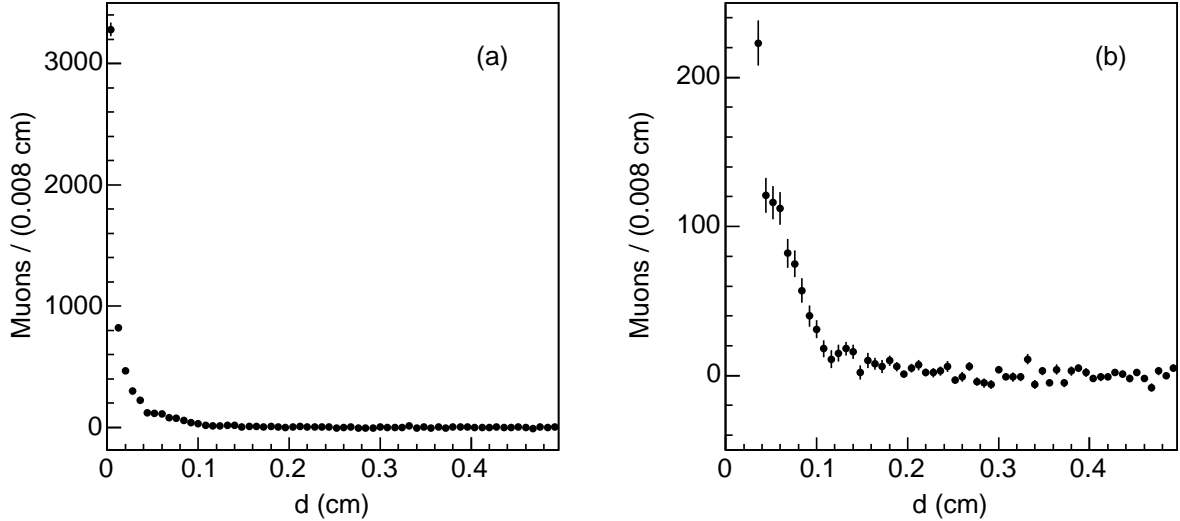}
 \caption[]{Impact parameter distribution of (a) additional muons found
            in events in which the initial muons are selected with tight
            SVX requirements. The same distribution is plotted in (b) with
            a magnified vertical scale. Additional muons are selected
            without SVX requirements.}
 \label{fig:fig_13bis}
 \end{center}
 \end{figure}

%%%%%%%%%%%%%%%%%%%%%%%%%%
 The number of additional muons contained in a cone of angle 
 $\theta \leq 36.8^{\deg}$ ($\cos \theta \geq 0.8$) around the direction
 of any initial muon is listed in Table~\ref{tab:tab_5}.
%%%%%%%%%%%%%%%%%%%%%%%%%%%%%%%%%%%% 
 \begin{table}
 \caption[]{Numbers of additional muons with an angle $\theta\leq 36.8^{\deg}$
            with respect to the direction of one of the initial muons. 
            We list separately the combination of additional and initial
            muons with opposite ($OS$) and same ($SS$) sign charge.}
 \begin{center}
 \begin{ruledtabular}
 \begin{tabular}{lcccc}
  Topology  &  All    &  SVX    &      QCD         &      Ghost      \\
  $OS$      &  83237  &  13309  & $54545 \pm 447$  & $28692 \pm 447$ \\
  $SS$      &  50233  &   7333  & $30053 \pm 246$  & $20180 \pm 246$ \\
 \end{tabular}
 \end{ruledtabular}
 \end{center}
 \label{tab:tab_5}
 \end{table} 
%%%%%%%%%%%%%%%%%%%%%%%%%%%%%%%%%%%%%%%%%%%%%%%%%%%%%%%%%%%%
 Figure~\ref{fig:fig_13} shows the two-dimensional distribution of the
 impact parameter of an initial muon versus that of all additional muons
 in a $\cos \theta \geq 0.8$ cone around its direction. The QCD contribution  
 has been removed using events in which the primary muons pass the tight SVX 
 requirement. The tail of the impact parameter distribution of additional
 muons in QCD events, shown in Fig.~\ref{fig:fig_13bis}, does not extend
 beyond 2 mm. In contrast, as shown in Fig.~\ref{fig:fig_13}, the impact
 parameter distribution of additional muons in ghost events extends to much
 larger values and is consistent with that of the initial muons. However,
 the impact parameters of the additional and initial muons are loosely 
 correlated (the correlation factor is $\rho_{d_{p}d_{s}}=0.03 $).  
%%%%%%%%%%%%%%%%%%%%%%%%%%
%%%%%%%%%%%%%%%%%%%%%%%%%
%%%%%%%%%%%%%%%%%%%%%%%%%%
%%%%%%%%%%%%%%%%%%%%%%%%%
 
 The contribution of fake muons is evaluated by weighting all tracks
 with $p_T \geq 2 \; \gevc$, $|\eta| \leq 1.1$, and contained in a
 $\cos \theta \geq 0.8$ cone, with the fake probabilities shown in 
 Fig.~\ref{fig:fig_6}. Table~\ref{tab:tab_6} lists the number of these 
 tracks for QCD and ghost events.
%%%%%%%%%%%%%%%%%%%%%%%%%
 \begin{table}
 \caption[]{Numbers of tracks with $p_T \geq 2 \; \gevc$, $|\eta| \leq 1.1$,
            and an angle $\theta \leq 36.8^{\deg}$ with respect to the 
            direction of one of the initial muons. We list separately the
            numbers of tracks with opposite ($OS$) and same ($SS$) charge as 
            the initial muon. Tracks  associated
            with a muon stub are excluded.}
 \begin{center}
 \begin{ruledtabular}
 \begin{tabular}{lcccc}
  Topology &  All     &    SVX  &         QCD       &       Ghost       \\
  $OS $    &  1315451 &  207344 & $849770 \pm 6965$ & $465860 \pm 6965$ \\
  $SS $    &   893750 &  140238 & $574745 \pm 4711$ & $318004 \pm 4711$ \\
 \end{tabular}
 \end{ruledtabular}
 \end{center}
 \label{tab:tab_6}
 \end{table} 
%%%%%%%%%%%%%%%%%%%%%%%%%%
  The QCD and ghost contributions have been previously determined to be
  1131090 and 295481 events, respectively. It follows that the average
  number of tracks contained in a $\theta \leq 36.8^{\deg}$ cone around
  the direction of one of the initial muons in ghost events is 1.58
  $OS$ and 1.08 $SS$, twice the values measured in QCD events (0.75 $OS$
  and 0.51 $SS$ tracks).

  Table~\ref{tab:tab_7} compares the observed number of additional muons 
  to the predicted number of additional fake muons in ghost events.
%%%%%%%%%%%%%%%%%%%%%%%%%
 \begin{table}
 \caption[]{Numbers of additional muons in ghost events are compared to fake
            muon expectations. The fake muon prediction is evaluated by 
            applying the fake probabilities shown in Fig.~\ref{fig:fig_6} to
            all tracks not associated to a muon stub and with 
            $p_T \geq 2 \; \gevc$, $|\eta| \leq 1.1$, and an angle 
            $\theta \leq 36.8^{\deg}$ with respect to the direction of one of
            the initial muons. We list separately the numbers of muons with
            opposite ($OS$) and same ($SS$)
            sign charge as the initial muon. $F_K$ and $F_\pi$ are the numbers
            of fake muons predicted assuming that hadronic tracks are all kaons
            or all pions, respectively. For kaon tracks, the rate of predicted
            fake muons should be increased by 10\% to also account for 
            in-flight-decay contributions. }
 \begin{center}
 \begin{ruledtabular}
 \begin{tabular}{lccc}
  Topology &  Observed         &     $F_K$        &      $F_\pi$    \\
  $OS$     &  $28692 \pm 447$  & $15447 \pm 210$  &  $9649 \pm 131$ \\
  $SS$     &  $20180 \pm 246$  & $10282 \pm 137$  &  $6427 \pm  81$ \\
 \end{tabular}
 \end{ruledtabular}
 \end{center}
 \label{tab:tab_7}
 \end{table} 
%%%%%%%%%%%%%%%%%%%%%%%%%%
 In ghost events, the fraction of real  muons with any charge is
 approximately four times larger than that of real muons in QCD events
 (9.4\% compared to 2.1\%, as obtained from the number of $OS-SS$ 
 dimuons listed in Table~\ref{tab:tab_5} and the number of $OS+SS$
 dimuons in ghost events after subtracting the average of the pion
 and kaon fake contributions listed in Table~\ref{tab:tab_7},
 respectively). In Table~\ref{tab:tab_7}, the ratio of real to background
 muons is approximately 1. This ratio is larger than that in
 QCD events (0.4) which are correctly modeled by the heavy flavor simulation.
As a cross-check that the difference in rates of additional muons between the
QCD and ghost sample is contributed by real muons, we restrict
ourselves to additional muons identified as CMUP muons.   In this case the
contribution of fake muons is significantly reduced and is expected to be 
negligible~\cite{bbxs} (see Appendix~A).
 The numbers of additional CMUP muons and expected
 fakes are listed in Table~\ref{tab:tab_7bis}.
%%%%%%%%%%%%%%%%%%%%%%%%%
 \begin{table}
 \caption[]{Numbers of additional CMUP muons in QCD and ghost events. $F_\pi$
            is the number of fake muons in ghost events, predicted assuming 
            that hadronic tracks are pions. If tracks are assumed to be kaons,
            the fake probability per track is approximately four times higher
            after including the in-flight-decay contribution. In
            QCD events, in which a large fraction of fake muons is due to kaons,
            the number of $SS$ 
            combinations underestimates the fake muon contribution
            to $OS$ combinations by approximately 10\%.}
 \begin{center}
 \begin{ruledtabular}
 \begin{tabular}{lcccc}
  Topology &  All     & QCD            &     Ghost     &    $F_\pi$   \\
  $OS$     &  $10812$ & $7380\pm 172$  &  $3432\pm 201$&  $216\pm 44$ \\
  $SS$     &  $4400$  & $2635\pm 104$  &  $1765\pm 123$&  $138\pm 35$ \\
 \end{tabular}
 \end{ruledtabular}
 \end{center}
 \label{tab:tab_7bis}
 \end{table} 
%%%%%%%%%%%%%%%%%%%%%%%%%%
 One notes that the fake contribution is much reduced at the expense
 of the muon acceptance that decreases by a factor of approximately five.
 The fraction of real additional CMUP muons is ($0.40\pm0.01$)\% in QCD
 events, and is four times larger $(1.64 \pm 0.08) $\% in ghost events. 
 This result is consistent with the previous determination that uses all
 muon detectors.
 
 To summarize, ghost events have the following features that differentiate
 them from QCD events. A $\cos \theta \geq 0.8$ cone around the direction
 of a primary muon contains twice as many tracks as QCD events. These cones  
 contain a number of additional real muons that is approximately four times
 larger than in QCD events. Since approximately 50\% of the ghost events 
 is accounted for by ordinary contributions, the remaining fraction
 contains a surprisingly large number of tracks and muons with
 $p_T \geq 2 \; \gevc$ per event. The shapes of the muon impact parameter
 distributions in QCD and ghost events are different. Additional and 
 initial muons in ghost events  have a tail that extends well
 beyond that observed in the QCD events. Impact parameters of muons
 contained in a 36.8$^{\deg}$ cone are loosely correlated.
 
\begin{figure}
 \begin{center}
 \vspace{-0.3in}
 \leavevmode
 \includegraphics*[width=\textwidth]{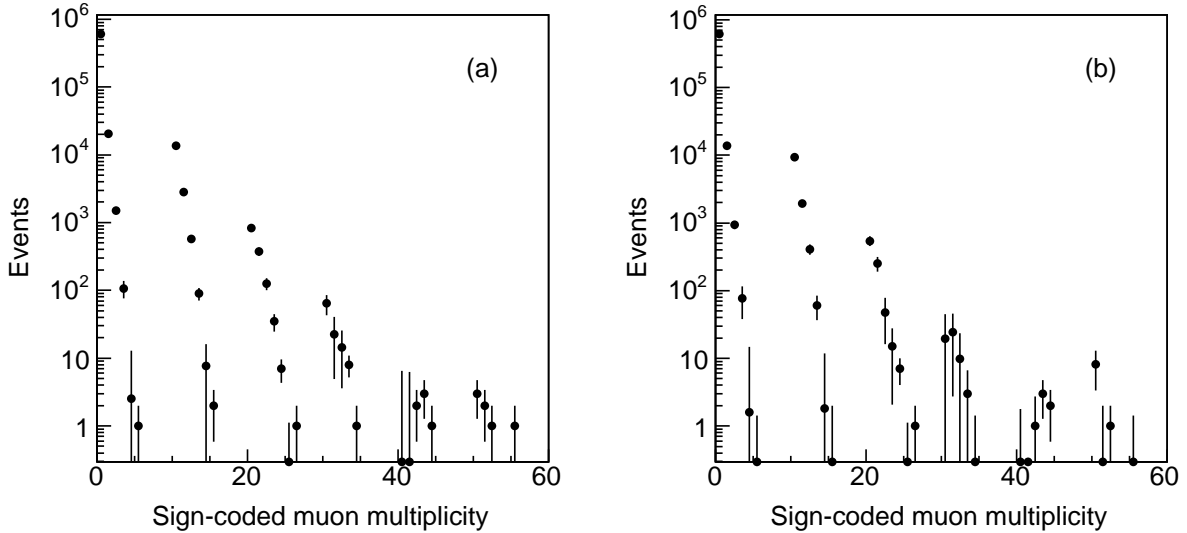}
 \caption[]{Sign-coded multiplicity distribution of additional muons found
            in a $\cos \theta \geq 0.8$ cone around the direction of a primary
            muon in ghost events before (a) and after (b) correcting for the
            fake muon contribution. An additional muon increases the 
            multiplicity by 1 when it  has the opposite sign and by 10 when 
            it has same sign charge as the initial muon. The background 
            subtracted distribution is also listed in
	    Table~\ref{tab:tab_mult}.}
 \label{fig:fig_14}
 \end{center}
 \end{figure}

 Figure~\ref{fig:fig_14}~(a) shows the distribution of the number of muons
 found in a $\cos\theta \geq 0.8$ cone around a primary muon due to ghost 
 events. In the plot, an additional muon increases the multiplicity by 1 
 when of opposite sign and by 10 when of the same sign charge as the initial
 muon~\footnote{
 As examples, the 3rd bin indicates cones with 3 muons with charge ($+--$)
 or ($-++$); and the 21st bin indicates cones with  3 muons with charge
 ($+++$) or ($-\:-\:-$).}.
 It is clear that a small fractions of ghost events contains a very large
 muon multiplicity.  The contribution of fake muons is  estimated assuming
 that the large majority of the tracks in a $\cos \theta \geq 0.8$ cone 
 are pions. We correct the distribution in Fig.~\ref{fig:fig_14}~(a) as
 follows. Given an event with $n$ muons, we loop over the tracks not 
 associated to a muon stub and with the same kinematic properties of muon
 candidates and randomly generate fake muons using the probability that
 a pion mimics a muon signal. If $m$ is the number of generated fake muons,
 we remove one event with $m+n$ muons in the distribution in 
 Fig.~\ref{fig:fig_14}~(a) and add one event to the bin with $n$ muons.
 The fake subtraction reduces the number of  $\cos \theta \geq 0.8$ cones 
 that contain one or more additional muons from 40409 to 27539.
 The resulting distribution is shown in Fig.~\ref{fig:fig_14}~(b).
 In conclusion, we are capable of predicting the number of additional
 muons in events in which the initial muons originate inside the beam pipe.
 In this case, the dominant sources of events are heavy flavor, $\Upsilon$
 and Drell-Yan production, and most of the additional muons arise from 
 sequential decays of single $b$ quarks.
 In contrast,  it seems difficult to account for the muon multiplicity
 distribution shown in Fig.~\ref{fig:fig_14}~(b)
 if the ghost events were  all due to ordinary sources, such as in-flight decays of pions and kaons,
 or hyperon decays in which the punchthrough of a hadronic prong mimics
 a muon signal.
 %%%%%%%%%%%%%%%%%%%%%%%%%%
%%%%%%%%%%%%%%%%%%%%%%%%% 
%%%%%%%%%%%%%%%%%%%%%%%%%
 \begin{table}
 \caption[]{Sign-coded, background subtracted, muon multiplicity in
            ghost events. Bins without entries are not shown. The
            multiplicity is not acceptance corrected because we do not know
            the mechanism producing ghost events. However, the detector
            acceptance for an additional muon with $p_T\geq 2\; \gevc$
            and $|\eta| \leq 1.1$ is $0.838 \pm 0.004$. The detector
            acceptance for an initial muon with $p_T\geq 3 \; \gevc$ and
            $|\eta|\leq 0.7$ is $0.506 \pm 0.003$. }
 \begin{center}
 \begin{ruledtabular}
 \begin{tabular}{lclc}
  Bin  &            Content    & Bin    &          Content       \\
   0   &  $ 620307 \pm 3413$   &  30    &  $   19.4  \pm   25.6$ \\
   1   &  $  13880 \pm  573$   &  31    &  $   24.2  \pm   21.5$ \\
   2   &  $    941 \pm  135$   &  32    &  $    9.8  \pm   13.8$ \\
   3   &  $     77 \pm   39$   &  33    &  $    3.0  \pm    3.6$ \\
   4   &  $    1.6 \pm   13.2$ &  34    &  $    0.00 \pm    1.4$ \\
   5   &  $      0.0 \pm  1.4$ &  40    &  $   -7.4  \pm    9.2$ \\
  10   &  $  9312  \pm  425$   &  41    &  $   -7.2  \pm    7.0$ \\
  11   &  $  1938  \pm  173$   &  42    &  $    1.0  \pm    1.7$ \\
  12   &  $   409  \pm   71$   &  43    &  $    3.0  \pm    1.7$ \\
  13   &  $    60  \pm   23$   &  44    &  $    2.0  \pm    1.4$ \\
  14   &  $     1.8\pm   10.1$ &  50    &  $    8.1  \pm    4.8$ \\
  15   &  $     0.0\pm    2.0$ &  51    &  $    0.0  \pm    2.0$ \\
  20   &  $   542  \pm   91$   &  52    &  $    1.0  \pm    1.0$ \\
  21   &  $   251  \pm   61$   &  55    &  $    0.0  \pm    1.4$ \\
  22   &  $    47  \pm   31$   &        &                        \\
  23   &  $    14.9\pm   12.8$ &        &                        \\
  24   &  $     7.0\pm    3.0$ &        &                        \\
  25   &  $    -3.1\pm    4.2$ &        &                        \\
  26   &  $     1.0\pm    1.0$ &        &                        \\
 \end{tabular}
 \end{ruledtabular}
 \end{center}
 \label{tab:tab_mult}
 \end{table} 
%%%%%%%%%%%%%%%%%%%%%%%%%%
\subsection{Robustness of the fake muon prediction}
\label{sec:ss-fkrob}
%%%%%%%%%%%%%%%%%%%%%%%%%%
 It is important to further verify that such a large number of muons
 contained in such a small angular cone is not a detector artifact.
 The display of the muon chamber hits in events that contain four
 or more muons did not yield any indication of a detector malfunction.
 However, there are events in which certain areas of the muon detectors
 have a dense clustering of dozens of hits. In these events, some muons
 correspond to tracks linked to muon stubs constructed in those clusters.
 We estimate the muon fake rate using the probability that pions and kaons
 from $D^0$ decays mimic a muon signal. After requiring that $D^0$ 
 candidates have an appreciable proper decay time in order to select $D^0$
 mesons from $b$-hadron decays, a 36.8$^{\deg}$ cone around the direction
 of these tracks  contains an average of 0.02 muons and 1.6 additional
 tracks with $p_T \geq 2 \; \gevc$. The muon fake probability does not
 increase at all when using $D^0$ prongs accompanied by at least two tracks.
 However, the  multiplicity in a 36.8$^{\deg}$ cone around the direction
 of the $D^0$ prongs does not have the high multiplicity tail of ghost 
 events. We do not possess a data set of a known process with a number
 of tracks  and muons as large as in ghost events that could be used to 
 verify the muon detector response to this type of event.
 
 One concern is that our procedure underestimates the fake rate in 
 multi-muon events in which hadronic tracks can take advantage of hits
 in the muon chambers produced by real muons or by hadronic punchthrough.
 Our muon selection criteria were not optimized for this type
 of event. A track is accepted as a muon if the distance of its 
 projection onto a muon detector from a muon stub is $\Delta x\leq$ 30, 40,
 and 30 cm for the CMU, CMP, and CMX detector, respectively. For CMX 
 or CMU muons with $p_T=2\; \gevc$, these $\Delta x $ cuts correspond to
 the requirement that the track extrapolation and the muon stub match
 within 3 $\sigma$ in the $r-\phi$ plane, where $\sigma$ is a rms 
 deviation that includes the effect of multiple scattering and energy loss.
 We have selected additional muons by adding the increasingly stricter
 requirements that track-stub matches are within 3 and 2 $\sigma$, 
 respectively. The latter requirement reduces the number of multiple-muon
 combinations by a factor of two, but does not affect the salient features
 of the multiplicity distribution in  Fig.~\ref{fig:fig_14}~(a). We have
 compared $\Delta x $ and  $\sigma$ distributions of muon-track matches
 for the different muon detectors in QCD and ghost events (see Appendix~A).
 These distributions, as well as the fractional usage of different muon
 detectors, in ghost events are not significantly different to those
 of QCD events. Since we are able to predict the rate of additional
 muons in QCD events, which have a larger fake muon background than
 ghost events, the present estimate of the fake muon contribution is an
 unlikely candidate to explain the large excess of additional muons in
 ghost events.

 As shown at the beginning of Sec.~\ref{sec:ss-3mu}, we have identified 
 $5348 \pm 225$  $K^0_S$ candidates in the dimuon data,  and $96 \pm 41$
 of them contain at least an additional muon in the event. By applying the 
 fake muon probability to all candidate tracks in events with a $K^0_S$
 candidate, we predict $86 \pm 30$  events with at least an  additional
 fake muon, consistent with the observation. 

\begin{figure}[]
 \begin{center}
 \vspace{-0.2in}
 \leavevmode
 \includegraphics*[width=\textwidth]{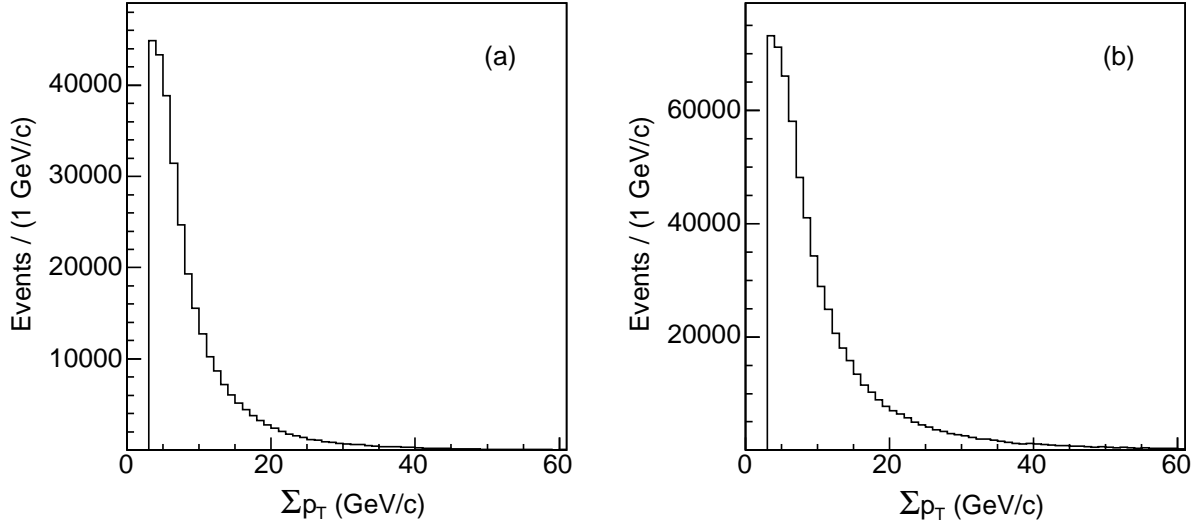}
 \caption[]{Distribution of the transverse momentum carried by all tracks
            with $p_T \geq 1 \; \gevc$ contained in a $36.8^{\deg}$ cone 
            around an initial muon in (a) QCD and (b) ghost events.}
 \label{fig:fig_sumpt}
 \end{center}
 \end{figure}

 Traditionally, searches for soft ($p_T\geq 2 \; \gevc$) muons performed 
 by the CDF collaboration estimate the fake muon contribution by using a
 fake probability per track~\cite{topev}. One could argue that the excess of
 muons in ghost events were due to a breakdown of this method when applied
 to high $E_T$ jets with many tracks that are not contained in the 
 calorimeter and muon absorber. This effect was not observed in previous
 analyses. We would also have observed the presence of multi-muons events
 in the QCD contribution because, as shown in Fig.~\ref{fig:fig_sumpt},
 the distributions of the transverse momentum carried by all tracks with
 $p_T \geq 1 \; \gevc$ and contained in a $36.8^{\deg}$ cone 
 are quite similar in ghost and QCD events.  
%%%%%%%%%%%%%%%%%%%%%%%%%%
%%%%%%%%%%%%%%%%%%%%%%%%%%%%%%%%%

\begin{figure}[]
 \begin{center}
 \vspace{-0.2in}
 \leavevmode
 \includegraphics*[width=\textwidth]{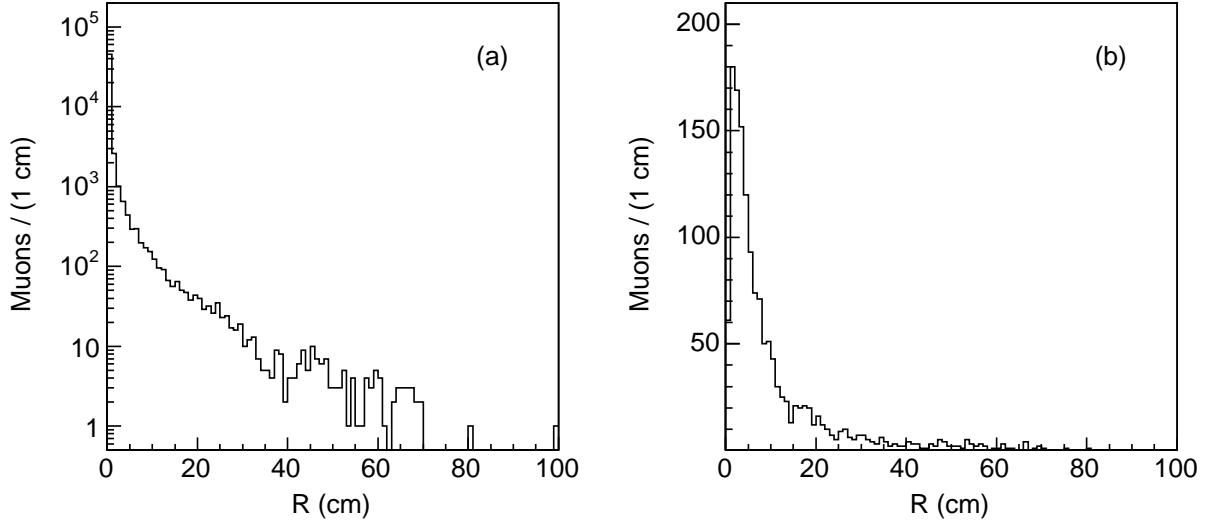}
 \caption[]{Distributions of $R$, the distance of dimuon vertices from
            the nominal beam line for initial muons with impact parameter
            (a) smaller and (b) larger than 0.3 cm.}
 \label{fig:fig_sint1}
 \end{center}
 \end{figure}

 The appearance of multi-muon events seems to be correlated with the
 presence of muons with large impact parameters. As discussed earlier,
 multi-track secondary interactions in the detector volume do not contribute
 significantly to the total number of ghost events. This does not exclude
 the possibility that the smaller number of multi-muon ghost events are 
 due to secondary interactions in the detector volume that point into
 calorimeter cracks. We search for secondary interactions by combining
 initial muons with small and large impact parameters with all additional
 muons in a 36.8$^{\deg}$ cone around the muon  direction. Dimuon 
 combinations are constrained to arise from a common space point. They are
 discarded if the three-dimensional vertex fit returns a $\chi^2$ 
 larger than 10. The distribution of $R$, the distance of a
 reconstructed secondary vertex from the detector origin in the plane
 transverse to the beam line, is shown in Fig.~\ref{fig:fig_sint1} for
 initial muons with small and large impact parameters. The absence of 
 spikes in the distributions shows that secondary interactions are not
 a significant source of multi-muon events.
%%%%%%%%%%%%%%%%%%%%%%%%%%
%%%%%%%%%%%%%%%%%%%%%%%%%%%%%%%%%
\section{Investigation of additional properties of multi-muon events}
\label{sec:ss-inter}
\begin{figure}
 \begin{center}
 \vspace{-0.3in}
 \leavevmode
 \includegraphics*[width=\textwidth]{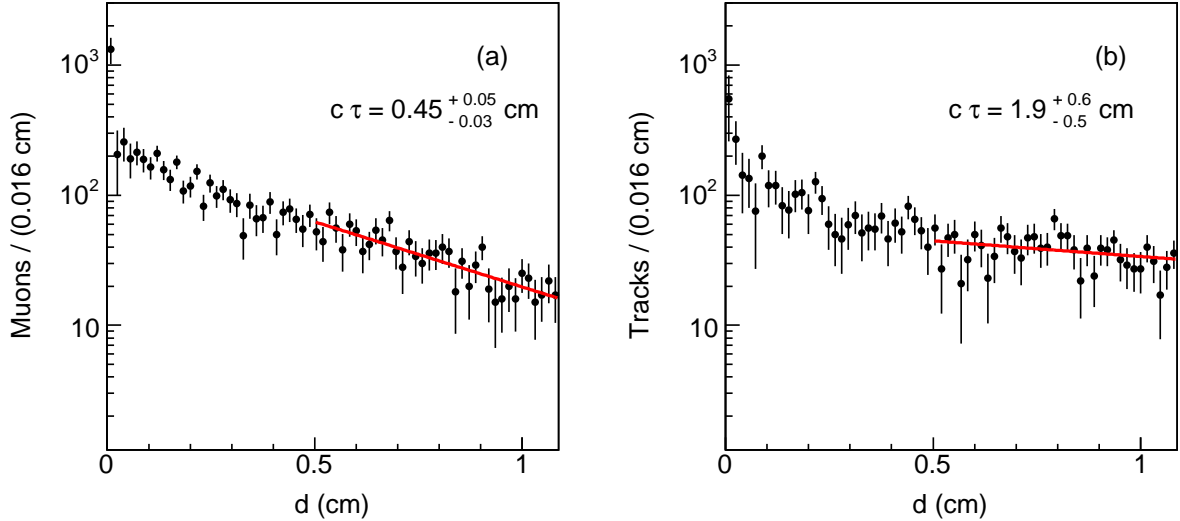}
 \caption[]{ Impact parameter distributions of (a) initial muons and (b)
             tracks of identified $K^0_S$ decays. The combinatorial
             background under the $K^0_S$ signal in  Fig.~\ref{fig:fake_3}
             has been removed using a sideband subtraction method.}
 \label{fig:fake_5}
 \end{center}
 \end{figure}

%%%%%%%%%%%%%%%%%%%%%%%%%%%%%%%%%
 We study the muon impact parameter distributions for the subset
 of ghost events in which a cone contains two or more muons.
 %%%%%%%%%%%%%%%%%%%%%%%%%%
 \begin{figure}
 \begin{center}
 \vspace{-0.3in}
 \leavevmode
 \includegraphics*[width=\textwidth]{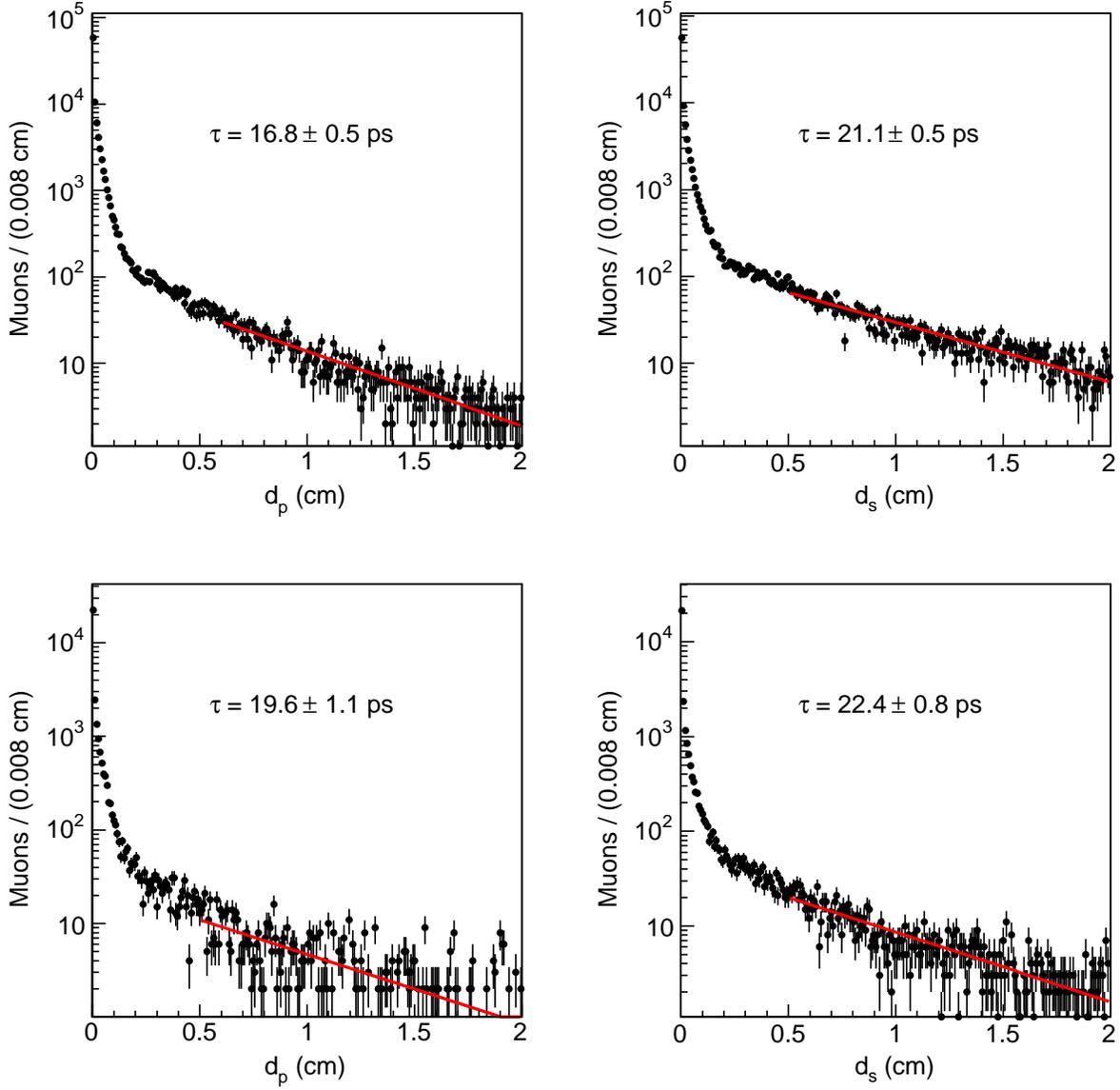}
 \caption[]{Muon impact parameter distributions for events containing (top) 
            only two muons or (bottom) more than two muons in a 
            $\cos \theta  \geq 0.8$ cone. We call $d_p$ and $d_s$ the
            impact parameter of  initial and additional muons, respectively.
            The solid lines represent fits to the data distribution with 
            an exponential function. The fit result is shown in each plot.}
 \label{fig:fig_15}
 \end{center}
 \end{figure}
%%%%%%%%%%%%%%%%%%%%%%%%%
 The impact parameter distribution of initial muons due to ghost events in
 Fig.~\ref{fig:fig_4bis} is derived using muon tracks that pass the loose
 SVX selection in order to minimize the possible contribution of interactions
 in the detector systems surrounding the SVXII detector. As mentioned earlier,
 this requirement sculpts the impact parameter distribution of muons arising
 from the decay of particles with a lifetime much larger than that of $b$
 quarks. The smaller number of events that contain two or more muons in a
 $\cos \theta \geq 0.8$ cone cannot be significantly contaminated by secondary
 interactions, and we select these muons without any SVX requirement.
 The corresponding impact parameter distributions are shown in
 Fig.~\ref{fig:fig_15}. In the assumption that the exponential tail
 at large impact parameter is produced by the decay of long-lived
 objects,  fits with an exponential function to the impact
 parameter distributions of additional muons in the range $0.5-2.0$ cm
 return a slope of approximately $21.4 \pm 0.5$ ps~\footnote{
 The error is statistical. A study of systematic effects due to possible
 background contributions is beyond the scope of this pioneering study.}.
 The fits to the impact parameters of initial muons yield smaller values
 of the slope. The difference is understood in term of kinematic and 
 trigger biases affecting the initial muons. As an example,
 Fig.~\ref{fig:fake_5} compares the result of fits to the impact parameter
 of muons and tracks corresponding to identified $K^0_S$ decays. 
 The fit to the track impact parameter yields a $K^0_S$ lifetime in
 agreement with the  PDG value of $\tau=89.6$ ps. In contrast, the
 lifetime measurement using initial muons yields a much smaller lifetime 
 value. The slope
 returned by the fits to the impact parameter tail of additional muons in
 ghost events is different from the lifetime of any known particle. 

\begin{figure}
 \begin{center}
 \vspace{-0.3in}
 \leavevmode
 \includegraphics*[width=\textwidth]{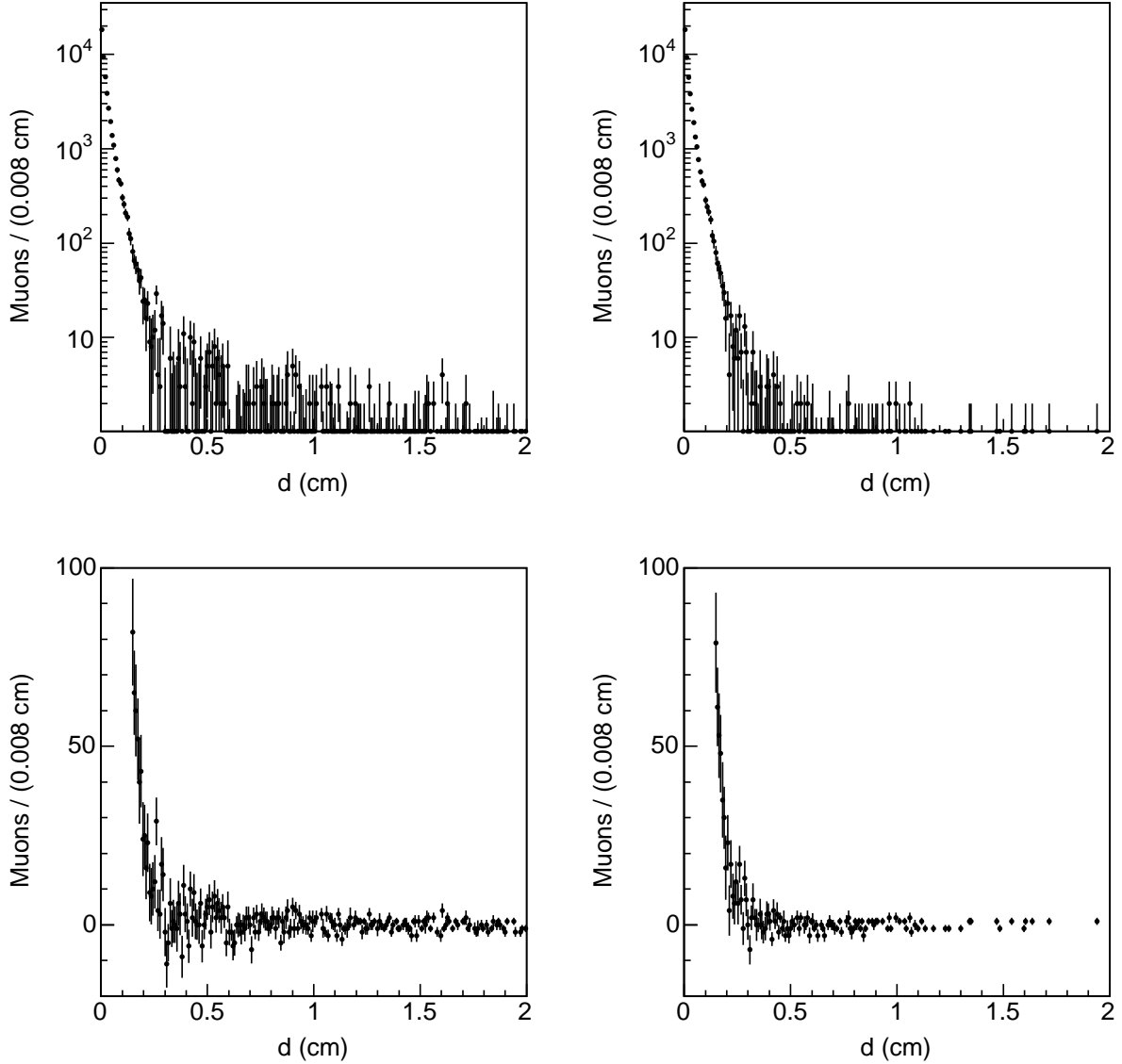}
 \caption[]{Impact parameter distributions of CMUP muons which are
            accompanied by a $D^0$ meson and are selected without
            (left) SVX  or with (right) loose SVX requirements.
            The bottom plots are magnified views to show distributions at 
            large impact parameters. The contribution of the combinatorial
            background under the $D^0$ signal has been removed with a
            sideband subtraction method. } 
 \label{fig:fig_15bis}
 \end{center}
 \end{figure}

\begin{figure}
 \begin{center}
 \vspace{-0.3in}
 \leavevmode
 \includegraphics*[width=\textwidth]{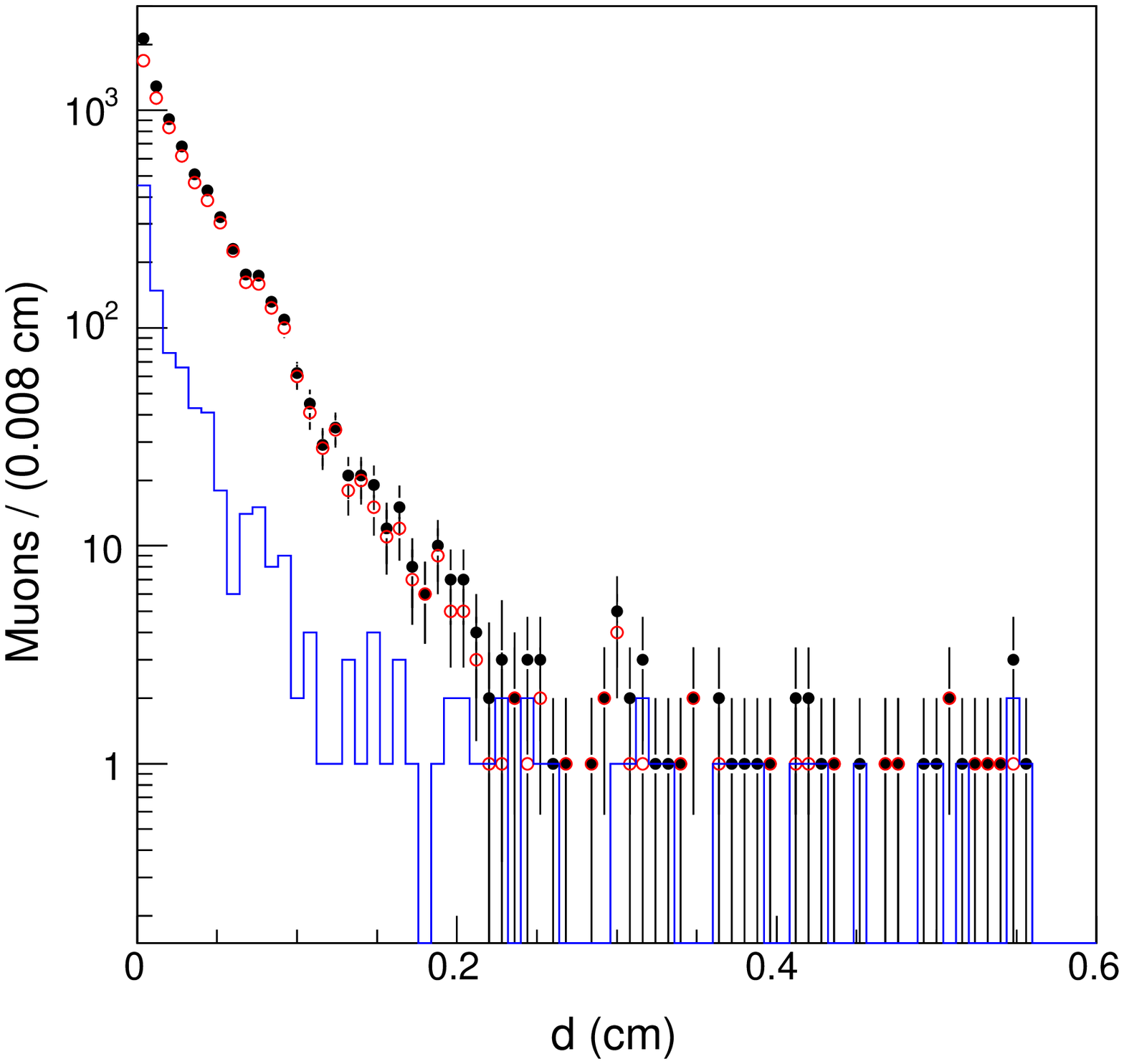}
 \caption[]{Impact parameter distributions of muons accompanied by a $D^0$
            meson and selected as the additional muons in this analysis.
            No SVX requirements are applied. All events ($\bullet$) are
            compared to $RS$  ($\circ$) and  $WS$ (histogram) combinations
            (see text). The contribution of the combinatorial background
            under the $D^0$ signal has been removed with a sideband
            subtraction method.}
 \label{fig:fig_15tris}
 \end{center}
 \end{figure}

 Conversely, one might wonder if the impact parameter tail is a detector
 effect that has not been noticed in  $t$- and $b$-quark studies
 performed by the CDF collaboration because these analyses customarily
 utilize muons and tracks with impact parameters smaller than $0.1-0.2$ cm.
 We study the impact parameter distributions of CMUP trigger muons
 accompanying a $D^0 \rightarrow \pi^+ K^-$ and charge conjugate
 candidates. We use events acquired with the $\mu$-SVT trigger and
 reconstruct $D^0$ candidates by attributing the kaon mass to the track
 with the same charge as the muon ($RS$ combinations as expected for
 $\mu+D^0$ systems produced by $b$ hadron decays). We retain combinations
 in which the muon plus two-track system has an invariant mass smaller
 than 5 $\gevcc$. No wrong-sign ($WS$) combinations are found. We use
 a sideband subtraction method to remove the combinatorial background
 in the invariant mass region corresponding to the $D^0$ signal. The
 impact parameter distribution of CMUP muons produced by $b$ hadron
 decays, shown in Fig.~\ref{fig:fig_15bis}, does not have the large
 tail at large impact parameters that is characteristic of initial
 muons in ghost events. Figure~\ref{fig:fig_15tris} is the analogous
 plot when muons are selected as the  additional muons in this
 analysis ($p_T \geq 2 \; \gevc$ and $|\eta| \leq 1.1$). No high
 impact parameter tails are observed. The fraction of fake muons,
 measured as the number of $WS$ combinations, is approximately 2\%. 
%%%%%%%%%%%%%%%%%%%%%%%%%%
%%%%%%%%%%%%%%%%%%%%%%%%%
%%%%%%%%%%%%%%%%%%%%%%%%%%
%%%%%%%%%%%%%%%%%%%%%%%%%
%%%%%%%%%%%%%%%%%%%%%%%%%%
%%%%%%%%%%%%%%%%%%%%%%%%%
\subsection{Lifetime}
\label{sec:ss-lifetime}
 The fact that multi-muon events have been isolated by the request that at 
 least one of the trigger muons originates outside of the beam pipe suggests
 that they could be associated with objects with  lifetime much larger than
 that of $b$ quarks. In the previous section, we have estimated the lifetime
 by using a small fraction of events in the tail of the muon impact parameter
 distribution. In the following, we search for a confirmation based on
 the entire sample of ghost data. We have seen in the previous section
 that  the impact parameters of muons contained in the same cone are not 
 strongly correlated. This would happen if each muon arise from the decay of
 a different object. Therefore, we search for secondary vertices produced by
 pairs of tracks with $p_T \geq 1 \; \gevc$ and opposite charge contained in
 a $36.8^{\deg}$ cone around the direction of each initial muon. Track pairs
 are constrained to arise from a common space point. Combinations are
 discarded if the three-dimensional vertex fit returns a $\chi^2$ larger
 than 10. If a track is associated with  more than one secondary vertex, 
 we discard those with lower fit probability. For each secondary vertex,
 we define $L_{xy}$ as the distance between the secondary and primary 
 event vertices projected onto the transverse momentum of the two-track
 system. Combinations of tracks arising from the primary vertex or from 
 the decay of different objects yield a $L_{xy}$ distribution symmetric
 around $L_{xy}=0$. An excess at positive $L_{xy}$ is a property of the
 decay of a long-lived object.

\begin{figure}
 \begin{center}
 \vspace{-0.3in}
 \leavevmode
 \includegraphics*[width=\textwidth]{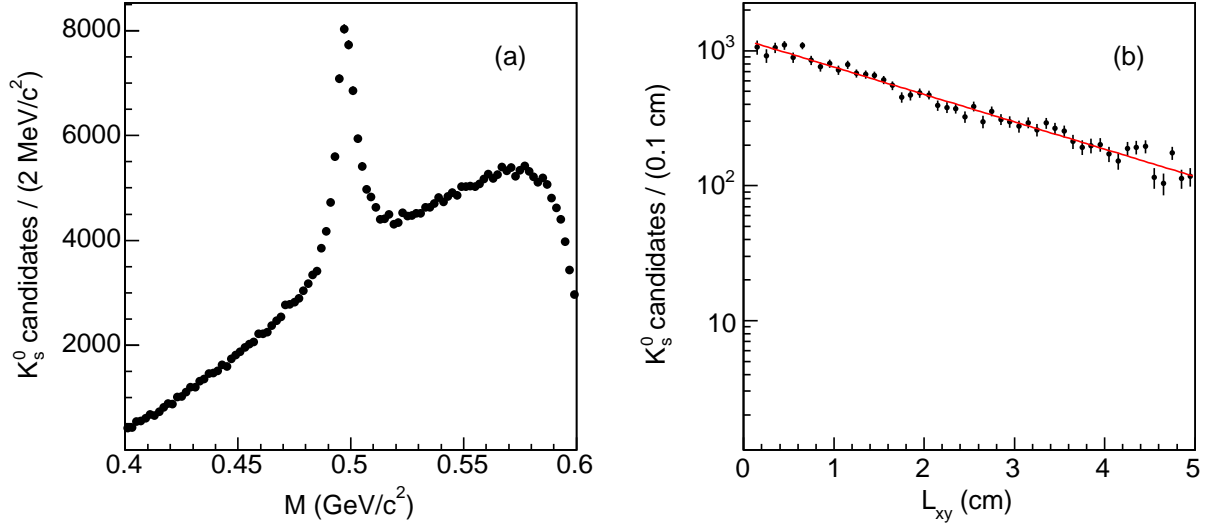}
 \caption[]{Invariant mass distribution (a) of $K_S^0 \rightarrow \pi^+ \pi^-$
            candidates. The background subtracted $L_{xy}$ distribution of
            $K_S^0$ mesons (b) is compared to the expectation based on the
            $K_S^0$ measured lifetime~\cite{pdg}.}
 \label{fig:fig_k0s}
 \end{center}
 \end{figure}

 We use $K_S^0 \rightarrow \pi^+ \pi^- $ decays to verify with data the 
 detector response in the impact parameter region populated by ghost events.
 We search for $K_S^0$ decays in the dimuon dataset used for this analysis
 by pairing tracks of opposite charge with $p_T \geq 0.5 \; \gevc$,
 $|\eta| \leq 1.1$, and opening angle smaller than 60$^{\deg}$. Track
 combinations are constrained to arise from a common space point.
 Combinations are discarded if the three-dimensional vertex fit returns
 a $\chi^2$ larger than 10 or the  $L_{xy}$ distance is smaller than 0.1 cm.
 In this case, the $L_{xy}$ distance is also corrected for the Lorentz
 boost of the two-track system. Figure~\ref{fig:fig_k0s}~(a) shows the
 invariant mass spectrum of the two-track systems passing this selection.
 The combinatorial background under the $K_s^0$ signal, integrated from
 0.486 to 0.510 $\gevcc$, is removed by subtracting the events contained
 in the side bands $0.474-0.486$ and $0.510-0.522$ $\gevcc$. The
 background subtracted $L_{xy}$ distribution, shown in
 Fig.~\ref{fig:fig_k0s}~(b) is consistent with the $K_S^0$ lifetime
 of 89.5 ps~\cite{pdg}.
 %%%%%%%%%%%%%%%%%%%%%%%%%%
%%%%%%%%%%%%%%%%%%%%%%%%% 

\begin{figure}
 \begin{center}
 \vspace{-0.3in}
 \leavevmode
 \includegraphics*[width=\textwidth]{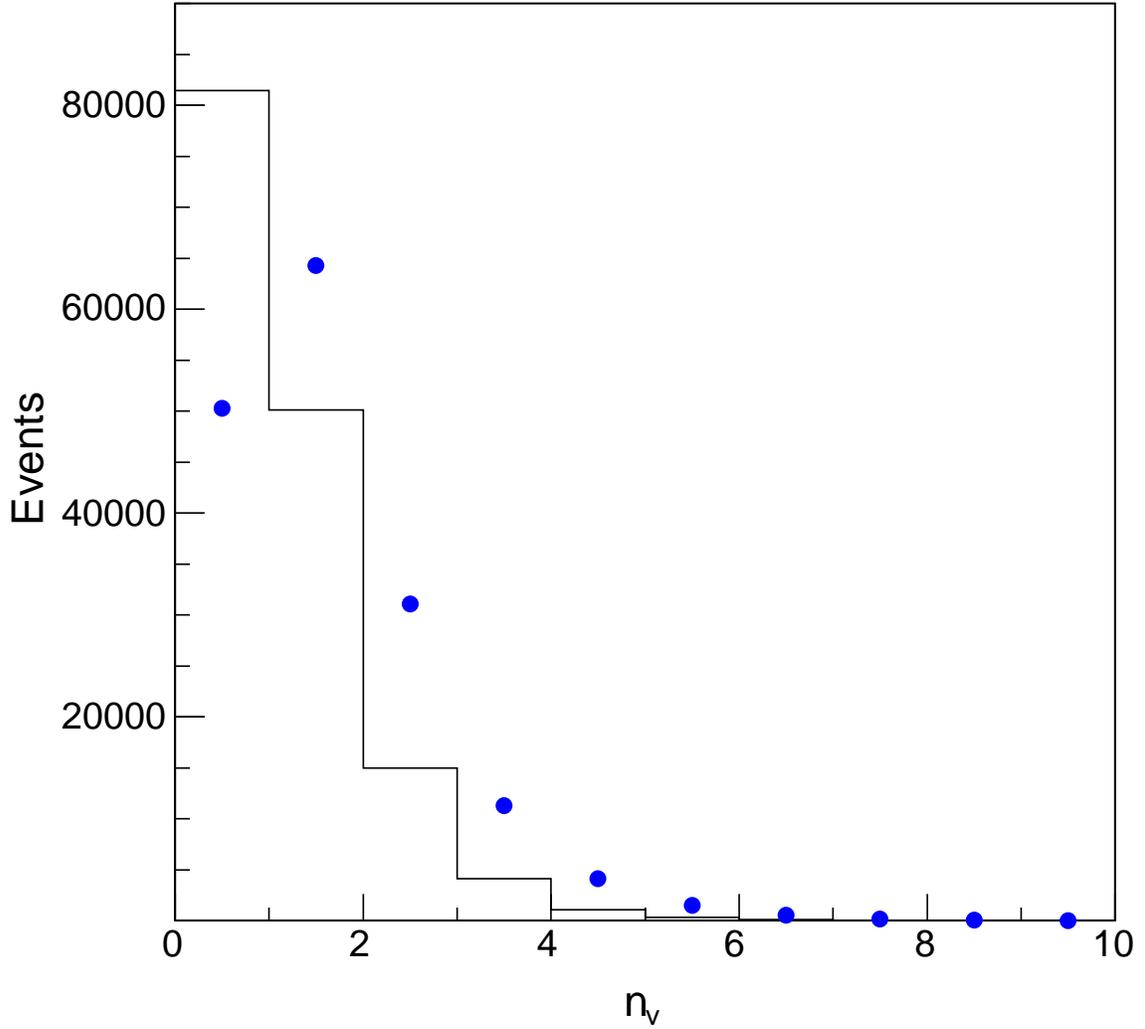}
 \caption[]{Distribution of $n_v$, the number of reconstructed secondary
            vertices of opposite sign track pairs in QCD (histogram) and
            ghost ($\bullet$) events. We use all tracks with
            $p_T \geq 1 \; \gevc$ contained in a $36.8^{\deg}$ cone around
            the direction of each initial muon.}
 \label{fig:fig_nvi}
 \end{center}
 \end{figure}

\begin{figure}
 \begin{center}
 \vspace{-0.3in}
 \leavevmode
 \includegraphics*[width=\textwidth]{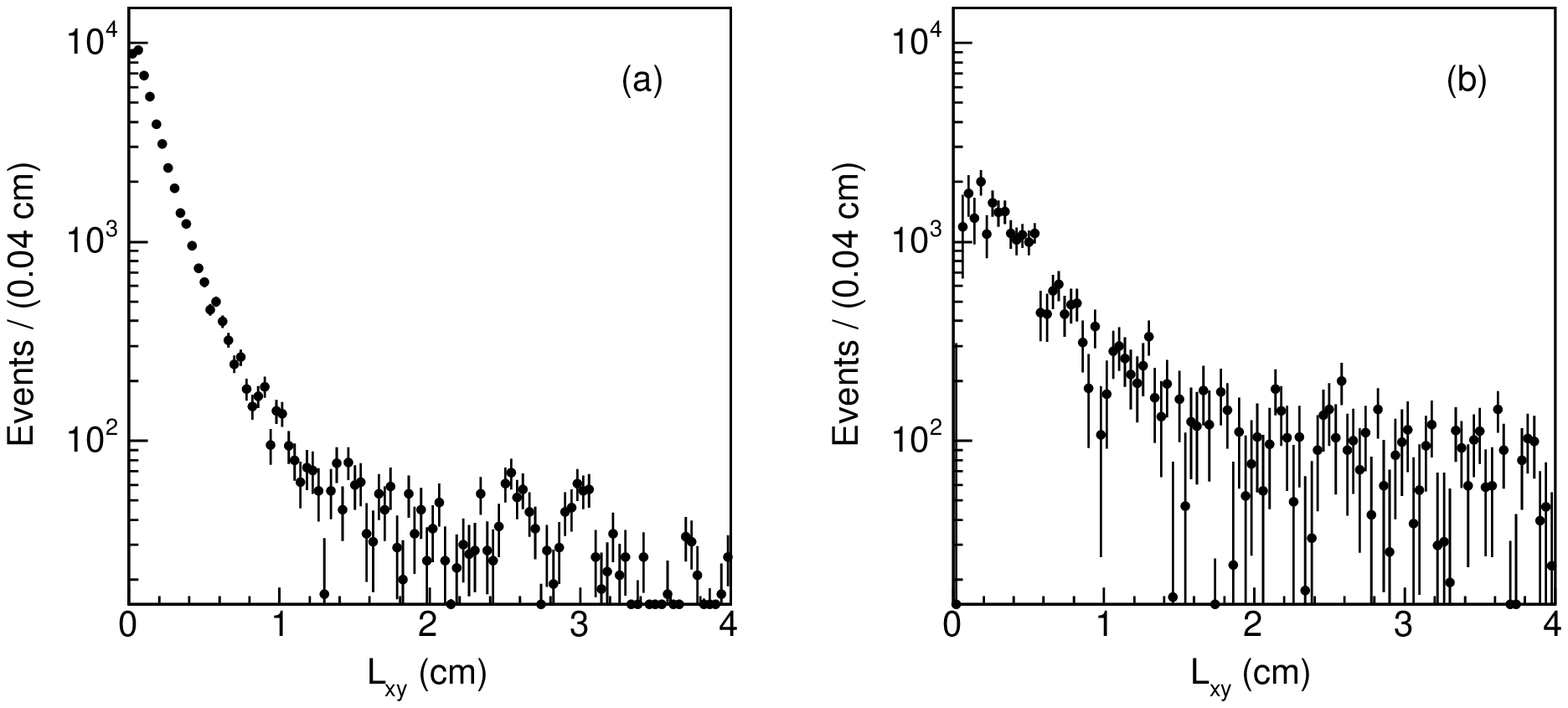}
 \caption[]{Distribution of the distance $L_{xy}$ of reconstructed secondary
            vertices due to long-lived decays in (a) QCD and (b) ghost events.
            The combinatorial background has been removed by subtracting the
            corresponding negative $L_{xy}$ distributions. The data correspond
            to an integrated luminosity of 742 pb$^{-1}$.}
 \label{fig:fig_nvlxy}
 \end{center}
 \end{figure}

 The distributions of the number of secondary vertices reconstructed in
 QCD and ghost events are shown in Fig.~\ref{fig:fig_nvi}.
 %%%%%%%%%%%%%%%%%%%%%%%%%%
%%%%%%%%%%%%%%%%%%%%%%%%% 
 Figure~\ref{fig:fig_nvlxy} shows the difference between the positive and
 negative $L_{xy}$ distributions of secondary vertices reconstructed in QCD
 and ghost events.
%%%%%%%%%%%%%%%%%%%%%%%%%%
%%%%%%%%%%%%%%%%%%%%%%%%%
 The shape of the distribution for ghost events is consistent with the
 hypothesis that a small but significant
  fraction of them arise from the production and
 decay of objects with a lifetime significantly larger than that of $b$
 hadrons and smaller than that of $K_S^0$ mesons. 
%%%%%%%%%%%%%%%%%%%%%%%%%%%%%%%%%
\subsection{Track multiplicity}
\label{sec:ss-trkmul}
\begin{figure}
 \begin{center}
 \vspace{-0.3in}
 \leavevmode
 \includegraphics*[width=\textwidth]{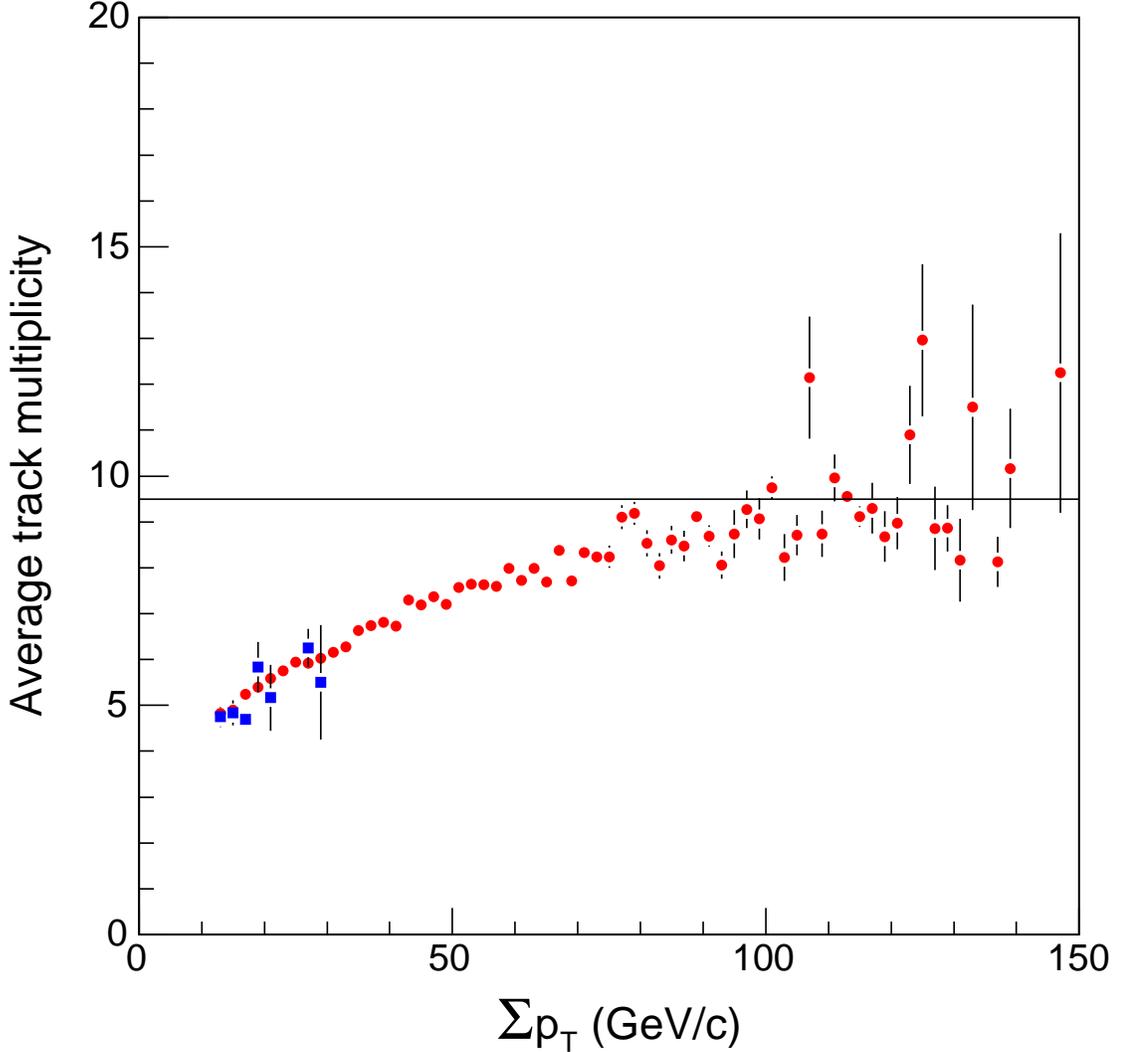}
 \caption[]{Average number of tracks in a $36.8^{\deg}$ cone around the
            direction of a primary muon as a function of $\sum p_T$, the
            transverse momentum carried by all the tracks. We use cones
            containing at least three muons. Data ($\bullet$) are compared
            to the QCD expectation ({\tiny $\blacksquare$}) based on the
            few events predicted by the heavy flavor simulation, normalized
            to the number of initial dimuons in the data and implemented
            with the probability that hadronic tracks mimic a muon signal.
            The detector efficiency for these tracks is close to unity.}
 \label{fig:fig_17}
 \end{center}
 \end{figure}

%%%%%%%%%%%%%%%%%%%%%%%%%%%%%%%%%
 As discussed in Sec.~\ref{sec:ss-a0}, ghost events include a sizable
 contribution from ordinary sources such as in-flight-decays, and $K^0_S$
 and hyperon decays. The average track multiplicity in ghost events is 
 a factor of two larger than in QCD events. In order to study the average
 multiplicity of multi-muon events, we use events that contain at least
 three muons in a $36.8^{\deg}$ cone. Figure~\ref{fig:fig_17} shows the
 average number of all tracks with $p_T \geq 2 \; \gevc$ contained in a
 $36.8^{\deg}$ cone around a primary muon as a function of the total
 transverse momentum of the tracks.
%%%%%%%%%%%%%%%%%%%%%%%%%%
%%%%%%%%%%%%%%%%%%%%%%%%%%%%%%%%%
\subsection{Cone correlations}
\label{sec:ss-cone}
%%%%%%%%%%%%%%%%%%%%%%%%%%%%%%%%%%%%%%%%%%%%%%%%%%%%%%%%%% 
 In the previous section, we have investigated the kinematics and topology
 of muons and tracks contained in a single $36.8^{\deg} $ cone around the
 direction of an initial muon. In this section, we extend the investigation
 to the rate and properties of events in which two $36.8^{\deg}$ cones
 contain a muon multiplicity larger than that of QCD events. After
 subtracting the QCD and fake muon contribution, in ghost events there are
 $27990\pm761$ cones that contain two or more muons, $4133 \pm 263$ cones
 that contain three or more muons, and $3016 \pm 60$ events in which both
 cones contain two or more muons. It follows that approximately 13\% of the
 ghost events in which one cone contains two or more muons also contain a
 second cone with the same feature. In events triggered by a central jet,
 the fraction of events also containing an additional central jet is
 $10-15$\% depending on the jet transverse energy~\cite{dijet}. Therefore,
 it is difficult to imagine detector effects that might produce a similar
 fraction of ghost events with two multi-muon cones.
 
\begin{figure}
 \begin{center}
 \vspace{-0.3in}
 \leavevmode
 \includegraphics*[width=\textwidth]{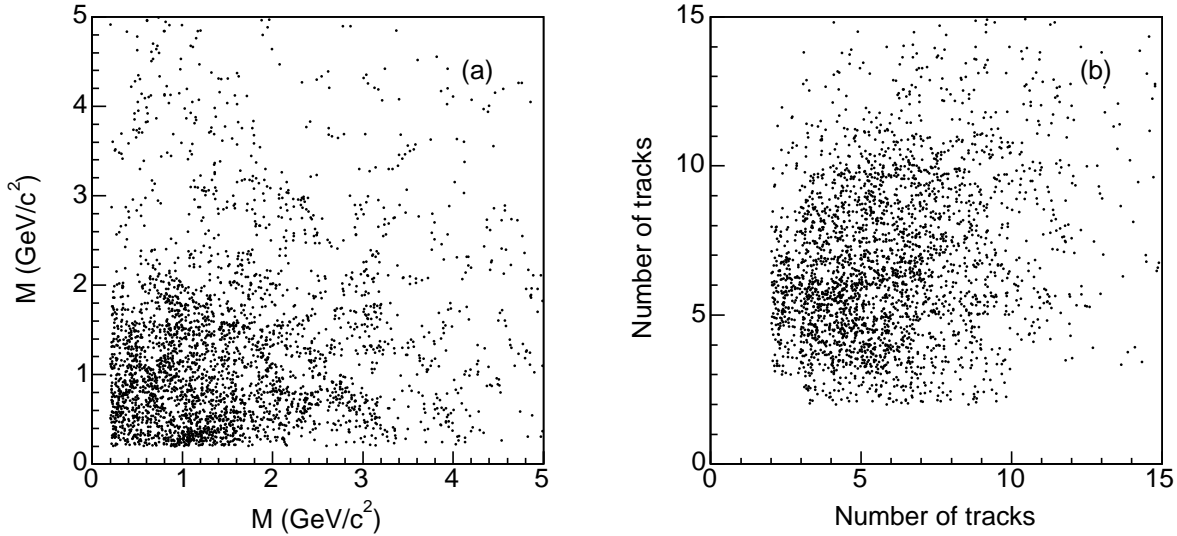}
 \caption[]{Two-dimensional distributions of (a) the invariant mass, $M$, 
            of all muons and (b) the total number of tracks contained in a 
            $36.8^{\deg}$ cone when both cones contain at least two muons.
            The QCD and fake muon contributions have been subtracted.}
 \label{fig:fig_19}
 \end{center}
 \end{figure}

\begin{figure}
 \begin{center}
 \vspace{-0.3in}
 \leavevmode
 \includegraphics*[width=\textwidth]{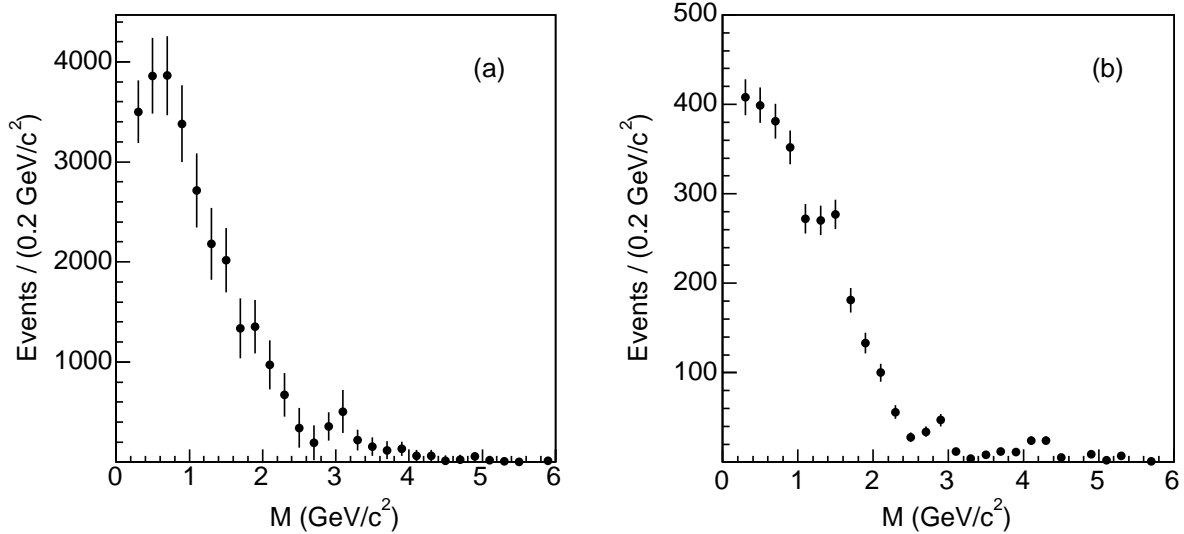}
 \caption[]{Distributions of invariant mass, $M$, of all muons contained
            in (a) the 27990 $36.8^{\deg}$ cones with two or more muons
            and (b) in each cone of  the 3016 events in which both cones
            contain two or more muons. The QCD and fake muon contributions
            have been subtracted. }
 \label{fig:fig_20}
 \end{center}
 \end{figure}

\begin{figure}
 \begin{center}
 \vspace{-0.3in}
 \leavevmode
 \includegraphics*[width=\textwidth]{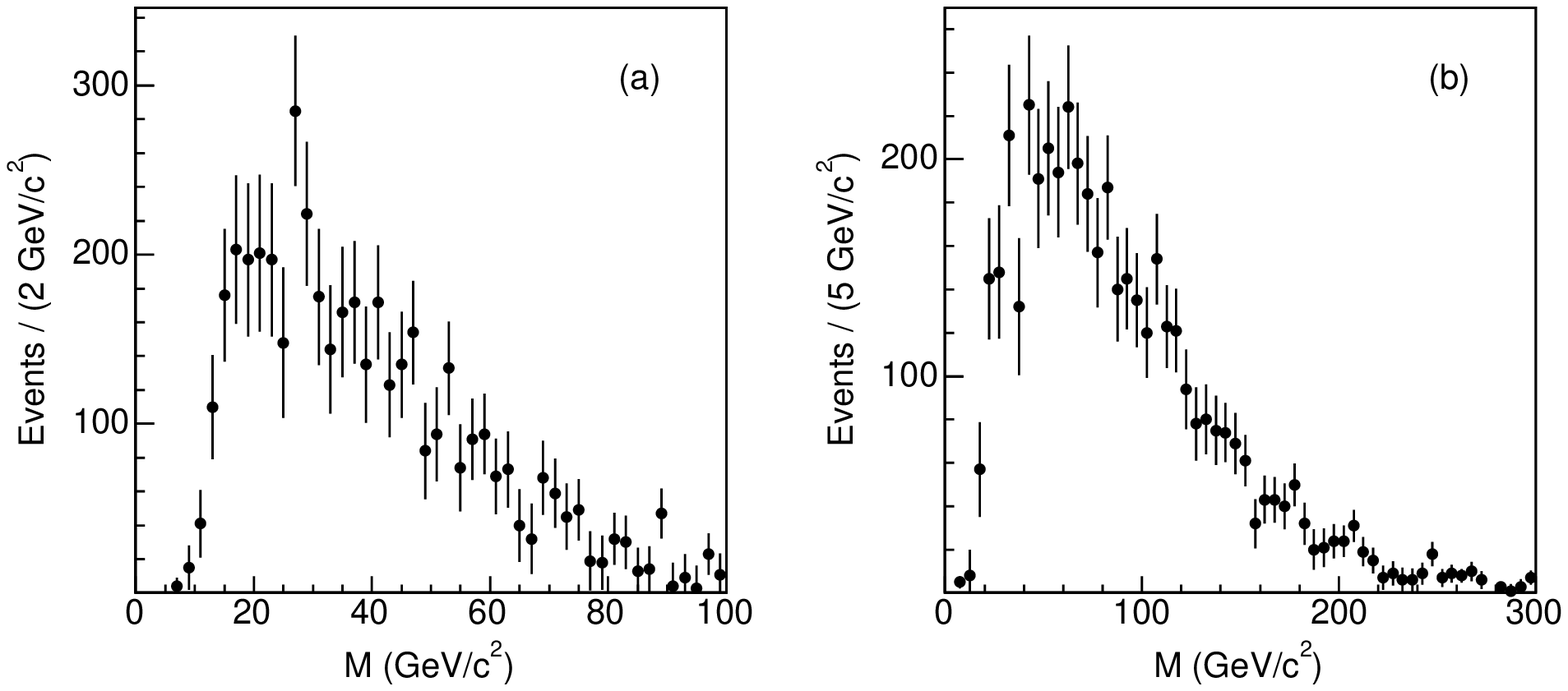}
 \caption[]{Invariant mass distribution of (a) all muons and (b) all tracks
            for events in which both cones contain at least two muons.
            The QCD and fake muon contributions are subtracted. The data
            correspond to an integrated luminosity of 2100 pb$^{-1}$.}
 \label{fig:fig_27}
 \end{center}
 \end{figure}

 The following distributions serve the purpose of 
 showing that, when a second cone containing multi muons is found,
 it has the same characteristics of the first found multi-muon cone.
 Figure~\ref{fig:fig_19} plots two-dimensional distributions of the
 invariant mass of all muons and of the number of tracks with
 $p_T \geq  2 \; \gevc$ contained in each cone for the 3016 events
 containing two cones with two or more muons.
%%%%%%%%%%%%%%%%%%%%%%%%%%
%%%%%%%%%%%%%%%%%%%%%%%%% 
 Figure~\ref{fig:fig_20} shows that the invariant mass distribution of
 all muons contained in the 27990 cones containing at least two muons
 is consistent with that of the 3016 events in which both cones contain
 at least two muons.  
%%%%%%%%%%%%%%%%%%%%%%%%%%
%%%%%%%%%%%%%%%%%%%%%%%%% 
 Figure~\ref{fig:fig_27} shows the invariant mass distribution of all muons
 and all tracks  with $p_T \geq  2 \; \gevc$ in events in which both cones
 contains two or more muons.
%%%%%%%%%%%%%%%%%%%%%%%%%%
%%%%%%%%%%%%%%%%%%%%%%%% 

\begin{figure}
 \begin{center}
 \vspace{-0.3in}
 \leavevmode
 \includegraphics*[width=\textwidth]{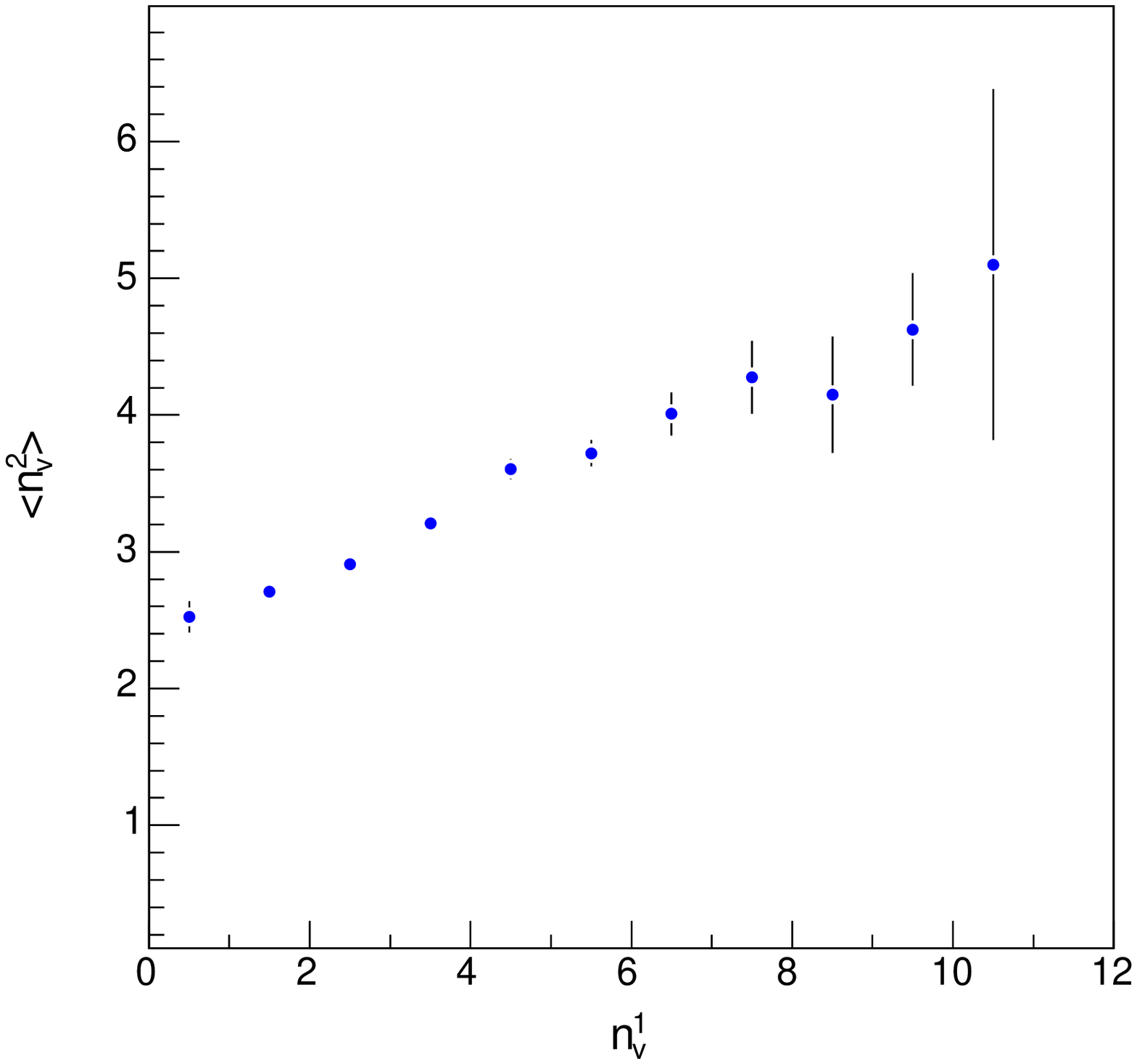}
 \caption[]{Average number of secondary vertices, $<n^2_v>$, in one cone as
            a function of the number of secondary vertices, $n^1_v$, observed
            in the recoiling cone. Both cones contain at least two muons.}
 \label{fig:fig_27bis}
 \end{center}
 \end{figure}

 Following the procedure outlined in Sec.~\ref{sec:ss-lifetime}, we count
 the number of secondary vertices of two-track systems in events with two
 cones containing at least two muons. Figure~\ref{fig:fig_27bis} shows the
 average number of secondary vertices in one cone as a function of the
 number of secondary vertices in the other cone.
%%%%%%%%%%%%%%%%%%%%%%%%%%
%%%%%%%%%%%%%%%%%%%%%%%%
\section{Conclusions}\label{sec:ss-concl} 
%%%%%%%%%%%%%%%%%%%%%%%%%
 We have studied a sample of events containing at least two central muons
 with $p_T \geq 3 \; \gevc$ and invariant mass
 $5 \leq m_{\mu\mu} \leq 80 \; \gevcc$. The data sets were collected with
 the CDF~II detector at the Fermilab Tevatron collider, and correspond to
 integrated luminosities up to 2100 pb$^{-1}$. Similar data samples have
 been previously used by the CDF and D\O~collaborations to derive measurements
 of the correlated $\sigma_{b\rightarrow\mu,\bar{b}\rightarrow \mu}$ cross
 section that are inconsistent with the NLO theoretical prediction.
 A similar data set was used by the CDF collaboration to extract a value
 of $\bar{\chi}$, the average time-integrated mixing probability of
 $b$-flavored hadrons, that is appreciably larger than that reported
 by the LEP experiments. This analysis extends a recent study~\cite{bbxs}
 by the CDF collaboration which has used a dimuon data sample to
 re-measure the correlated $\sigma_{b\rightarrow\mu,\bar{b}\rightarrow \mu}$
 cross section. In Ref.~\cite{bbxs}, the value of 
 $\sigma_{b\rightarrow\mu,\bar{b}\rightarrow \mu}$ is measured using the
 sample composition as determined by fitting the impact parameter distribution
 of these primary muons with the expected shapes from all known sources. 
 The data are well described by contributions from the following QCD
 processes: semileptonic heavy flavor decays, prompt quarkonia decays,
 Drell-Yan production, and instrumental backgrounds from hadrons mimicking
 the muon signal. Reference~\cite{bbxs} reports 
 $\sigma_{b\rightarrow\mu,\bar{b}\rightarrow \mu}= 1549 \pm 133$ pb
 for muons with $p_T \geq 3 \; \gevc$ and $|\eta| \leq 0.7$.
 That result is in good agreement with the NLO prediction as well as
 with analogous measurements that identify $b$ quarks via secondary vertex
 identification~\cite{ajets,shears}. The study in Ref.~\cite{bbxs}
 uses a subset of dimuon events in which each muon track is reconstructed
 in the SVX with hits in the two inner layers and in at least four of the
 inner six layers. These tight SVX requirements select events in which both
 muons originate within 1.5 cm from the nominal beam line. According to the
 simulation, approximately 96\% of the dimuon events contributed by known
 QCD processes satisfy this condition. This study reports the presence of
 a much larger than expected sample of events, referred to as ghost events,
 that does not satisfy this condition. This component was present in previous 
 $\sigma_{b\rightarrow\mu,\bar{b}\rightarrow \mu}$~\cite{d0b2,2mucdf} and 
 $\bar{\chi}$~\cite{bmix} measurements in which this decay-radius requirement
 was not made. When applying the tight SVX criteria to initial muons,
 the invariant mass spectrum of combinations of an initial muon with an
 additional accompanying muon is well described by known QCD sources and 
 is dominated by sequential semileptonic heavy flavor decays. In contrast,
 without any SVX requirement the invariant mass spectrum is not well modeled
 by the QCD simulation and the inconsistencies at low invariant mass
 reported in Ref.~\cite{dilb} are reproduced. Our study shows that ghost
 events offer a plausible resolution to these long-standing inconsistencies
 related to $b\bar{b}$ production and decay. A large portion of these events
 is due to muons arising from in-flight-decays of pions and kaons or 
 punchthrough of hadronic prongs of $K^0_S$ and hyperon decays. However, 
 a significant fraction of these events has features that cannot be explained
 with our present understanding of the CDF~II detector, trigger and event
 reconstruction. The nature of these events is characterized by the 
 following properties. Impact parameters of initial muons are distributed
 differently from those of QCD events. After subtracting the contribution
 of hadrons mimicking a muon signal, an angular cone of 36.8$^{\deg}$ around
 the direction of an initial muon contains a rate of additional muon
 candidates that is approximately four times larger than that of cascade
 semileptonic decays of $b$ quarks. In contrast with sequential semileptonic
 decays of $b$ hadrons, initial and additional muon candidates have the same or
 opposite charge with equal probability. The impact parameter distribution
 of additional muon candidates, as well as that of secondary vertices
 reconstructed using tracks contained in a 36.8$^{\deg}$, have shapes
 different from what is expected if they were produced by known long-lived
 particles. The average number of tracks contained in a 36.8$^{\deg}$ cone
 is also two times larger than that of QCD events. We have verified these
 findings using stricter analysis selections and several control samples of
 data. We are continuing detailed studies with longer timescales for
 completion to better understand the cause of these effects.
%%%%%%%%%%%%%%%%%%%%%%%%%%%%%%%%%%%%%%%%%%%
 \section{Acknowledgments}
%%%%%%%%%%%%%%%%%%%%%%%%%%%%%%%%%%%%%%%%%%%
  We thank the Fermilab staff and the technical staffs of the participating
  institutions for their vital contributions. This work was supported by the
  U.S. Department of Energy and National Science Foundation; 
  the Italian Istituto Nazionale di Fisica Nucleare; the Ministry of Education,
  Culture, Sports, Science and Technology of Japan;  the National Science Council of the
  Republic of China; the Swiss National Science Foundation; the A.P. Sloan 
  Foundation; 
  %the Bundesministerium f\"ur Bildung und Forschung, Germany; 
  the Korean Science and Engineering Foundation and the Korean Research 
  Foundation; the Science and Technology Facilities Council and the Royal
  Society, UK;  the Institut National de Physique Nucleaire et Physique
  des Particules/CNRS; the Russian Foundation for Basic Research; 
  the Ministerio de Ciencia e Innovaci\'{o}n, Spain; the European Community's
  Human Potential Programme; the Slovak R\&D Agency;
  and the Academy of Finland.
%%%%%%%%%%%%%%%%%%%%%%%%%%%%%
\appendix
%%%%%%%%%%%
\section{Detector level distributions in QCD and ghost events}
%%%%%%%%%%%
 This appendix presents a few of many detector-level distributions that
 have been investigated looking for pathologies in track reconstruction,
 muon reconstruction, detector response, and in the observed properties
 of the ghost events. The assumption is that detector and pattern 
 recognition failures are not an issue if detector-level distributions
 for ghost events are similar to those for QCD events, which in turn
 are correctly modeled by a simulation based on the {\sc herwig} and 
 {\sc geant} Monte Carlo programs. 
%%%%%%%%%%%
\subsection{Quality of reconstructed tracks}
\begin{figure}
 \begin{center}
 \vspace{-0.3in}
 \leavevmode
 \includegraphics*[width=\textwidth]{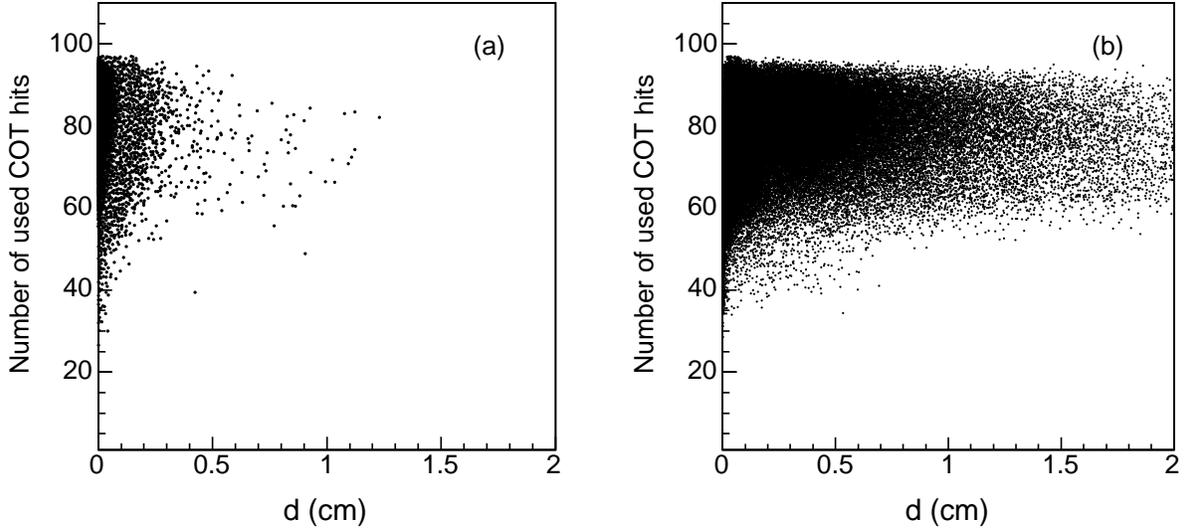}
 \caption[]{Number of COT hits associated with initial muon tracks
            as a function of the track impact parameter for (a) QCD
            and (b) ghost events.} 
 \label{fig:fig_nip}
 \end{center}
 \end{figure}

%%%%%%%%%%%
 A visual investigation of the display of reconstructed muon tracks and  
 associated COT and SVX hits has not shown any indication of detector or
 track-reconstruction program failures. COT tracks reconstructed using 
 hits in at least 20 COT layers are considered well measured tracks and 
 are used in most CDF analyses. Figure~\ref{fig:fig_nip} shows the number
 of COT hits used to reconstruct initial muon tracks as a function of
 the track impact parameter. In both QCD and ghost events, muon tracks
 are associated with an average of 75 hits, and the average number of
 associated hits does not depend on the impact parameter value.
 %%%%%%%%%%%%%%%%%%%%%%%%%%
%%%%%%%%%%%%%%%%%%%%%%%%%

\begin{figure}
 \begin{center}
 \vspace{-0.3in}
 \leavevmode
 \includegraphics*[width=\textwidth]{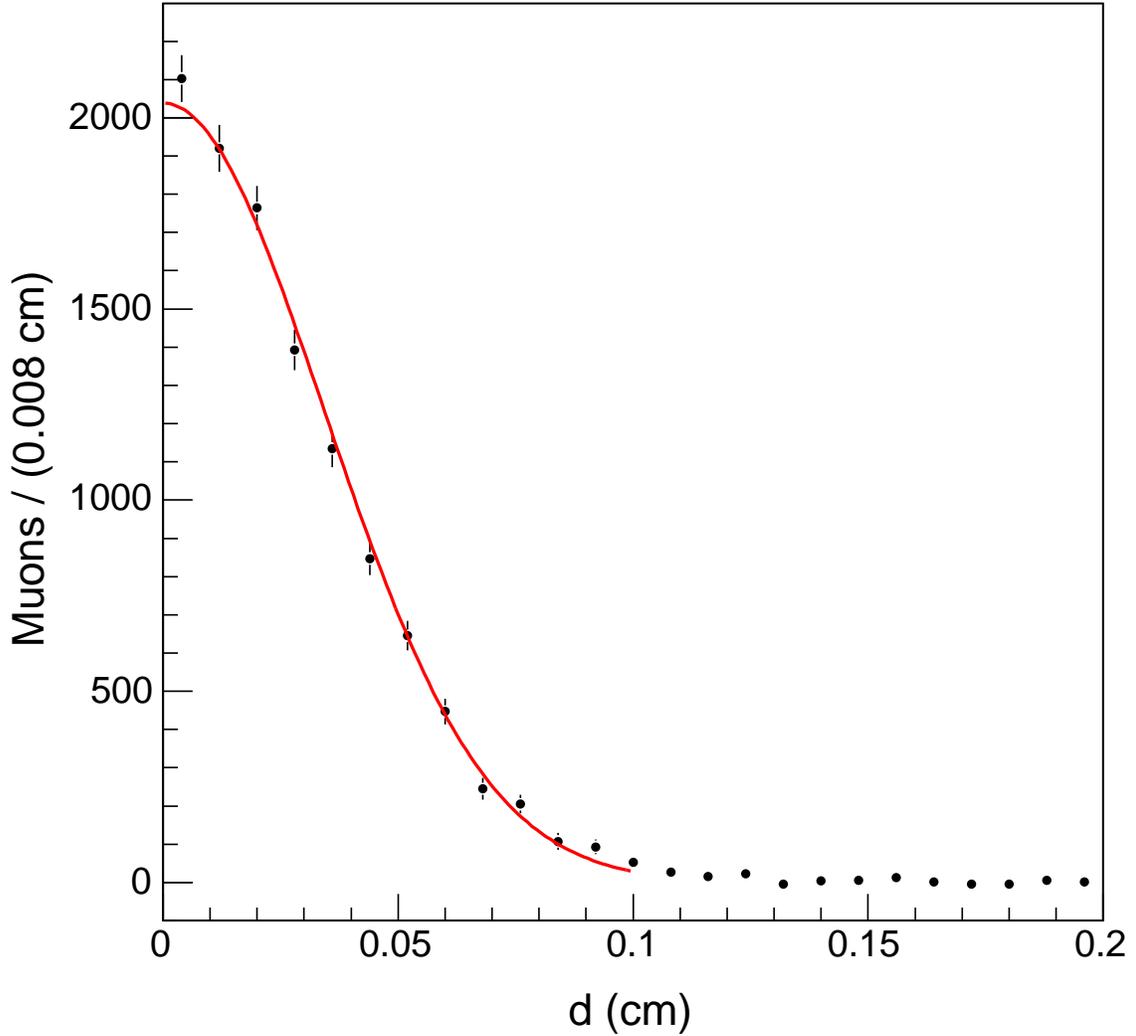}
 \caption[]{Impact parameter distribution of tracks corresponding to muons
            from $\Upsilon$ decays. Tracks are not associated with silicon
            hits. The combinatorial background under the $\Upsilon$ signal
            has been removed with a sideband subtraction technique. The 
            solid line is a  fit to the data with a Gaussian function.} 
 \label{fig:fig_yd0}
 \end{center}
 \end{figure}

 As also shown by cosmic muons in Fig.~\ref{fig:fig_cosm}~(b), the impact
 parameter of COT tracks associated with at least three silicon hits
 is measured with a rms resolution of approximately $30\;\mu$m~\cite{bbxs}.
 We have studied the impact parameter resolution of COT tracks without
 silicon hits, which populate ghost but not QCD events, by using muons
 from $\Upsilon$ decays included in our data sample. The impact parameter
 distribution is shown in Fig.~\ref{fig:fig_yd0}. The rms resolution is
 approximately 230 $\mu$m, and the impact parameter distribution is
 exhausted beyond 0.15 cm. Therefore, the large impact parameter tail
 characteristic of muons in ghost events is not due to tracks reconstructed
 without silicon hits. We have studied a large sample of $K^0_S$ mesons 
 reconstructed in the dimuon sample by using COT tracks with and without
 silicon hits, and with small or very large impact parameters (see
 Fig.~\ref{fig:fig_k0s}). The observed $L_{xy}$ distribution is correctly
 modeled by the value of the $K^0_S$ lifetime~\cite{pdg}. As shown in 
 Figs.~\ref{fig:fig_40}, \ref{fig:fig_15bis}, and~\ref{fig:fig_15tris},
 initial muons in ghost events are not accompanied by $D^0$ mesons and
 muons in events acquired with the request of a $D^0$ meson do not exhibit
 any large impact parameter tail. It is therefore unlikely that a significant
 fraction of ghost events arises from detector or pattern recognition
 failures in standard QCD events.
 %%%%%%%%%%%%%%%%%%%%%%%%%%
%%%%%%%%%%%%%%%%%%%%%%%%%
 \subsection{Quality of reconstructed muons}
\begin{figure}
 \begin{center}
 \vspace{-0.3in}
 \leavevmode
 \includegraphics*[width=\textwidth]{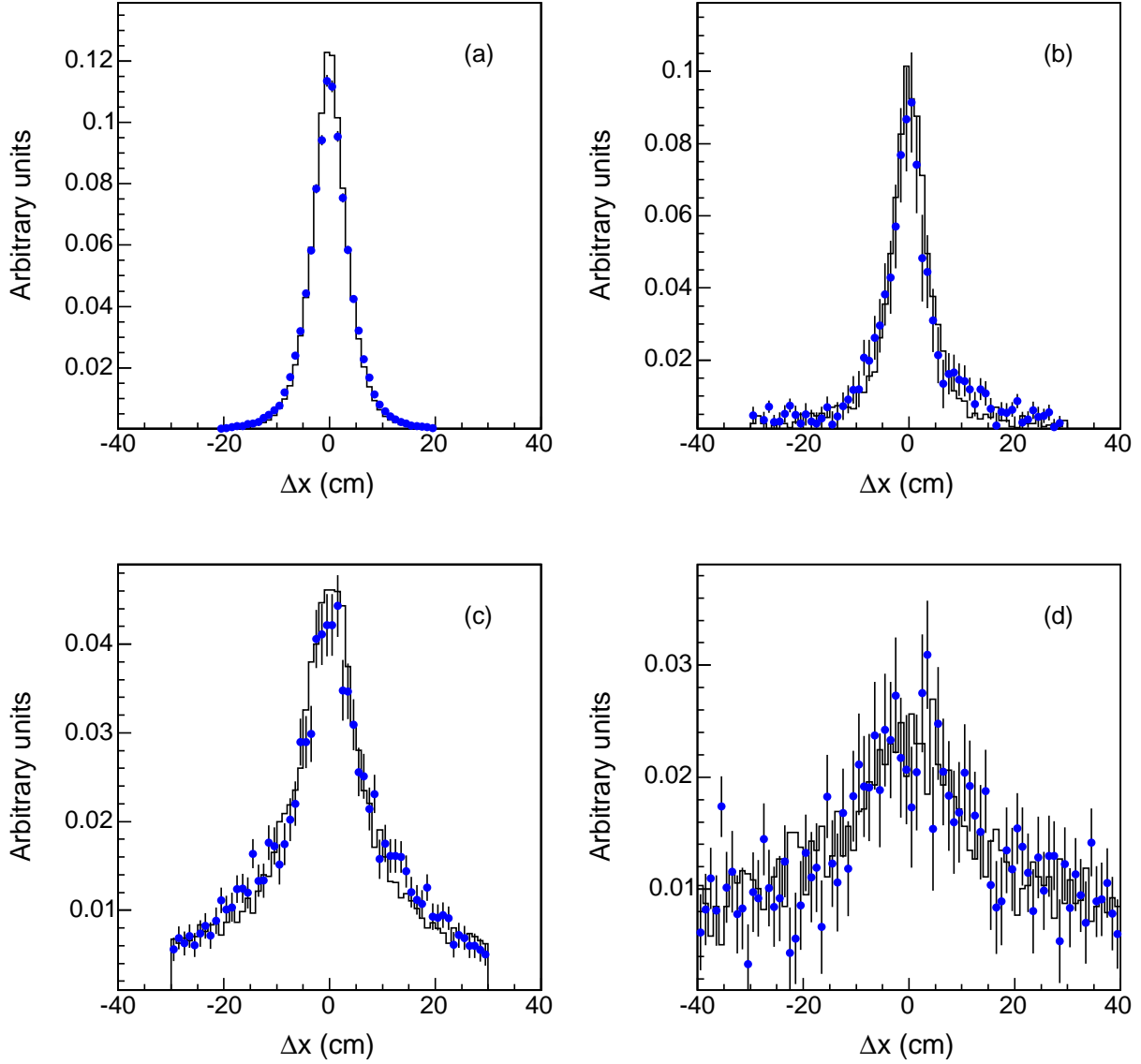}
 \caption[]{Distributions of $\Delta x$ (see text) for (a) initial and (b)
            additional CMUP muons, and additional (c) CMU or (d) CMP muons
            in QCD (histogram) and ghost ($\bullet$) events.}
 \label{fig:fig_dx}
 \end{center}
 \end{figure}

\begin{figure}
 \begin{center}
 \vspace{-0.3in}
 \leavevmode
 \includegraphics*[width=\textwidth]{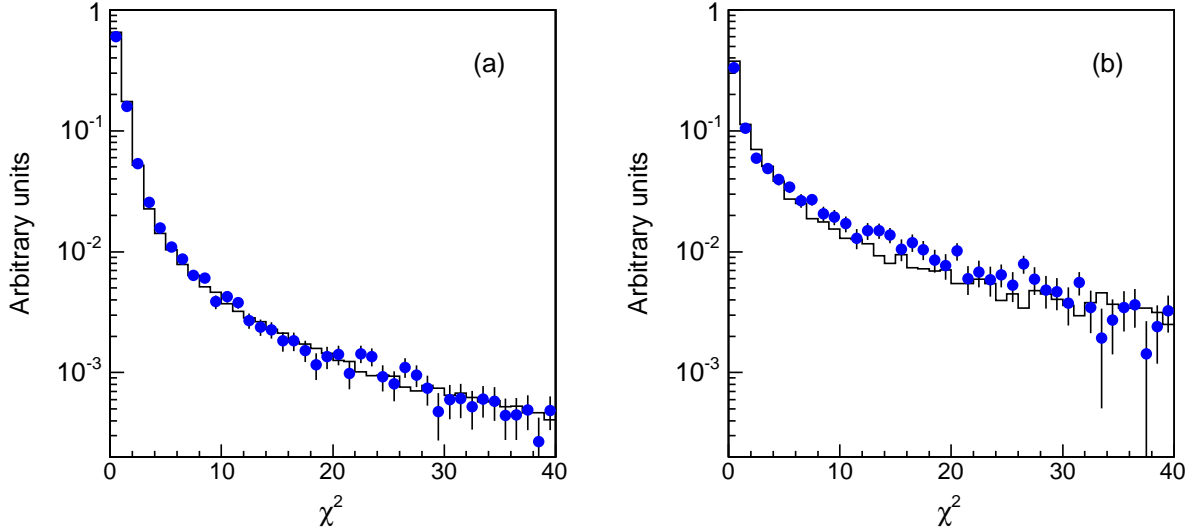}
 \caption[]{Distributions of $\chi^2$ (see text) for (a) initial and (b)
            additional muons in QCD (histogram) and ghost($\bullet$) events.}
 \label{fig:fig_chi2}
 \end{center}
 \end{figure}

 A track is accepted as a muon if the $r-\phi$ distance between its 
 projection onto a muon detector and a muon stub is $\Delta x\leq$ 30,
 40, and 30 cm for the CMU, CMP, and CMX detector, respectively. For
 CMX or CMU muons, we also construct the quantity
 $\chi^2= (\Delta x /\sigma)^2$, where $\sigma$ is a rms deviation that
 includes the effect of muon multiple scattering and energy loss. These
 quantities are compared in Figs.~\ref{fig:fig_dx} and~\ref{fig:fig_chi2}
 for initial and additional muons in QCD and ghost events.
 %%%%%%%%%%%%%%%%%%%%%%%%%%
%%%%%%%%%%%% 
%%%%%%%%%%%%%%%%%%%%%%%%%%
%%%%%%%%%%%%%%%%%%%%%%%%%%%%%%%%%%%%%%
 Table~\ref{tab:tab_a1} shows the fraction of additional muons identified 
 by the different detectors in QCD and ghost events.
%%%%%%%%%%%%%%%%%%%%%%%%%
 \begin{table}
 \caption[]{Fractional contributions (\%) to additional muons of different
            detectors in QCD and ghost events.}
 \begin{center}
 \begin{ruledtabular}
 \begin{tabular}{lcccc}
  Sample &  CMUP          &  CMU           &     CMP         &   CMX   \\
   QCD   & $17.0 \pm 0.4$ & $53.0 \pm 0.7$ &  $26.0 \pm 0.5$ & $4.0 \pm 0.2$ \\
   Ghost & $14.0 \pm 0.8$ & $60.0 \pm 1.4$ &  $24   \pm 1$   & $2.0 \pm 0.4$ \\
 \end{tabular}
 \end{ruledtabular}
 \end{center}
 \label{tab:tab_a1}
 \end{table} 
%%%%%%%%%%%%%%%%%%%%%%%%%% 
 These matching distributions, as well as the fractional usage of
 different muon detectors, in ghost events are not significantly 
 different to those of QCD events. Since we are able to predict 
 the rate of additional muons in QCD events, the response of the
 muon detector is an unlikely candidate to explain the large excess
 of additional muons in ghost events.

 The $\Delta x$ distributions for CMU and CMP muons in 
 Fig.~\ref{fig:fig_dx} show a significant quasi-flat contribution due
 to random track-stub matches under the Gaussian signal of real muons.
 This contribution is negligible for CMUP muons. These features are
 consistent with the fake muon prediction based on the fake probability
 per track derived using the decay products of $D^0$ mesons. For CMUP muons,
 the fake probability has been verified using the data in Ref.~\cite{bbxs}. 
 Ref.~\cite{bbxs} estimates the fraction of dimuons due to heavy flavor
 production that are faked by hadrons from heavy flavor decays in two
 complementary ways. This fraction is estimated by applying the fake
 probability per track to simulated hadrons from heavy flavor decays. 
 This fraction is also estimated by simultaneously fitting the impact
 parameter distributions of dimuon events selected with loose and tight
 $\chi^2$ requirements, and therefore containing different fractions of
 fake muons. The fit result shows that the fraction of fake muons is
 negligible, and slightly overestimated by the fake probability prediction.
 This conclusion is also supported by the fact that, when using initial
 CMUP muons no wrong-sign $\mu D^0$ candidates are observed in
 Fig.~\ref{fig:fig_40}. The fit to the muon impact parameters in
 Ref.~\cite{bbxs} yields the rate of dimuons due to $b\bar{b}$ and
 $b g$ production ($BB$ and $BP$ component in Table~\ref{tab:tab_1},
 respectively). In the latter case, the muon signal is mimicked by a
 prompt hadron in the gluon jet. The ratio of these components returned
 by the fit is $0.194 \pm 0.013$. When applying the fake probability per
 track to simulated $b g$ events normalized to the observed $b \bar{b}$ 
 cross section, Ref.~\cite{bbxs} predicts this ratio to be $0.21 \pm 0.01$.
 These comparisons show that the fake CMUP probability per track cannot be
 underestimated by more than 10\%. Since the rate of fake CMUP muons
 predicted in Table~\ref{tab:tab_7bis} is approximately 4\% of the signal,
 it seems unlikely that the additional CMUP signal in ghost events can be
 explained by an underestimate of the fake rate. In turn, this supports
 the main findings of our study that uses all muon detectors since they
 are consistent with the result based on CMUP muons only. As mentioned
 in Sec.~\ref{sec:ss-fkrob}, the multi-muon signal in ghost events is not
 affected by selections based on stricter track-stub matching, whereas the
 fake probability per track models correctly the number of additional
 muons observed in events in which one initial muon is mimicked by the
 hadronic leg of a $K^0_S$ decay. 
%%%%%%%%%%%%%%%%%%%%%%%%%%%%%%%%%%%%%%%%%%%%%%%%%%%%  
\section{Additional data distributions}
\begin{figure}
 \begin{center}
 \vspace{-0.3in}
 \leavevmode
 \includegraphics*[width=\textwidth]{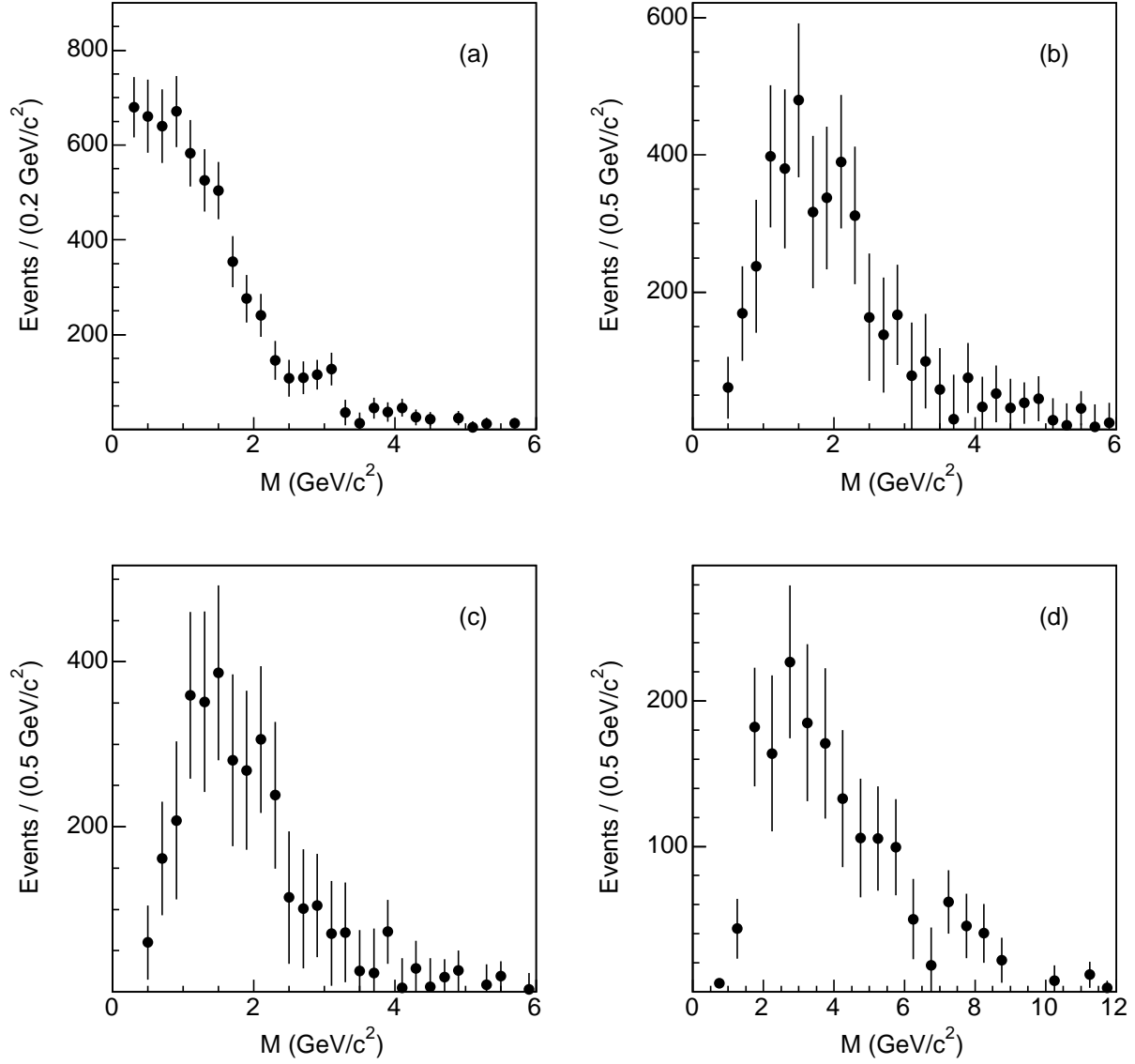}
 \caption[]{Invariant mass, $M$, distributions of all muons in a $36.8^{\deg}$
            cone when (a) both cones contain at least two muons, (b) a cone
            contains three or more muons, (c) a cone contains three muons,
            and (d) of muons and tracks for cones containing 5 to 6 tracks
            and three or more muons. QCD and fake muon contributions have been
            subtracted.} 
 \label{fig:fig_18}
 \end{center}
 \end{figure}

 Figure~\ref{fig:fig_18} shows the invariant mass distribution of several combinations of
 muon and tracks with $p_T \geq 2 \; \gevc$ and contained in a $36.8^{\deg}$
 cone around an initial muon.
%%%%%%%%%%%%%%%%%%%%%%%%%%
%%%%%%%%%%%%%%%%%%%%%%%%% 

\begin{figure}
 \begin{center}
 \vspace{-0.3in}
 \leavevmode
 \includegraphics*[width=\textwidth]{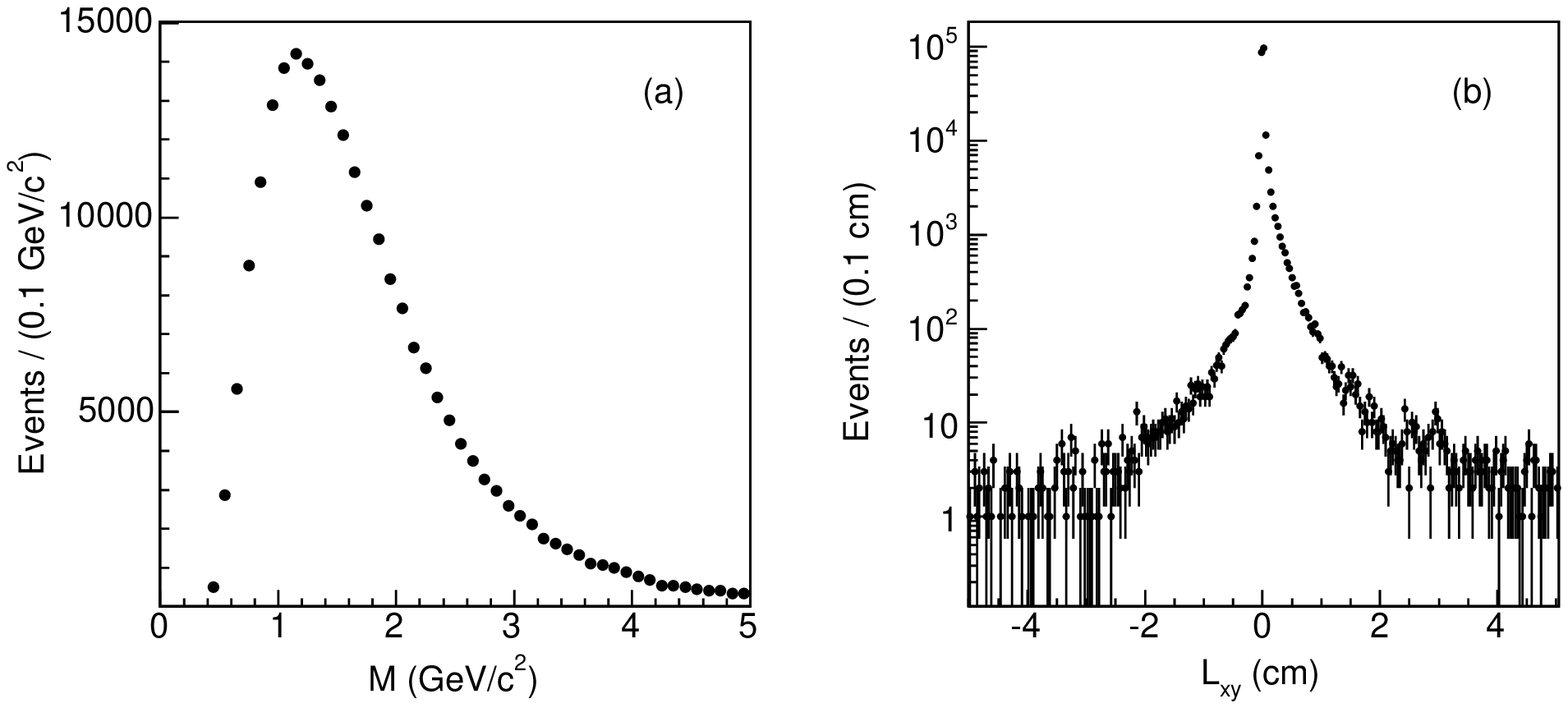}
 \caption[]{Distributions of  (a) the invariant mass and (b) the distance
            $L_{xy}$ of three-track systems in QCD events. } 
 \label{fig:fig_34}
 \end{center}
 \end{figure}

\begin{figure}
 \begin{center}
 \vspace{-0.3in}
 \leavevmode
 \includegraphics*[width=\textwidth]{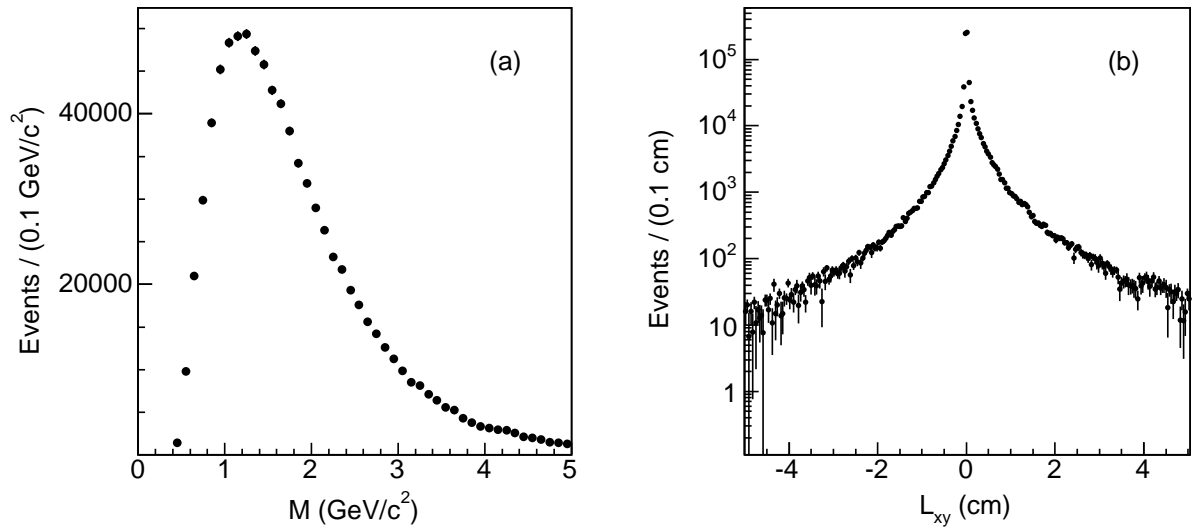}
 \caption[]{Distributions of (a) the  invariant mass and (b) the distance 
            $L_{xy}$ of three-track systems in ghost events. } 
 \label{fig:fig_35}
 \end{center}
 \end{figure}

 Figures~\ref{fig:fig_34} and~\ref{fig:fig_35} show the $L_{xy}$
 distributions of three-track systems in ghost and QCD events,
 respectively. We search for tracks with $p_T \geq 1.0 \; \gevc$ and
 $|\eta| \leq 1.1$ in a $36.8^{\deg}$ cone around the direction of 
 each initial muon. Track systems with total charge of $\pm 1$ are 
 constrained to arise from a common space point. Three-track combinations 
 are discarded if the three-dimensional vertex fit returns a $\chi^2$
 divided by three degree of freedom larger than five.
%%%%%%%%%%%%%%%%%%%%%%%%%%
%%%%%%%%%%%%%%%%%%%%%%%%%  
%%%%%%%%%%%%%%%%%%%%%%%%%%
%%%%%%%%%%%%%%%%%%%%%%%%%% 

\begin{figure}
 \begin{center}
 \vspace{-0.3in}
 \leavevmode
 \includegraphics*[width=\textwidth]{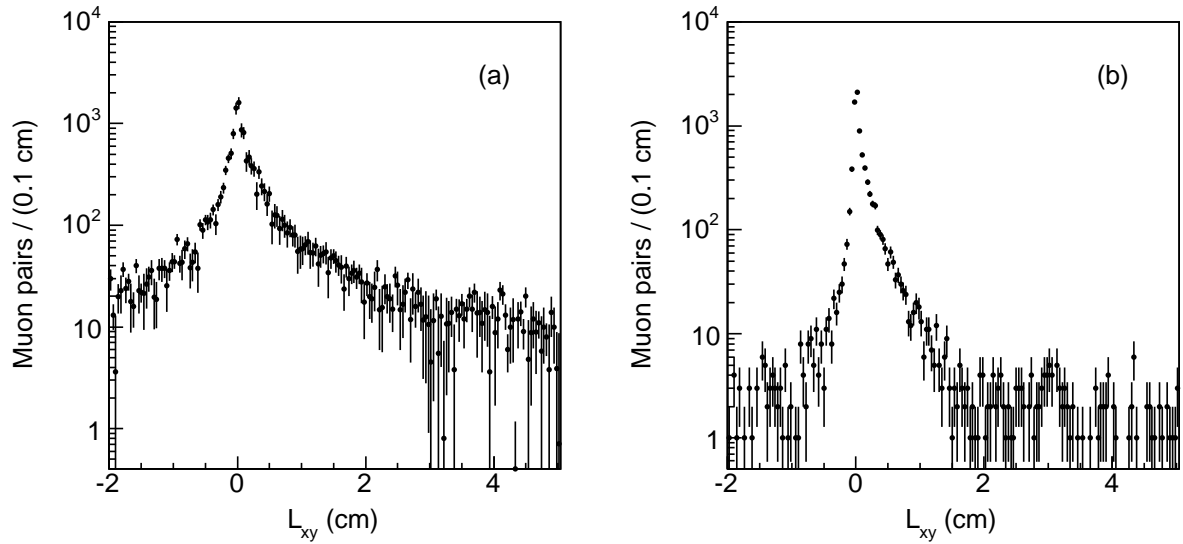}
 \caption[]{Distribution of the distance $L_{xy}$ of the fit-constrained
            vertices of muon pairs contained in a $36.8^{\deg}$ cone for
            (a) ghost and (b) QCD events.} 
 \label{fig:fig_36}
 \end{center}
 \end{figure}

 Similar distributions can be constructed by pairing initial muons with any
 additional muon contained in a $36.8^{\deg}$ cone. The two-track systems are 
 constrained to arise from a common space point, and combinations are 
 discarded if the three-dimensional vertex fit returns a $\chi^2$  larger
 than 10. The $L_{xy}$ distributions for ghost and QCD events are shown in
 Fig.~\ref{fig:fig_36}.
%%%%%%%%%%%%%%%%%%%%%%%%%%
%%%%%%%%%%%%%%%%%%%%%%%%%%%%%%%%%%%%%%%%%%%%%%%

\begin{figure}
 \begin{center}
 \vspace{-0.3in}
 \leavevmode
 \includegraphics*[width=\textwidth]{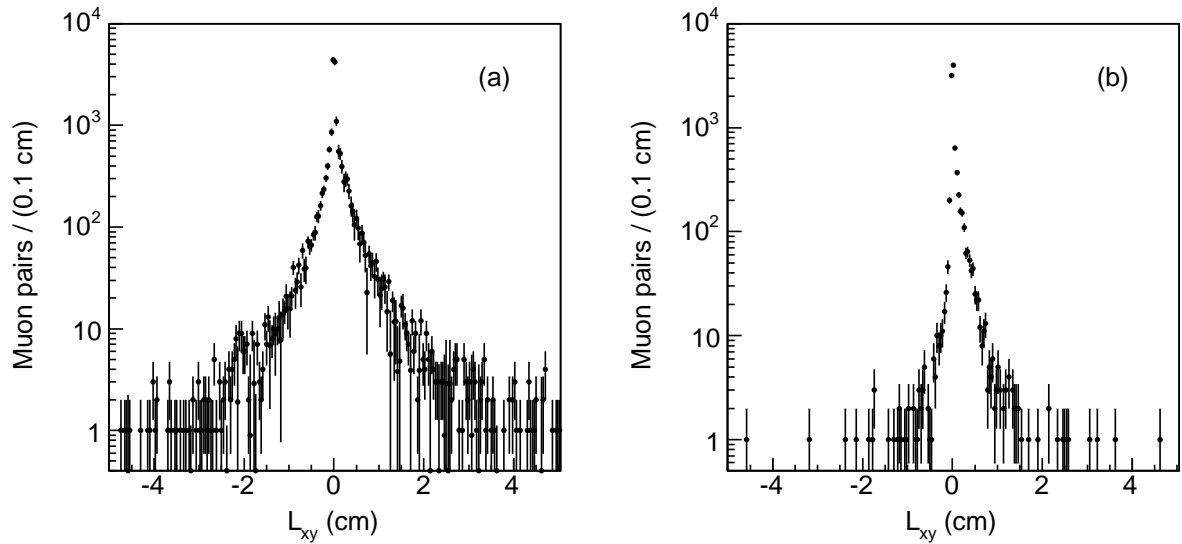}
 \caption[]{Distribution of the distance  $L_{xy}$ of fit-constrained vertices
            of three-track systems contained in a $36.8^{\deg}$ cone around  
            the direction of an initial muon for (a) ghost and (b) QCD events.
            We select cases in which angular cones contain only three tracks.}
 \label{fig:fig_37}
 \end{center}
 \end{figure}

\begin{figure}
 \begin{center}
 \vspace{-0.3in}
 \leavevmode
 \includegraphics*[width=\textwidth]{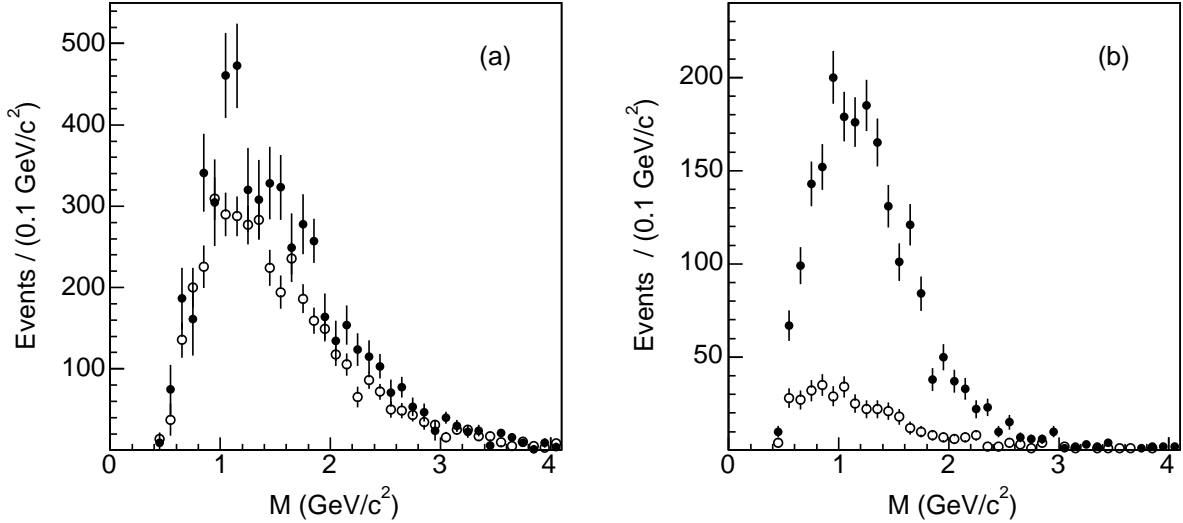}
 \caption[]{Distributions of the invariant mass, $M$, of three-track systems
            in (a) ghost and (b) QCD events. Systems with distance 
            $L_{xy} \geq 0.04$ cm ($\bullet$) are compared to those with 
            $L_{xy} \leq -0.04$ cm ($\circ$).}
 \label{fig:fig_38}
 \end{center}
 \end{figure}

 We also select events in which a cone around the direction of an initial
 muon contains only three tracks with $p_T \geq 1 \; \gevc$. Three-track
 systems with total charge of $\pm 1$ are constrained to arise from a
 common space point. Three-track combinations are discarded if the
 three-dimensional vertex fit returns a $\chi^2$ divided by three degree
 of freedom larger than five. Figure~\ref{fig:fig_37} shows the resulting $L_{xy}$
 distribution for ghost and QCD events. Figure~\ref{fig:fig_38} compares
 the invariant mass distribution  of the three-track systems for positive and negative
 $L_{xy}$ values.
%%%%%%%%%%%%%%%%%%%%%%%%%%
%%%%%%%%%%%%%%%%%%%%%%%%%%
%%%%%%%%%%%%%%%%%%%%%%%%%%
%%%%%%%%%%%%%%%%%%%%%%%%
%%%%%%%%%%%%%%%%%%%%%%%%%%%%%%%

%%%%%%%%%%%%%%%%%%%%%%%%%%%%%%%
%%%%%%%%%%%%%%%%%%%%%%%%%%%%%%%
 \end{document}